\documentclass[12pt,a4paper]{article}

\usepackage{ifthen} 
\newboolean{articletitles}
\setboolean{articletitles}{true} 
\newboolean{uprightparticles}
\setboolean{uprightparticles}{false} 
\newboolean{inbibliography}
\setboolean{inbibliography}{false} 

\usepackage{microtype}
\usepackage{lineno}  
\usepackage{xspace} 

\usepackage{graphicx}  
\usepackage{color}
\usepackage{colortbl}
\graphicspath{{./figs/}} 

\usepackage{amsmath} 
\usepackage{amssymb}
\usepackage{amsfonts}
\usepackage{mathrsfs,bm,url,color}
\usepackage{upgreek} 
\usepackage{blindtext} 

\newcommand*\patchAmsMathEnvironmentForLineno[1]{%
\expandafter\let\csname old#1\expandafter\endcsname\csname #1\endcsname
\expandafter\let\csname oldend#1\expandafter\endcsname\csname
end#1\endcsname
 \renewenvironment{#1}%
   {\linenomath\csname old#1\endcsname}%
   {\csname oldend#1\endcsname\endlinenomath}%
}
\newcommand*\patchBothAmsMathEnvironmentsForLineno[1]{%
  \patchAmsMathEnvironmentForLineno{#1}%
  \patchAmsMathEnvironmentForLineno{#1*}%
}
\AtBeginDocument{%
\patchBothAmsMathEnvironmentsForLineno{equation}%
\patchBothAmsMathEnvironmentsForLineno{align}%
\patchBothAmsMathEnvironmentsForLineno{flalign}%
\patchBothAmsMathEnvironmentsForLineno{alignat}%
\patchBothAmsMathEnvironmentsForLineno{gather}%
\patchBothAmsMathEnvironmentsForLineno{multline}%
}

\usepackage{hyperref}    
\usepackage[all]{hypcap} 

\def\ux85 {\mbox{UX85}\xspace}

\ifthenelse{\boolean{uprightparticles}}%
{

 \def\PDelta      {\ensuremath{\Delta}\xspace}                 
 \def\PXi      {\ensuremath{\Xi}\xspace}                 
 \def\PLambda      {\ensuremath{\Lambda}\xspace}                 
 \def\PSigma      {\ensuremath{\Sigma}\xspace}                 
 \def\POmega      {\ensuremath{\Omega}\xspace}                 
 \def\PUpsilon      {\ensuremath{\Upsilon}\xspace}                 
 
 \def\PB      {\ensuremath{\mathrm{B}}\xspace}                 
                  
 \def\PD      {\ensuremath{\mathrm{D}}\xspace}

 \def\PK      {\ensuremath{\mathrm{K}}\xspace}

 \def\Pi      {\ensuremath{\mathrm{i}}\xspace}

}
{

 \mathchardef\PDelta="7101
 \mathchardef\PXi="7104
 \mathchardef\PLambda="7103
 \mathchardef\PSigma="7106
 \mathchardef\POmega="710A
 \mathchardef\PUpsilon="7107
                  
 \def\PB      {\ensuremath{B}\xspace}                 
                  
 \def\PD      {\ensuremath{D}\xspace}

 \def\PK      {\ensuremath{K}\xspace}

 \def\Pi      {\ensuremath{i}\xspace}

}

\def\kaon  {\ensuremath{\PK}\xspace}
  \def\Kbar  {\kern 0.2em\overline{\kern -0.2em \PK}{}\xspace}

\def\Kz    {\ensuremath{\kaon^0}\xspace}
\def\Kzb   {\ensuremath{\Kbar^0}\xspace}
\def\KzKzb {\ensuremath{\Kz \kern -0.16em \Kzb}\xspace}
\def\Kp    {\ensuremath{\kaon^+}\xspace}
\def\Km    {\ensuremath{\kaon^-}\xspace}

\def\KpKm  {\ensuremath{\Kp \kern -0.16em \Km}\xspace}

\def\Dbar    {\kern 0.2em\overline{\kern -0.2em \PD}{}\xspace}
\def\D       {\ensuremath{\PD}\xspace}

\def\Dz      {\ensuremath{\D^0}\xspace}
\def\Dzb     {\ensuremath{\Dbar^0}\xspace}
\def\DzDzb   {\ensuremath{\Dz {\kern -0.16em \Dzb}}\xspace}
\def\Dp      {\ensuremath{\D^+}\xspace}
\def\Dm      {\ensuremath{\D^-}\xspace}

\def\DpDm    {\ensuremath{\Dp {\kern -0.16em \Dm}}\xspace}

\def\Bbar    {\ensuremath{\kern 0.18em\overline{\kern -0.18em \PB}{}}\xspace}

  \def\Y#1S{\ensuremath{\PUpsilon{(#1S)}}\xspace}

\def\Lbar {\ensuremath{\kern 0.1em\overline{\kern -0.1em\PLambda}}\xspace}

\def\to                 {\ensuremath{\rightarrow}\xspace}

\def\AT#1     {\ensuremath{A_{\mathrm{T}}^{#1}}\xspace}           

\def\C#1      {\ensuremath{\mathcal{C}_{#1}}\xspace}                       
\def\Cp#1     {\ensuremath{\mathcal{C}_{#1}^{'}}\xspace}                    
\def\Ceff#1   {\ensuremath{\mathcal{C}_{#1}^{\mathrm{(eff)}}}\xspace}        
\def\Cpeff#1  {\ensuremath{\mathcal{C}_{#1}^{'\mathrm{(eff)}}}\xspace}       
\def\Ope#1    {\ensuremath{\mathcal{O}_{#1}}\xspace}                       
\def\Opep#1   {\ensuremath{\mathcal{O}_{#1}^{'}}\xspace}                    

\newcommand{\tev}{\ensuremath{\mathrm{\,Te\kern -0.1em V}}\xspace}
\newcommand{\gev}{\ensuremath{\mathrm{\,Ge\kern -0.1em V}}\xspace}
\newcommand{\mev}{\ensuremath{\mathrm{\,Me\kern -0.1em V}}\xspace}
\newcommand{\kev}{\ensuremath{\mathrm{\,ke\kern -0.1em V}}\xspace}
\newcommand{\ev}{\ensuremath{\mathrm{\,e\kern -0.1em V}}\xspace}
\newcommand{\gevc}{\ensuremath{{\mathrm{\,Ge\kern -0.1em V\!/}c}}\xspace}
\newcommand{\mevc}{\ensuremath{{\mathrm{\,Me\kern -0.1em V\!/}c}}\xspace}
\newcommand{\gevcc}{\ensuremath{{\mathrm{\,Ge\kern -0.1em V\!/}c^2}}\xspace}
\newcommand{\gevgevcccc}{\ensuremath{{\mathrm{\,Ge\kern -0.1em V^2\!/}c^4}}\xspace}
\newcommand{\mevcc}{\ensuremath{{\mathrm{\,Me\kern -0.1em V\!/}c^2}}\xspace}

\def\invfb   {\ensuremath{\mbox{\,fb}^{-1}}\xspace}
\def\invab   {\ensuremath{\mbox{\,ab}^{-1}}\xspace}

\def\gsim{{~\raise.15em\hbox{$>$}\kern-.85em
          \lower.35em\hbox{$\sim$}~}\xspace}
\def\lsim{{~\raise.15em\hbox{$<$}\kern-.85em
          \lower.35em\hbox{$\sim$}~}\xspace}

\def\tell1  {TELL1\xspace}
\def\ukl1   {UKL1\xspace}

\def\invfb   {\ensuremath{\mbox{\,fb}^{-1}}\xspace}
\def\invab   {\ensuremath{\mbox{\,ab}^{-1}}\xspace}

\def\DeltaE     {\Delta {\rm{E}}}

\def\GeV{\ifmmode{\mathrm{Ge\kern -0.1em V}}\else
                  \textrm{Ge\kern -0.1em V}\fi}
\def\MeV{\ifmmode{\mathrm{Me\kern -0.1em V}}\else
                  \textrm{Me\kern -0.1em V}\fi} 

\usepackage{cite} 
\usepackage{mciteplus}
\usepackage{overpic}
\usepackage{subfigure}

\usepackage[noabbrev,capitalise]{cleveref}

\usepackage{floatrow}

\newfloatcommand{capbtabbox}{table}[][\FBwidth]

\usepackage{caption}
\usepackage{longtable}
\usepackage{array}
\usepackage{amsmath}
\usepackage{amsbsy}
\usepackage{bm}
\usepackage{xltabular}
\usepackage{ifpdf}

\newcommand{\re}[2][()] {\ifthenelse{\equal{#1}{()}}{{\ensuremath{{\rm \, Re}}\left(#2\right)}}
                                                    {{\ensuremath{{\rm \, Re}}\left[#2\right]}}}
\newcommand{\im}[2][()] {\ifthenelse{\equal{#1}{()}}{{\ensuremath{{\rm \, Im}}\left(#2\right)}}
                                                    {{\ensuremath{{\rm \, Im}}\left[#2\right]}}}

\definecolor{orange}{rgb}{1,0.5,0}

\usepackage{booktabs}
\usepackage{rotating}
\usepackage{multirow}
\usepackage{mathtools}

\newcommand\snowmass{\begin{center}\rule[-0.2in]{\textwidth}{0.01in}\\\rule{\textwidth}{0.01in}\\
\vskip 0.1in Submitted to the  Proceedings of the US Community Study\\ 
on the Future of Particle Physics (Snowmass 2021)\\ 
\rule{\textwidth}{0.01in}\\\rule[+0.2in]{\textwidth}{0.01in} \end{center}}

\begin{document}

\renewcommand{\thefootnote}{\fnsymbol{footnote}}
\setcounter{footnote}{1}
\begin{titlepage}


\snowmass

{\bf\boldmath\huge
\begin{center}
Snowmass 2021 White Paper:
Charged lepton flavor violation \\
in the tau sector
\end{center}
}

\begin{center}

Swagato Banerjee$^{1}$\footnote{Corresponding author: swagato.banerjee@louisville.edu},
Vincenzo Cirigliano$^{2,3}$\footnote{Corresponding author: cirigv@uw.edu},
Mogens Dam$^{4}$,
Abhay Deshpande$^{5,6,7}$,
Luca Fiorini$^{8}$,
Kaori Fuyuto$^{2}$,
Ciprian Gal$^{5,9}$,
Tom\'a\v{s} Husek$^{10}$,
Emanuele Mereghetti$^{2}$,
Kevin Mons\'alvez-Pozo$^{8}$,
Haiping Peng$^{11}$,
Francesco Polci$^{12}$,
Jorge Portol\'es$^{8}$,
Armine Rostomyan$^{13}$,
Michel Hern\'andez Villanueva$^{13}$,
Bin Yan$^{2}$,
Jinlong Zhang$^{14}$,
Xiaorong Zhou$^{11}$
 
\bigskip

{\it\footnotesize

$^1$ University of Louisville, Louisville KY 40292\\
$^2$ Theoretical Division, Los Alamos National Laboratory, Los Alamos, NM 87545\\
$^3$ Institute for Nuclear Theory, University of Washington, Seattle WA 98195-1550\\
$^4$ Niels Bohr Institute, Copenhagen University, Copenhagen, Denmark\\
$^{5}$ Center for Frontiers in Nuclear Science, Stony Brook University, NY 11764\\
$^{6}$ Stony Brook University, Stony Brook, NY 11794-3800\\
$^{7}$ Brookhaven National Laboratory, Upton, NY 11973-5000\\
$^{8}$ Instituto de Física Corpuscular (IFIC), Universidad de Valencia -- CSIC, Valencia, Spain \\
$^{9}$ Mississippi State University, MS 39762\\
$^{10}$ Department of Astronomy and Theoretical Physics, Lund University, Lund, Sweden\\
$^{11}$ University of Science and Technology of China, Hefei 230026, China \\
$^{12}$ LPNHE, Sorbonne Université, Paris Diderot Sorbonne Paris Cité, CNRS/IN2P3, Paris, France\\
$^{13}$ Deutsches Elektronen--Synchrotron, Hamburg 22607, Germany\\
$^{14}$ Shandong University, Qingdao, Shandong, China\\
}
\end{center}

\vspace{\fill}

\begin{abstract}
\noindent Charged lepton flavor violation has long been recognized as unambiguous signature of New Physics. Here we describe the physics capabilities and discovery potential of New Physics models with charged lepton flavor violation in the tau sector as its experimental signature. Current experimental status from the B-Factory experiments BaBar, Belle and Belle II, and future prospects at Super Tau Charm Factory, LHC, EIC and FCC-ee experiments to discover New Physics via charged lepton flavor violation in the $\tau$ sector are discussed in detail.

\end{abstract}

\vspace{\fill}

\end{titlepage}

\pagestyle{empty}  


\renewcommand{\thefootnote}{\arabic{footnote}}
\setcounter{footnote}{0}
\tableofcontents
\cleardoublepage
\pagestyle{plain} 
\setcounter{page}{1}
\pagenumbering{arabic}


\graphicspath{{figs/}}
\allowdisplaybreaks

\section{Executive summary}

The discovery of charged lepton flavor violation (CLFV) will be an unambiguous manifestation of physics beyond the Standard Model (SM), with the potential to shed light on unsolved problems in the SM, first and foremost the origin of neutrino masses. 
CLFV is thus an area of intense experimental and theoretical activity.

\begin{figure}[!h]
\centering
\includegraphics[width=\textwidth]{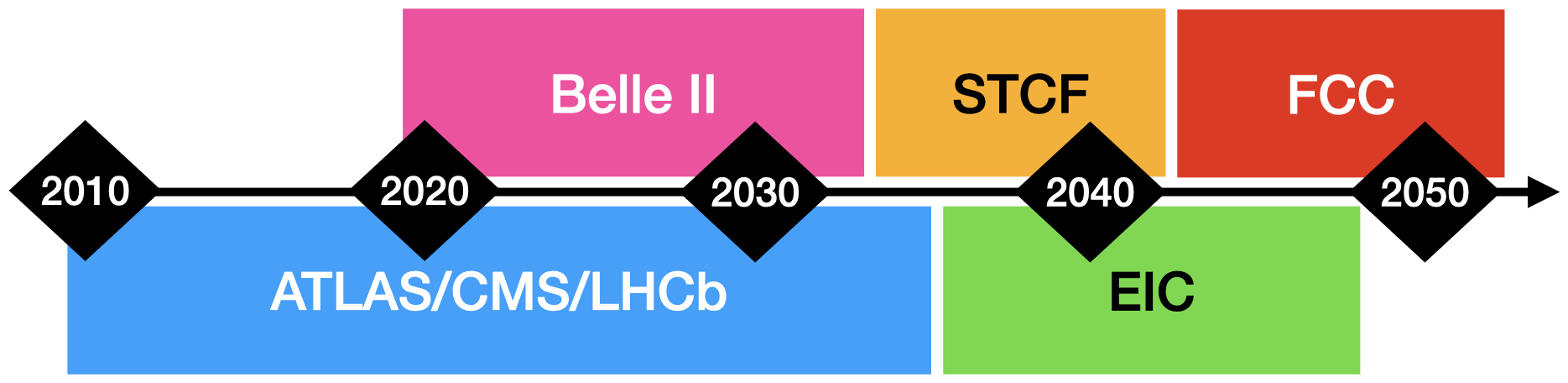}
\caption{Tentative timeline for data-taking at different experiments probing CLFV in the $\tau$ sector.}
\label{fig:exptimeline}
\end{figure}

Focusing on the $\tau$ sector,  the experimental landscape will undergo tremendous progress in the next ten years, with Belle II working towards its 50 ab$^{-1}$ goal, with the LHC collecting 300 fb$^{-1}$ of data in Run 3 and starting its high luminosity runs, and with the EIC coming online. 
On a longer time scale, the Super $\tau$-Charm Facility (STCF), the Electron-Ion Collider (EIC) and the Future Circular Collider (FCC) will also play a major role. 
A very approximate timeline for data-taking at different experiments searches for CLFV in the $\tau$ sector is shown in Figure~\ref{fig:exptimeline}.

All these experiments 
will be sensitive to CLFV predicted in many BSM models, from supersymmetric scenarios to leptoquarks, and  
offer complementary probes of CLFV at different energy scales, 
crucial to identify the underlying sources of LFV and the underlying mediation mechanism. 

\section{Introduction}

Charged lepton flavor violating (CLFV)  processes have long been recognized as very powerful tools to search for  
new physics beyond the Standard Model (BSM)  for a number of reasons: 
(i) the observation of CLFV  at experiments in the foreseeable future  would immediately point to new physics beyond the minimal extension of the SM that only 
includes neutrino mass (so-called $\nu$SM).  This is  because in the  $\nu$SM, CLFV amplitudes are proportional to $(m_\nu/m_W)^2$~\cite{Petcov:1976ff,Marciano:1977wx,Lee:1977qz,Lee:1977tib}, where $m_\nu$ and $m_W$ are the masses of neutrinos and W boson, respectively,  leading to rates forty orders of magnitude below current sensitivity; 
(ii)   current and future CLFV  experiments probe new mediator particles with 
masses that can be well above the scales directly accessible at  high-energy colliders 
(see for example supersymmetric scenarios~\cite{Lee:1984kr,Lee:1984tn,Borzumati:1986qx,Barbieri:1995tw}), 
in certain cases reaching the PeV scale~\cite{Altmannshofer:2013lfa}; 
(iii) CLFV processes probe an accidental symmetry of the Standard Model (corresponding to lepton family number) and 
therefore  play a special role in probing  models of neutrino mass generation. 
Examples of  studies of the  correlations between minimal neutrino mass models and   CLFV processes  
can be found in Refs.~\cite{Abada:2008ea,Abada:2007ux,Alonso:2012ji,Cirigliano:2005ck}. 
There is a vast literature on the subject and for reviews we refer the reader to 
Refs.~\cite{Raidal:2008jk,deGouvea:2013zba,Bernstein:2013hba,Calibbi:2017uvl}.  

CLFV can be probed by a number of processes, spanning many energy scales. 
At low-energy  one has the decays of the   $\mu$ and $\tau$ leptons and 
decays of the $B$ and $K$ mesons and quarkonia, 
which can be probed at a number of experiments. 
At  high-energy one has searches for SM-forbidden events  
such as $p p \to \ell_\alpha \bar \ell_\beta + X$ 
(where $\ell_{\alpha, \beta} = e^-, \mu^-, \tau^-$ and $X$ denotes other final state particles) at the Large Hadron Collider (LHC) 
or  $ e p \to \ell + X$ at 
fixed target experiment such as NA64 
or electron-hadron colliders such as HERA and the future Electron-Ion Collider (EIC) and
Large Hadron-Electron Collider (LHeC). 
Currently, the most stringent limits on CLFV in the $\mu  \leftrightarrow  e$, $\tau \leftrightarrow \mu$, and $\tau \leftrightarrow e$ sector 
come from low-energy searches such as the decays $\mu \to e \gamma$, $\tau \to \mu \gamma$, etc. 
Within low-energy processes,  
the strongest constraints are in the $\mu \leftrightarrow e$ sector, 
with branching ratios at the level of $10^{-13}$, e.g.\  ${\rm BR} (\mu^+ \to  e^+ \gamma)  < 4.2 \times 10^{-13}$ at 90\% CL~\cite{TheMEG:2016wtm}.  
The constraints on $\tau \leftrightarrow e$ 
transitions are a few orders of magnitude weaker but still impressive, e.g.\ 
${\rm BR} (\tau^\pm \to  e^\pm Y)  <  {\rm few}  \times 10^{-8}$~\cite{Tanabashi:2018oca}, with $Y \in \{ \gamma, \pi \pi, ...\}$, and crucial to understand the origin of lepton flavor. 
As outlined below, in the next decade great progress is expected 
in the CLFV $\tau$ decay sensitivity at Belle II. 

The multiplicity of probes is essential to infer information on  (i) the underlying sources of lepton family violation {\it and} (ii) the underlying mediation mechanism. 
The first problem is explored by studying CLFV transitions among different families 
($\mu  \leftrightarrow  e$, $\tau \leftrightarrow \mu$, and $\tau \leftrightarrow e$), 
while the second is best probed by studying different CLFV processes within the same two families (e.g.\ 
$\mu \to e \gamma$ vs $\mu \to e$ conversion or $\mu \to 3 e$ or 
$\tau \to e \gamma$ vs $\tau \to e \pi \pi$, etc.). 
Therefore, various CLFV probes are highly complementary and should be vigorously pursued. 
In what follows, we will focus on the $\tau \leftrightarrow \mu$ and $\tau \leftrightarrow e$ transitions, 
first describing the theory framework and then  discussing the experimental status and prospects.

\section{Theoretical Overview}
To  assess the impact of CLFV searches across various energy scales,   the most efficient theoretical framework is provided by the   
Standard Model Effective Field Theory (SMEFT)
\cite{Weinberg:1979sa,Wilczek:1979hc,Buchmuller:1985jz,Grzadkowski:2010es,Jenkins:2013zja,Jenkins:2013wua,Alonso:2013hga,Crivellin:2013hpa}, 
which captures new potential sources of CLFV  above the electroweak scale 
 $v = 1/(\sqrt{2} G_\text{F})^{1/2}  \simeq 246$~GeV  in a model-independent way.
SMEFT encodes  new physics originating at energies higher than $v$   
in  operators of dimension greater than four  built out of SM fields,  suppressed by inverse powers of heavy scale $\Lambda$ 
\begin{equation}
 {\cal L}_{\rm eff} = {\cal L}_{\rm SM}   +  \sum_{n,\ D \geq 5}  \frac{C_n^{(D)}}{ \Lambda^{D-4} } \, O_n^{(D)} ~.
 \label{eq:SMEFT}
 \end{equation}
The scale $\Lambda$ represents generically the mass of the lowest-lying new particles appearing in the underlying new theory. 
The  Wilson Coefficients  $C_n^{(D)}$  encode information about the underlying model (couplings, ratio of masses, etc). 
If the underlying model is known, the Wilson coefficients can be calculated in terms of the model parameters. 
Therefore, the above effective Lagrangian describes the low-energy limit of any UV  extension of the SM. 
 The leading CLFV operators appear at dimension $D=6$ and therefore are suppressed by $1/\Lambda^2$. 

The SMEFT framework is  applicable to processes in which the center-of-mass energy is well below the expected scale of new physics. 
This means that $\tau$ and $B$-meson decays can be analyzed in this framework. 
Moreover, given the null results so far for new physics searches at LHC, the SMEFT is  applicable 
with minimal caveats to the analysis of LHC processes and with no caveats to an EIC with center-of-mass energy $\sqrt{S} < v \sim 200$~GeV. 
Therefore, the SMEFT  provides a common framework to assess the relative sensitivity,  
discovery potential and model diagnosing power of various CLFV probes, 
from lepton and meson decays  all the way to EIC and LHC processes. 

CLFV processes involving $\tau$ leptons have been studied in the recent literature~\cite{Husek:2021isa,Cirigliano:2021img,Antusch:2020vul,Husek:2020fru,Gninenko:2018num,Aaboud:2016hmk,Abada:2016vzu,Takeuchi:2017btl,Hazard:2016fnc,Celis:2014asa,Celis:2013xja,Petrov:2013vka,Daub:2012mu,Han:2010sa,Gonderinger:2010yn,Dassinger:2007ru,Matsuzaki:2007hh,Black:2002wh} 
both within specific models and in the SMEFT framework.
Within the SMEFT approach,  Refs.~\cite{Husek:2020fru,Cirigliano:2021img}  provide the most comprehensive study of all leading (dimension-six) CLFV operators, including heavy quark operators. 
Ref.~\cite{Cirigliano:2021img}   also considers the broadest set of CLFV processes: 
decays of  the $\tau$ lepton ($\tau \to e Y$) and $B$ meson ($B \to X \tau \bar{\ell}$) at B-factories and LHCb, 
the EIC process  $ep\to\tau X$, and the  $pp \to e \tau$ process  at the LHC. 
We will therefore use the results of  Ref.~\cite{Husek:2020fru,Cirigliano:2021img}  as the baseline for our analysis, updating and extending them as necessary. 

\section{Experimental Status}

\subsection{Belle II}
The first generation B-Factory experiments, BaBar,  at the PEP-II B-meson factory located at SLAC US, and Belle, at the KEKB accelerator in Tsukuba Japan, recorded a data sample with an integrated luminosity of about 0.5~ab$^{-1}$ and 1~ab$^{-1}$, respectively.  They shared characteristics such as large acceptance detectors with sophisticated particle identification systems, high vertex resolution, excellent calorimetry, and precise muon detectors with a high number of collisions, providing the ideal environment for precision measurements and searches of new Physics.  
The KEK accelerator was upgraded to SuperKEKB and the Belle detector to the next generation B-Factory experiment, Belle II, with the aim of increasing the integrated luminosity by a factor of 50 and improving the detector performance in the high luminosity environment. 

The $e^+e^-$ annihilation experiments at the B-factories also serve as $\tau$ factories, owing to the large production cross-section~\cite{Banerjee:2007is} of $\tau^-\tau^+$ pairs at center-of-mass energy $\sqrt{s}$ = 10.58 $\GeV$ at the $\Upsilon(4S)$ resonance, with a well-defined initial state up to radiative effects. 
Many models predict LFV in $\tau$ decays at $10^{-10}$--$10^{-8}$ levels, which will be probed by $e^-e^+ \to \tau^-\tau^+$ events at Belle II. 
In total, 52 LFV $\tau$ decay modes have been searched in B-Factory experiments, as listed in the Table~\ref{tab:taudecays} in Section~\ref{expsummary}. 
Belle and BaBar experiments also  searched for LFV processes in the decays of heavy particles such as B mesons, and $\Upsilon(nS)$ ($n=1-3$) resonances, into $\tau$ leptons (see Table~\ref{tab:taufinalstate} in Section~\ref{expsummary}). The Belle~II experiment will continue to improve the sensitivity of those searches further in the future. 

The LFV $\tau$ decay modes can be classified as neutrinoless 2-body or 3-body decays to final states containing:
\begin{itemize}
\item a light lepton and a photon: $\tau^- \to \ell^- \gamma$ with $\ell = e, \mu$ (2 modes);
\item a light lepton and a pseudoscalar meson: $\tau^- \to \ell^- P^0$ with $P^0 = \pi^0, \; K^0, \; \eta, \; \eta^\prime$ (8 modes); 
\item a light lepton and a scalar meson: $\tau^- \to \ell^- S^0$ with $S^0 = f_0(980)$ (2 modes);
\item a light lepton and a vector meson: $\tau^- \to \ell^- V^0$ with $V^0 = \rho, \; \omega, \; K^{\star 0}, \; \bar{K}^{\star 0}, \;  \phi$ (10 modes);
\item three light leptons: $\tau^- \to \ell^- \ell^+ \ell^-, \; \ell^- \ell'^+ \ell''^-$ and $\ell^+ \ell'^- \ell''^- $ (6 modes). 
\item a light lepton and two mesons: $\tau^- \to \ell^- h^+h^-, \;
\ell^+ h^- h^-$, \; $\ell^- h^0 h^0$ (16 modes);
\item a $\Lambda$ / $\bar{\Lambda}$: $\tau^- \to \pi^- \Lambda$ and $\tau^- \to \pi^- \bar{\Lambda}$ (2 modes);
\item a proton / anti-proton: $\tau^- \to \bar{p}^- \ell^+ \ell^-$ and $\tau^- \to p^+ \ell^- \ell^-$ (6 modes).
\end{itemize}
Several of these decay modes violate simultaneously the lepton number conservation~\cite{LopezCastro:2012udb} or baryon number conservation~\cite{Hou:2005iu}. 
Wrong sign decays, e.g.\ $\tau^-\to\ell_j^-\ell_j^-\ell_i^+$ decays are expected at rates only one order of magnitude below present bounds in some SM extensions~\cite{Pacheco:2021djh}.

Searches for these full set of measurements of LFV processes are necessary, because there are strong correlations between the expected rates of the different channels in various models. 
For example, in some supersymmetric seesaw model~\cite{Babu:2002et, Sher:2002ew}, the relative rates of 
${\cal{B}}(\tau^\pm \to \mu^\pm \gamma)$ :
${\cal{B}}(\tau^\pm \to \mu^\pm \mu^+\mu^-)$ :
${\cal{B}}(\tau^\pm \to \mu^\pm \eta)$
are predicted to have specific ratios, 
depending on the model parameters. 
In the unconstrained minimal supersymmetric model, which includes various correlations between the $\tau$ and $\mu$ LFV rates, the LFV branching fractions of the $\tau$ lepton can be as high as $10^{-8}$~\cite{Brignole:2004ah, Goto:2007ee}), while respecting the strong experimental bounds on LFV decays of the $\mu$ lepton. Thus, it is critical to probe all possible LFV modes of the $\tau$ lepton, because any excess in a single channel will not provide sufficient information to identify an underlying theory of the LFV mechanism. 
 
The characteristic feature of LFV decays is that both the energy and the mass of the $\tau$-daughters are known in $e^+e^-\to\tau^+\tau^-$ annihilation environment. An LFV decay candidate is reconstructed dividing the event into two hemispheres in the center-of-mass system (CMS), with reconstructed energy $\mathrm{E}_{\tau}$ of the signal $\tau$-daughters in CMS expected to be equal to $\sqrt{s}/2$ and the invariant mass to the mass of a $\tau$ lepton, $m_\tau$. 
The signal is then clustered around $(m_\tau,0)$ in the two dimensional plane of invariant mass versus the $\DeltaE$,
where ${\DeltaE} = \mathrm{E}_{\tau} - \sqrt{s}/2$. 
Signal yield is optimized using MC simulations to give the smallest expected upper limits on the branching fractions in the background-only hypothesis, with data-driven corrections from control channels. 

Over the decade of their operation, Belle and BaBar experiments improved the sensitivity of LFV $\tau$ decay modes by $\sim$2 orders of magnitude w.r.t.\ CLEO experiment at CESR $e^+e^-$ collider.  
Stringent bounds on LFV decays are set, the most recent result being the ones reported by Belle in search for the decays $\tau^- \to \ell^- \gamma$ ($\ell = e, \mu$). No significant excess over background predictions was observed and upper limits were set on LFV branching fractions ranging between $10^{-7}-10^{-8}$ at the 90\% confidence level.

 \begin{figure}[!h]
    \centering
    \includegraphics[width=\linewidth]{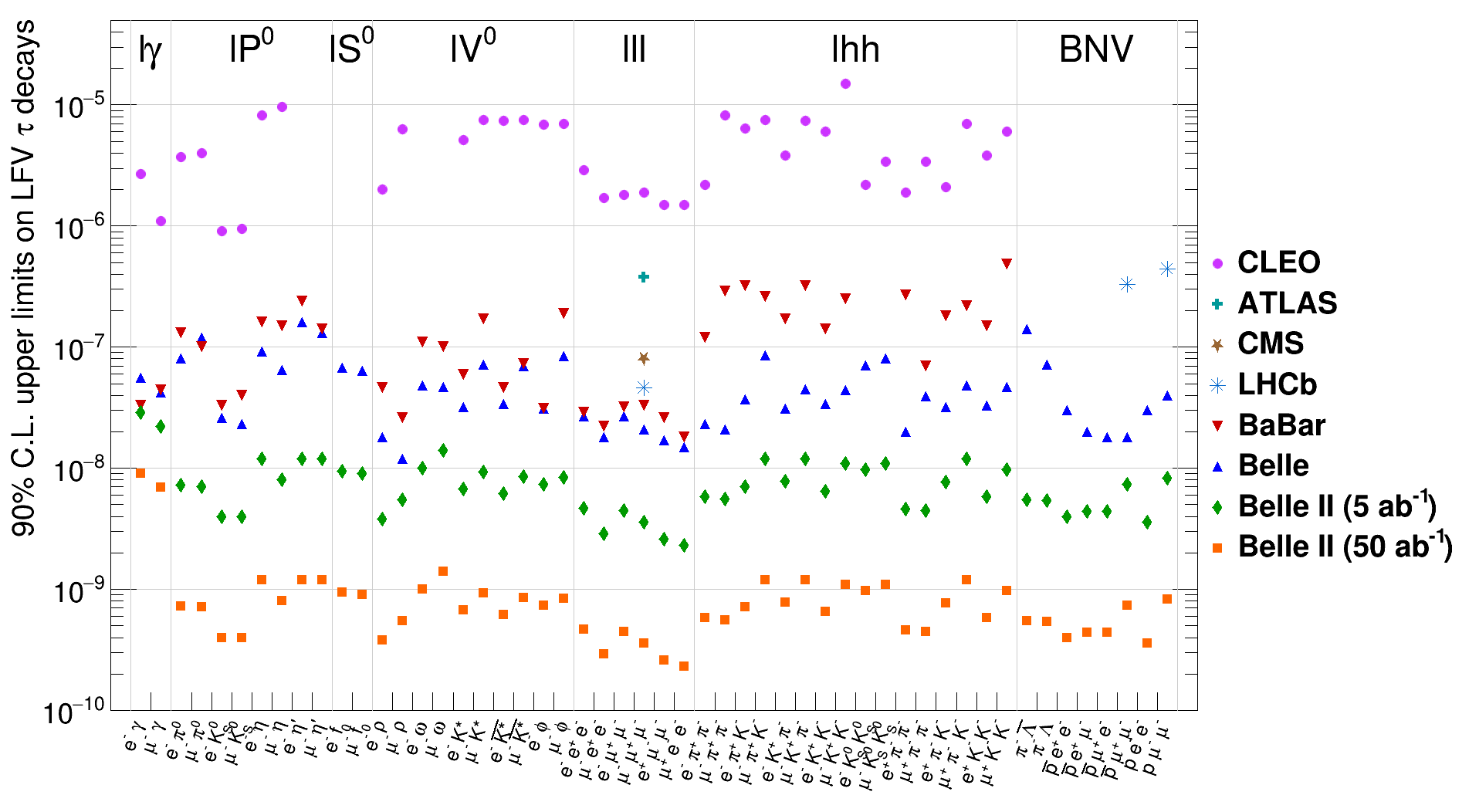} 
    \caption{Projection of expected upper limits at the Belle II experiment~\cite{Belle2WP} and current status of observed  upper limits at CLEO, BaBar, Belle, ATLAS, CMS and LHCb experiments~\cite{Amhis:2019ckw} on LFV, LNV and BNV processes in $\tau$ decays.}
    \label{fig:TauLFV}
\end{figure}

Current experimental status on the observed bounds on LFV in the 52 benchmark $\tau$ decay channels are shown in Figure~\ref{fig:TauLFV}.  Belle II will collect an immense amount of data from $e^+e^-$ annihilation at the upgraded SuperKEKB facility. This will be one of the factors pushing up the sensitivity of LFV probes at Belle II. Equally important is the increase of the signal detection efficiency which directly translates into enhancement in sensitivity. At Belle and BaBar, the signal efficiencies lied between $3\%$ and $12\%$ depending on the decay channel. At Belle II an increase in the signal efficiency will be achieved due to anticipated higher trigger efficiencies; improvements in the vertex reconstruction, charged track and neutral meson reconstructions, particle identification; as well as from a better understanding of the physics backgrounds and refinements in the analysis techniques.  

Projections for two illustrative scenarios of luminosity ${\cal{L}}$ = 5~\invab and 50~\invab for Belle II are shown in Figure~\ref{fig:TauLFV},
and listed in the Table~\ref{tab:taudecays} in Section~\ref{expsummary}.
The extrapolations are done from the expected limits obtained at the Belle experiment, assuming similar efficiencies of the individual channels. The presence of irreducible backgrounds for $\tau^- \to \ell^-\gamma$ decays is assumed, thus approximating the projection to be proportional to $1/\sqrt{\cal{L}}$, and the presence of accidental backgrounds only for all other channels, in which case the projection is proportional to $1/{\cal{L}}$, as discussed in~\cite{Raidal:2008jk}. Belle II limits will improve current bounds by more than two orders of magnitude in the next decade, probing LFV in $\tau$ decays down to a few parts in $10^{-9}-10^{-10}$~\cite{Belle-II:2018jsg, Belle2WP}.

\subsection{LHC}
\subsubsection{Experimental Status at LHCb }

At the LHC, $\tau$ leptons are produced almost entirely from the decays of $b$ and $c$ hadrons. Using the $b\bar{b}$ and $c\bar{c}$ cross-sections measured by LHCb~\cite{LHCb:2011zfl, LHCb:2013xam} and the inclusive $b \to  \tau$ and $c \to  \tau$ branching fractions~\cite{ParticleDataGroup:2014cgo}, the inclusive $\tau$  cross-section has been estimated to be 85 $\mu b$ at 7 TeV.

The  LHCb collaboration has taken advantage of this large cross section for producing the first limit on a search for LFV $\tau^- \to \mu^- \mu^+ \mu^-$  decays at a hadron collider, using a data sample corresponding to the first collected 1 fb$^{-1}$ from proton-proton collisions at a centre-of-mass energy of 7 TeV~\cite{LHCb:2013fsr}. This search profits also of the fact that muons provide an extremely clean signature for the trigger in LHCb.  This search was subsequently updated using new analysis techniques and  adding  2 fb$^{-1}$ of proton-proton collisions at a centre-of-mass energy of 8 TeV. No evidence has been found for a signal, and a limit has been set at 90$\%$ confidence level (CL) on the branching ratio: $B(\tau^- \to \mu^- \mu^+ \mu^-)  < 4.6 \times 10^{-8}$ ~\cite{LHCb:2014kws}.  This result is statistically dominated, and the collaboration is currently analysing an additional data sample of  6  fb$^{-1}$ already collected by the experiment. 

In addition to  the $\tau^- \to \mu^- \mu^+ \mu^-$ decays, LHCb has also searched for $\tau^- \to \bar{p} \mu^+ \mu^-$  and $\tau^- \to p \mu^- \mu^-$ decays, exploiting the  excellent proton identification  provided by its ring-imaging Cherenkov (RICH) detectors. These two additional decays,  searched for the first time at LHCb experiment, 
imply not only LFV but also BNV, with $|\Delta(B - L)|$ = 0.  With the first 1 fb$^{-1}$ of data collected at a centre-of-mass energy of 7 TeV,  the following limits were set: $B(\tau^- \to \bar{p} \mu^+ \mu^-) < 3.3 \times  10^{-7} $ and $ B(\tau^- \to p \mu^- \mu^-) < 4.4 \times 10^{-7}$ at 90$\%$ confidence level~\cite{LHCb:2013fsr}.

It should be noted that these limits are given for the phase-space model of $\tau$ decays. However, the physical processes that introduce LFV would affect the kinematic properties of the decay. Considering the approach in \cite{Dassinger:2007ru} of  a model-independent analysis of the decay distributions in an effective field-theory approach  including BSM operators with different chirality structures,  the observed limit on $B(\tau^- \to \mu^- \mu^+ \mu^-)$ varies within the range $(4.1 - 6.8) \times 10^{-8}$ at 90$\%$ CL, depending on the choice of operator. 
Current LHCb results for $\tau$ LFV searches are summarized in table~\ref{tab:taudecays}. 

LHCb has also searched for lepton flavor violating  $b$-hadrons decays in final states containing a muon and a $\tau$ lepton:  $B^0 \to \mu^\pm \tau^\mp$, $B_s^0 \to \mu^\pm \tau^\mp$ (using 3 fb$^{-1}$ of proton-proton collisions at a centre-of-mass energy of 7 and 8 TeV)~\cite{Aaij:2019okb} and $B^+ \to K^+ \mu^- \tau^+$ (using 9 fb$^{-1}$ of proton-proton collisions at a centre-of-mass energy of 7, 8 and 13 TeV)~\cite{LHCb:2020khb}. A search for the decays $H \to   \mu^\pm \tau^\mp$ has also been performed using 2 fb$^{-1}$ of proton-proton collisions at a centre-of-mass energy of 8 TeV~\cite{LHCb:2018ukt}, complementing the angular coverage of the analogous search done by  the ATLAS~\cite{ATLAS:2019pmk} and CMS~\cite{CMS:2021rsq} experiments.   The  upper limits set on these decays are shown in table~\ref{tab:taufinalstate}. 

\subsubsection{Experimental Status at ATLAS and CMS }

The main sources of $\tau$-leptons at LHC are decays of $D$ mesons ($>$ 70\%), $B$ mesons ($\sim$ 25\%)  and $W$ bosons ($\sim$ 0.01\%). Dedicated channels are employed for $\tau$ decays from heavy flavour (HF) mesons decays and from $W$ decays. The ATLAS and CMS collaboration have exploited their high luminosity interaction points and trigger rate to search for Lepton Flavor Violating (LFV) processes.

The CMS collaboration has performed a search for $\tau^- \to \mu^- \mu^+ \mu^-$ decay using 33.2~fb$^{-1}$ of Run 2 data at a center of mass energy of 13~TeV~\cite{cmst3m}. In the HF channel, the high rate of hadronic particles produced in proton-proton collisions is one of the main challenges. Dedicated online triggers are used for the event selection. A multivariate analysis based on a Boosted Decision Tree (BDT) is used to separate the signal from the background.
In the $W$ channel, the expected background is significantly lower than for the HF channel. The final state of this channel is characterized by isolated and high transverse momentum ($p_{\mathrm{T}}$) muons and large missing transverse momentum ($E^{\mathrm{miss}}_{\mathrm{T}}$).
A different BDT is used to separate the signal from the background and events are divided into two categories.
The branching fraction $\mathcal{B}(\tau^- \to \mu^- \mu^+ \mu^-)$ is extracted from a simultaneous unbinned maximum likelihood fit to the trimuon invariant mass distribution in the 1.6--2.0~GeV mass range of each category of the two channels.
The observed (expected) upper limit at 90\% confidence level (CL) on  $\mathcal{B}(\tau^- \to \mu^- \mu^+ \mu^-)$ using all events categories is $8.0\times 10^{-8}$
($6.9\times 10^{-8}$). Fitting the $W$ boson and HF channels separately returns observed (expected)
90\% CL upper limits of $20\times 10^{-8}$ ($13\times 10^{-8}$) and $9.2\times 10^{-8}$ ($10.0\times 10^{-8}$), respectively.

The ATLAS collaboration performed a search for $\tau^- \to \mu^- \mu^+ \mu^-$ decay using 20.3 fb$^{-1}$ of $pp$ collision data at a centre-of-mass energy of 8~TeV collected during Run 1 of the LHC~\cite{ATLAS:2016jts}. The search exploits the production of $\tau$ leptons via $W\to \tau \nu$ decays. The observed (expected) upper limit on the branching fraction $\mathcal{B}(\tau^- \to \mu^- \mu^+ \mu^-)$ is $3.76\times 10^{-7}$ ($3.94\times 10^{-7})$ at 90\% confidence level. \\

CMS has performed searches for LFV decays of the Higgs boson in the $H\to e\tau$ and $H\to\mu\tau$ final states. The searches involve  decays with a $\tau$-lepton and are performed  with $pp$ collisions at a centre-of-mass energy of 13 TeV and corresponding to a total integrated luminosity of $137\,\mathrm{fb}^{-1}$~\cite{CMS:2021rsq}. 
For each of the two searches, the events are separated in two channels depending on whether the $\tau$ decay includes a charged lepton ($e$ or $\mu$) or not. Events of each channel are further divided into four categories: one VBF category focusing on the Higgs boson production via the vector boson fusion and three categories based on the jets multiplicity (0-jet, 1-jet and 2-jet).
No significant excess above the expected background from SM processes is observed, hence upper limits on the LFV branching fractions are set for a Higgs boson with $m_H =125$\,GeV. The observed (median expected) 95\% confidence level (CL) upper limits are $0.22\%$ ($0.16\,\%$) and $0.15\%$
($0.15\,\%$) for the $H\to e\tau$ and $H\to\mu\tau$ searches, respectively.
The upper limits are computed assuming $\mathcal{B}(H\to\mu\tau) = 0$ for the $H\to e\tau$ search and $\mathcal{B}(H\to e\tau) = 0$ for the $H\to\mu\tau$ search. \\

ATLAS has performed searches for LFV decays of the Higgs boson in the following final states $H\to e\tau$ and $H\to\mu\tau$. The two searches are performed   with the $pp$ collisions corresponding to a total integrated luminosity of $36.1\,\mathrm{fb}^{-1}$  at a centre-of-mass energy of 13 TeV~\cite{ATLAS:2019pmk}. 
For each of the $H\to e\tau$ and $H\to\mu\tau$ searches, the events are separated into two channels depending on whether the $\tau$ decay include a charged lepton ($e$ or $\mu$) or not. Events are further divided into a VBF category, focusing on the Higgs boson production via the vector boson fusion and a non-VBF category. The VBF selection is   based on the kinematics of the two jets with the highest $p_\mathrm{T}$, where $\mathrm{j}_1$ and $\mathrm{j}_2$ denote the leading and subleading jet in $p_\mathrm{T}$, respectively.
The non-VBF category contains events failing the VBF selection, but still passing further selection criteria described in Ref.~\cite{ATLAS:2019pmk}.
No significant excess above the
expected background from SM processes is observed and upper limits
on the LFV branching fractions are set for a Higgs boson with $m_H =
125$\,GeV. The observed (median expected) 95\% CL upper limits are
$0.47\%$ ($0.34^{+0.13}_{-0.10}\,\%$) and $0.28\%$
($0.37^{+0.14}_{-0.10}\,\%$) for the $H\to e\tau$ and $H\to\mu\tau$ searches, respectively.
The upper limits are computed assuming $\mathcal{B}(H\to\mu\tau) = 0$ for the $H\to e\tau$ search and $\mathcal{B}(H\to e\tau) = 0$ for the $H\to\mu\tau$ search. \\

ATLAS performed searches for $Z\to e\tau$ and $Z\to \mu\tau$  with $139\,\mathrm{fb}^{-1}$ of $pp$ collisions at a centre-of-mass energy of 13 TeV~\cite{EXOT-2020-28,EXOT-2018-36}. The searches are performed independently for the case that the $\tau$ decay includes hadrons or not and then they are combined. Different $\tau$ polarization hypothesis are considered (unpolarized, left-handed $\tau$ and right-handed $\tau$). The polarization of the $\tau$-lepton affects the energy of its visible decay products and thus the acceptance for signal events. In the scenario where the $\tau$-leptons are unpolarized, results are combined with a previous analysis of Run 1 data\cite{HIGG-2015-09} and the observed upper limits at 95\% CL on $\mathcal{B}(Z\to e\tau)$ and $\mathcal{B}(Z\to \mu\tau)$ are $5.0\times 10^{-6}$ and $6.5\times 10^{-6}$, respectively.
These results supersede the limits from the Large Electron–Positron Collider experiments conducted more than two decades ago.

\section{Future Prospects}

\subsection{STCF}
A Super $\tau$-Charm Facility~(STCF)~\cite{STCF} is a symmetric double ring electron-positron collider designed to operate at c.m.~energies between $\sqrt{s}=2\sim7$~GeV, at a peak luminosity of $0.5\times10^{35}$~cm$^{-2}$s$^{-1}$ or higher.
The proposed STCF would leave space for higher luminosity upgrades and for the implementation of a polarized electron beam in a phase-II project~\cite{Luo:2019gri}.
It is expected to deliver more than 1~ab$^{-1}$ data per year that brings about $3.5\times10^{9}$
$\tau^{-}\tau^{+}$ pairs at $\sqrt{s}=4.26$~GeV.
At the production threshold there could be as many as $10^{8}$ $\tau^{-}\tau^{+}$ events per year. 
The $\tau^{-}\tau^{+}$ pairs produced near-threshold enables a better control of systematic uncertainties by using data just below the threshold. These near-threshold $\tau$ pairs are primarily in an S-wave, and thus can be longitudinally polarized similar in magnitude to that of the incident electron beam. 

The STCF detector is designed to meet the stringent performance requirements for physics. It includes a nearly $4\pi$ solid angle coverage for both charged and neutral particles; excellent momentum and angular resolution for charged particles, with $\sigma_{p}/p=0.5$\% at $p=1$~GeV/c;
high resolution of energy and position reconstruction for photons, with 
    $\sigma_{E}/E\approx2.5\%$ and $\sigma_{\rm pos}\approx 5$~mm at $E_{\gamma}=1$~GeV; superior PID ability and high detection efficiency for low momentum particles; and tolerance to high background environment.

The sensitivity study of two benchmark CLFV processes, $\tau^{-}\to\mu^{-}\mu^{+}\mu^{-}$ and $\tau^{-}\to\mu^{-}\gamma$, are performed with
$\tau$ produced in $e^{-}e^{+}\to\tau^{-}\tau^{+}$ at $\sqrt{s}=4.26$~GeV~\cite{STCF:clfv}, utilizing a fast simulation software package that can model the STCF detector responses and optimize it in turn~\cite{Shi:2020nrf}.
For $\tau^{-}\to\mu^{-}\mu^{+}\mu^{-}$, the single-tag $\tau^{+}$s are reconstructed 
from $\tau^{+}\to l^{+}\nu_{l}\bar{\nu}_{\tau}$~$(l=e,\mu)$ and $\tau^{+}\to\pi^{+}\bar{\nu}_{\tau}+n\pi^{0}$~$(n=0,1,2...)$. 
It is almost background-free for $\tau^{-}\to\mu^{-}\mu^{+}\mu^{-}$ after selection and
the energy and mass constraints of signal side are used to estimate the sensitivities. 
With $3.5\times10^{9}$ $\tau$ pairs collected one year at STCF, the 
sensitivity of $\tau^{-}\to\mu^{-}\mu^{+}\mu^{-}$ is estimated to be of $1.4\times10^{-9}$ at 90\% CL.
For $\tau^{-}\to\mu^{-}\gamma$, the single-tag $\tau^{+}$s are reconstructed from
$\tau^{+}\to e^{+}\nu_{e}\bar{\nu}_{\tau}$, $\tau^{+}\to\pi^{+}\bar{\nu}_{\tau}$ 
and $\tau^{+}\to\pi^{+}\pi^{0}\bar{\nu}_{\tau}$. 
There are severe background processes from $e^{-}e^{+}\to\tau^{-}\tau^{+}$ with both $\tau$ goes to SM decay modes, that are selected due to photon mis-identification or $\pi/\mu$ mis-identification, such as $\tau^{-}\to\pi^{-}\pi^{0}\nu_{\tau}$. 
The radiative process $e^{-}e^{+}\to\gamma_{ISR}\tau^{-}\tau^{+}$, however, is not 
a dominant background in this energy region anymore that can be
easily removed by a certain energy requirement without much efficiency loss.
Both cut-based and multi-variate-analysis are applied to further suppress the backgrounds in the selection of $\tau^{-}\to\mu^{-}\gamma$,
and the sensitivity of $\tau^{-}\to\mu^{-}\gamma$
is found to be consistently within the range $(1.2\sim 1.8)\times10^{-8}$ at 90\% CL for these attempts. 
In the study of the CLFV processes, the detector responses are optimized where
a $\mu/\pi$ suppression power of 30 for $\mu$ with momentum from 500~MeV/c to 2~GeV/c is required, along with a high particle
identification efficiency for $\mu$, {\it i.e.} larger than 95\% at $p=1$~GeV/c. 
Moreover, the cLFV decays $\tau^{-} \to l^{-} P_1P_2$~$(P_i = \pi,~K)$
can also be studied at STCF with a more stringent sensitivity 
with the excellent $\pi/K$ identification power at STCF, with a mis-identification rate less than 2\% and efficiency higher than 97\% up to a momentum of 2~GeV/c.
It is worth noting that, at STCF, with nearly 3 trillion $J/\psi$ samples produced at STCF one year,
the CLFV decay can studied via the process $J/\psi\to l\tau$ 
an expected sensitivity of $4.0\times10^{-9}$ at 90\% CL or better. 

\subsection{HL-LHC}\label{HLHC}
The High-Luminosity LHC (HL-LHC) is planned to start delivering $p$--$p$ collisions in 2029. The energy in the center of mass of the proton collisions is expected to be $\sqrt{s}=14$~TeV and the peak instantaneous luminosity is planned to reach up to  $7.5\times10^{34}$~cm$^{-2}$s$^{-1}$.

\subsubsection{Prospects for LHCb}

The current experimental limit  $B(\tau^- \to \mu^- \mu^+ \mu^-) < 1.2 \times  10^{-8}$, obtained by combining the results from LHCb and the B-factories, reaches the upper limit of the range predicted for this decays by theories beyond the SM. 
As already noted above, the LHC proton proton collisions at 13 TeV produce $\tau$  leptons primarily in the decay of heavy flavour hadrons. The cross-section is five orders of magnitude larger than at Belle II. This compensates for the higher background levels and lower integrated luminosity.  As pointed out in ~\cite{LHCb:2018roe}, during the HL-LHC era,  the LHCb Upgrade II detector will allow to collect 300 fb$^{-1}$. With this large data sample, LHCb   will be able  to probe the branching ratio down to $O(10^{-9}$), and either independently confirm any Belle II discovery or significantly improve the limit.  

\subsubsection{Prospects for ATLAS and CMS}
The number of $\tau$ leptons that will be produced during the lifetime of the HL-LHC is of the order of O($10^{15}$). This a compelling scenario for the search of $\tau^- \to \mu^- \mu^+ \mu^-$ decay.

CMS performed a study of the expected sensitivity for the search for the $\tau^- \to \mu^- \mu^+ \mu^-$ decay at HL-LHC with a dataset corresponding to an integrated luminosity of 3000~fb$^{-1}$~\cite{cmsupt3m}. The study was performed in the context of the technical design report of the CMS muon detector. It considers the presence of additional muon chambers as part of the muon system upgrade, which extend the CMS muon coverage in the
first muon station from $|\eta| <$ 2.4 to 2.82, hence increasing the signal fiducial acceptance by a factor of two.
A dedicated muon identification algorithm for low momentum muons is exploited. Events are separated in two categories: 
\begin{enumerate}
    \item  For Category 1, the L1 trigger requires two tracker muons ($p_{\mathrm{T}} >$ 2 GeV) and one track segment in the first muon endcap station.
    \item For Category 2, the trigger requires one tracker muon and two segments in the first muon endcap station, allowing for segments in the $|\eta|$ = 2.4--2.8 range.
\end{enumerate}

 The projections to HL-LHC conditions of the expected exclusion limits at 90\% CL on  $\mathcal{B}(\tau^- \to \mu^-\mu^+\mu^-)$ are  $3.7\times 10^{-9}$ and  $4.3\times 10^{-9}$ in the case of no additional muon coverage.\\ 

ATLAS performed a simulation-based analysis of the expected sensitivity which the experiment can achieve in the search for the $\tau^- \to \mu^-\mu^+\mu^-$ decay with the HL-LHC data-taking campaign corresponding to an integrated luminosity of 3000~fb$^{-1}$~\cite{atlast3m}. Both the $W$ boson and HF channels are considered.
For the $W$ channel, three scenarios are considered:
\begin{enumerate}
    \item  Non-improved scenario, where only integrated luminosity and higher production cross section at $\sqrt{s}=14$~TeV are considered with respect to the Run 1 analysis.
    \item Intermediate scenario, where the improvements in triggering and reconstruction of low $p_{\mathrm{T}}$ muons estimated from Run 2 Monte Carlo (MC) are also included in the projection.
    \item Improved scenario, where the signal search window is tightened, taking into account expected improvements at the HL-LHC in mass resolution.
\end{enumerate}

For the HF channel, the three scenarios taken into account are the High, Medium and Low background scenarios, where the background levels are rescaled from the Run 1 $W$ channel analysis based on the integrated luminosity and higher cross section of the HL-LHC and an additional penalty factor of ten, three and one is applied, respectively. The projections to HL-LHC conditions of the expected exclusion limits at 90\% CL on  $\mathcal{B}(\tau^- \to \mu^-\mu^+\mu^-)$ are  $5.4\times 10^{-9}$ for the improved scenario of the $W$ boson channel and  $1.0\times 10^{-9}$ for the Low background scenario of the HF channel.

\subsection{EIC}\label{EIC}
The Electron Ion Collider (EIC) will be the first collider providing collisions between a polarized electron beam with a wide range of ions, ranging from polarized proton, helium-3 to unpolarized heavier ions up to uranium. These collisions can happen at a variable center of mass energy between $\sqrt{s}=20$~GeV (5~GeV electron on 41~GeV protons)  and $\sqrt{s}=140$~GeV (18~GeV electron on 275~GeV protons). This versatility makes it an ideal machine to explore Quantum Chromo Dynamics (QCD). To be able to properly explore the QCD phase space and to perform luminosity hungry measurements of the 3D distributions of the quarks inside the proton, the EIC is planned to have instantaneous luminosity up to $10^{34}$~cm$^{-2}$~s$^{-1}$. This last point as well as the ability to select the polarization direction for both electrons and protons in the source opens the door for precision studies that can significantly test the Standard Model. A leading observable in this arena is the electron to $\tau$ transition.

The current constraints on the $e^- \leftrightarrow \tau^-$ transition is much weaker compared to the transition limits already set for $e\leftrightarrow \mu$~\cite{Gonderinger:2010yn,Cirigliano:2021img}. The former limits are set in $e-\tau$ couplings space through searches for $e +  p\to \tau +X$, $\tau\to e\gamma$, and $p+p\to e+\tau + X$ at HERA~\cite{ZEUS:2005nsy,H1:2007dum,ZEUS2012,ZEUS2019}, BaBar~\cite{BaBar:2009hkt}, and the LHC~\cite{ATLAS:2018mrn} respectively. A simulation study with the ECCE detector configuration was undertaken to evaluate the potential for such a measurement at the EIC. The leptoquark generator LQGENEP~\cite{Bellagamba:2001fk} (version 1.0) with a default 1.9~TeV leptoquark mass, the Djangoh generator, and the Pythia generator were used to produce the leptoquark signal, background DIS NC and CC, and background photoproduction Monte-Carlo events, respectively. The leptoquark candidate events were identified by ensuring they contain a high $p_{\rm T}$ quark initiated jet along with an isolated and high-$p_{\rm T}$ $\tau$ which replaces the scattered electron in the NC DIS events. After being produced, the $\tau$ will decay into stable particles after flying a short distance, of the order of millimeters. For this study only the 3-prong decay was thoroughly investigated ($\tau^- \to \pi^-\pi^+\pi^- \nu_\tau$), although we note that the 1-prong decays are also under investigation. These last decay modes are expected to have a worse signal to background at the EIC.

We estimate the 3-sigma exclusion limit on leptoquark cross sections to be 11.4\,fb and 1.7\,fb for the case where the decay channels not in the 3-prong mode are not detected and when they are detected with the same efficiency as the "3-prong" mode presented here, respectively.  Assuming 100~fb$^{-1}$ of luminosity for the $18\times 275$~GeV energy configuration, we estimate that the EIC with the 3-prong decay channel will be able to improve on previous limits set by HERA by up to a factor of 10.

A discussion of the EIC reach in the context of the SMEFT is 
presented in Section~\ref{sect:global}.

\subsection{FCC-ee}
The FCC-ee is the first stage of the integrated Future Circular Colliders (FCC) program to be based on a novel research infrastructure hosted in a $\sim$100-km tunnel in the neighbourhood of CERN. 
The FCC-ee program~\cite{Abada:2019lih,FCC:2018evy,Blondel:2021ema}
includes four major phases with precision measurements of the four heaviest particles of the Standard Model: 
\begin{itemize}
\item[\emph{i})]~the Z boson, with $5\times 10^{12}$ Z decays collected around the Z pole \mbox{(4 years)}, 
\item[\emph{ii})]~the W boson, with $10^8$ WW pairs collected close to threshold \mbox{(2 years)}, 
\item[\mbox{\emph{iii})}]~the Higgs boson, with $1.2\times 10^6$ \mbox{e$^-$e$^+ \to$ HZ} events produced at the  cross-section maximum \mbox{(3 years)}, and
\item[\emph{iv})]~the top quark, with $10^6$ $\mathrm{t\bar{t}}$ pairs produced at and slightly above threshold \mbox{(5 years)}. 
\end{itemize}

The collider will have two (possibly four) interaction points, each equipped with a powerful, state-of-the-art detector system. Detector concepts being studied feature a solenoidal magnetic field, a small-pitch, thin-layers vertex detector providing an excellent impact parameter resolution for lifetime measurements, a highly transparent tracking system providing a superior momentum resolution, a finely segmented calorimeter system with excellent energy resolution for $\text{e}/\gamma$, isolated hadrons, and jets, and a very efficient muon system. At least one of the detector systems will be equipped with efficient particle identification (PID) capabilities allowing $\pi/\text{K}/\text{p}$ separation over a wide momentum range. 

At an extremely high instantaneous luminosity exceeding 10$^{36}$\,cm$^{-2}$\,s$^{-1}$, 150\,ab$^{-1}$ of data will be collected during four years of scan of the Z pole~\cite{FCC:2018evy}. This corresponds to the production of about $5\times 10^{12}$ Z decays, out of which $1.7 \times 10^{11}$ will decay to tau pairs, \mbox{Z $\to \tau^-\tau^+$}, hence exceeding the LEP statistics~\cite{ALEPH:2005ab} by more than five orders of magnitude.
As was the situation at LEP, the experimental conditions will be clean and favourable with the $\tau$-lepton having a sizeable and well-defined boost-factor of $\beta\gamma \simeq 26$. These favourable experimental conditions allow for the optimal exploitation of the large statistics and open the door to a very rich $\tau$-physics program, including 
searches of LFV in $\tau$ decays~\cite{Pich:2020qna}.

A broad palette of 52 LFV $\tau$ decays modes have been searched for by the Belle collaboration, here summarised in \mbox{Fig.~\ref{fig:TauLFV}}. Properly equipped with PID capabilities, it is reasonable to believe that FCC-ee detectors will be able to cover the same palette of channels via highly efficient analyses with no or small backgrounds levels depending on the channel.  Hence, sensitivities are expected at the 10$^{-10}$--10$^{-9}$ level depending on channel.

A first simulation study~\cite{Dam:2018rfz} has been carried out of $\tau^- \to \mu^-\mu^+\mu^-$ and $\tau^-\to\mu^-\gamma$ as benchmark modes. The analysis strategy employed a \emph{tag-side} to identify a
clear Standard Model $\tau$ decay and a \emph{signal-side} where LFV
decays were searched for. Search variables employed were the total energy and the invariant mass of the final-state system.
No backgrounds were identified for the $\tau^- \to \mu^-\mu^+\mu^-$ mode, and a sensitivity of 
$\mathcal{O}(10^{-10})$ seems certainly within reach.
For the $\tau^-\to\mu^-\gamma$ mode, the study involved the non-negligible background from radiative events, e$^-$e$^+ \to \tau^-\tau^+\gamma$, which is believed to be dominant. The resolution on the search variables, and hence the search sensitivity, was found to depend primarily in the ECAL energy resolution. Assuming, conservatively, a resolution of $16.5\%/\sqrt{E\,(\text{GeV})}$, typical for a CALICE-like silicon-based calorimeter~\cite{Aleksa:2021ztd}, the search was found to be sensitive down to branching fractions of $2\times 10^{-9}$. The sensitivity was found to scale slightly stronger than linear in the ECAL resolution, allowing the sensitivity to reach well below $10^{-9}$ for a potential crystal-based ECAL with a resolution of typically $3\%/\sqrt{E\,(\text{GeV})}$.

\section{Experimental Summary}
\label{expsummary}
A summary of observed and expected limits at 90\% confidence level (CL) on different LFV processes in $\tau$ decays from the different experimental efforts are presented in Table~\ref{tab:taudecays}.

{\footnotesize
\begin{longtable}{l|rrc|rrc}

\caption{Current status of observed (obs) and expected (exp) upper limits (UL).} \label{tab:taudecays} \\

\toprule

\multicolumn{1}{c|}{} & \multicolumn{3}{c|}{\textbf{Observed Limits}} & \multicolumn{3}{c}{\textbf{Expected Limits}} \\ \hline 

$\tau^-\to$            &  Experiment                   & Luminosity       & UL (obs)& Experiment & Luminosity  & UL (exp)        \\  \midrule\hline
\endfirsthead

\multicolumn{4}{l}%
{{\bfseries \tablename\ \thetable{}} --- Continued from previous page} \\
\toprule \multicolumn{1}{c|}{\textbf{}} & \multicolumn{3}{c|}{\textbf{Observed Limits}} & \multicolumn{3}{c}{\textbf{Expected Limits}} \\ \hline 
$\tau^-\to$            &  Experiment                   & Luminosity       & UL (obs)& Experiment & Luminosity  & UL (exp)        \\  \midrule
\endhead

\midrule \multicolumn{7}{r}{{Continued on next page}} \\ \bottomrule
\endfoot

\bottomrule
\endlastfoot

$e^-\gamma$            & Belle~\cite{Belle:2021ysv}   & 988~\invfb & 5.6$\times10^{-8}$ & Belle II~\cite{Belle2WP}   & 50 \invab & 9.0$\times10^{-9}$  \\ 
     	               & BaBar~\cite{BaBar:2009hkt}   & 516~\invfb & 3.3$\times10^{-8}$ &                            &           &                     \\\hline
$\mu^-\gamma$	       & Belle~\cite{Belle:2021ysv}   & 988~\invfb & 4.2$\times10^{-8}$ & Belle II~\cite{Belle2WP}   & 50 \invab & 6.9$\times10^{-9}$  \\ 
                       & BaBar~\cite{BaBar:2009hkt}   & 516~\invfb & 4.4$\times10^{-8}$ &                            &           &                     \\
                       &                              &            &                    & STCF~\cite{STCF:clfv}      &  1~\invab & 1.2$\times10^{-8}$  \\
                       &                              &            &                    & FCC-ee~\cite{FCC:2018evy,Dam:2018rfz}      
                                                                                                                     &150~\invab &${\cal{O}}(10^{-9})$ \\\hline
$e^-\pi^0$	           & Belle~\cite{Belle:2007cio}   & 401~\invfb & 8.0$\times10^{-8}$ & Belle II~\cite{Belle2WP}   & 50~\invab & 7.3$\times10^{-10}$ \\ 
                       & BaBar~\cite{BaBar:2006jhm}   & 339~\invfb & 1.3$\times10^{-7}$ &                            &           &                     \\\hline
$\mu^-\pi^0$	       & Belle~\cite{Belle:2007cio}   & 401~\invfb & 1.2$\times10^{-7}$ & Belle II~\cite{Belle2WP}   & 50~\invab & 7.1$\times10^{-10}$ \\ 
                       & BaBar~\cite{BaBar:2006jhm}   & 339~\invfb & 1.1$\times10^{-7}$ &                            &           &                     \\\hline
$e^- K_S^0$	           & Belle~\cite{Belle:2010rxj}   & 671~\invfb & 2.6$\times10^{-8}$ & Belle II~\cite{Belle2WP}   & 50~\invab & 4.0$\times10^{-10}$ \\  
                       & BaBar~\cite{BaBar:2009qra}   & 469~\invfb & 3.3$\times10^{-8}$ &                            &           &                     \\\hline
$\mu^- K_S^0$	       & Belle~\cite{Belle:2010rxj}   & 671~\invfb & 2.3$\times10^{-8}$ & Belle II~\cite{Belle2WP}   & 50~\invab & 4.0$\times10^{-10}$ \\  
                       & BaBar~\cite{BaBar:2009qra}   & 469~\invfb & 4.0$\times10^{-8}$ &                            &           &                     \\\hline
$e^- \eta$	           & Belle~\cite{Belle:2007cio}   & 401~\invfb & 9.2$\times10^{-8}$ & Belle II~\cite{Belle2WP}   & 50~\invab & 1.2$\times10^{-9}$  \\                              & BaBar~\cite{BaBar:2006jhm}   & 339~\invfb & 1.6$\times10^{-7}$ &                            &           &                     \\\hline
$\mu^-\eta$	           & Belle~\cite{Belle:2007cio}   & 401~\invfb & 6.5$\times10^{-8}$ & Belle II~\cite{Belle2WP}   & 50~\invab & 8.0$\times10^{-10}$ \\ 
                       & BaBar~\cite{BaBar:2006jhm}   & 339~\invfb & 1.5$\times10^{-7}$ &                            &           &                     \\\hline
$e^-\eta^\prime$	   & Belle~\cite{Belle:2007cio}   & 401~\invfb & 1.6$\times10^{-7}$ & Belle II~\cite{Belle2WP}   & 50~\invab & 1.2$\times10^{-9}$  \\ 
                       & BaBar~\cite{BaBar:2006jhm}   & 339~\invfb & 2.4$\times10^{-7}$ &                            &           &                     \\\hline
$\mu^-\eta^\prime$	   & Belle~\cite{Belle:2007cio}   & 401~\invfb & 1.3$\times10^{-7}$ & Belle II~\cite{Belle2WP}   & 50~\invab & 1.2$\times10^{-9}$  \\ 
                       & BaBar~\cite{BaBar:2006jhm}   & 339~\invfb & 1.4$\times10^{-7}$ &                            &           &                     \\\hline
$e^- f_0(980)$	       & Belle~\cite{Belle:2008pdf}   & 671~\invfb & 6.8$\times10^{-8}$ & Belle II~\cite{Belle2WP}   & 50~\invab & 9.5$\times10^{-10}$ \\\hline
$\mu^-f_0(980)$	       & Belle~\cite{Belle:2008pdf}   & 671~\invfb & 6.4$\times10^{-8}$ & Belle II~\cite{Belle2WP}   & 50~\invab & 9.1$\times10^{-10}$ \\\hline
$e^- \rho^0$	       & Belle~\cite{Belle:2011ogy}   & 854~\invfb & 1.8$\times10^{-8}$ & Belle II~\cite{Belle2WP}   & 50~\invab & 3.8$\times10^{-10}$ \\  
                       & BaBar~\cite{BaBar:2009wtb}   & 451~\invfb & 4.6$\times10^{-8}$ &                            &           &                     \\\hline
$\mu^-\rho^0$	       & Belle~\cite{Belle:2011ogy}   & 854~\invfb & 1.2$\times10^{-8}$ & Belle II~\cite{Belle2WP}   & 50~\invab & 5.5$\times10^{-10}$ \\  
                       & BaBar~\cite{BaBar:2009wtb}   & 451~\invfb & 2.6$\times10^{-8}$ &                            &           &                     \\\hline
$e^-\omega$	           & Belle~\cite{Belle:2011ogy}   & 854~\invfb & 4.8$\times10^{-8}$ & Belle II~\cite{Belle2WP}   & 50~\invab & 1.0$\times10^{-9}$  \\                              & BaBar~\cite{BaBar:2007amy}   & 384~\invfb & 1.1$\times10^{-7}$ &                            &           &                     \\\hline
$\mu^-\omega$          & Belle~\cite{Belle:2011ogy}   & 854~\invfb & 4.7$\times10^{-8}$ & Belle II~\cite{Belle2WP}   & 50~\invab & 1.4$\times10^{-9}$  \\ 
                       & BaBar~\cite{BaBar:2007amy}   & 384~\invfb & 1.0$\times10^{-7}$ &                            &           &                     \\\hline
$e^-K^{\ast 0}$	       & Belle~\cite{Belle:2011ogy}   & 854~\invfb & 3.2$\times10^{-8}$ & Belle II~\cite{Belle2WP}   & 50~\invab & 6.7$\times10^{-10}$ \\  
                       & BaBar~\cite{BaBar:2009wtb}   & 451~\invfb & 5.9$\times10^{-8}$ &                            &           &                     \\\hline
$\mu^-K^{\ast 0}$	   & Belle~\cite{Belle:2011ogy}   & 854~\invfb & 7.2$\times10^{-8}$ & Belle II~\cite{Belle2WP}   & 50~\invab & 9.3$\times10^{-10}$ \\  
                       & BaBar~\cite{BaBar:2009wtb}   & 451~\invfb & 1.7$\times10^{-7}$ &                            &           &                     \\\hline
$e^-\bar{K}^{\ast0}$   & Belle~\cite{Belle:2011ogy}   & 854~\invfb & 3.4$\times10^{-8}$ & Belle II~\cite{Belle2WP}   & 50~\invab & 6.2$\times10^{-10}$ \\  
                       & BaBar~\cite{BaBar:2009wtb}   & 451~\invfb & 4.6$\times10^{-8}$ &                            &           &                     \\\hline
$\mu^-\bar{K}^{\ast0}$ & Belle~\cite{Belle:2011ogy}   & 854~\invfb & 7.0$\times10^{-8}$ & Belle II~\cite{Belle2WP}   & 50~\invab & 8.5$\times10^{-10}$ \\  
                       & BaBar~\cite{BaBar:2009wtb}   & 451~\invfb & 7.3$\times10^{-8}$ &                            &           &                     \\\hline
$e^-\phi$	           & Belle~\cite{Belle:2011ogy}   & 854~\invfb & 3.1$\times10^{-8}$ & Belle II~\cite{Belle2WP}   & 50~\invab & 7.4$\times10^{-10}$ \\  
                       & BaBar~\cite{BaBar:2009wtb}   & 451~\invfb & 3.1$\times10^{-8}$ &                            &           &                     \\\hline
$\mu^-\phi$	           & Belle~\cite{Belle:2011ogy}   & 854~\invfb & 8.4$\times10^{-8}$ & Belle II~\cite{Belle2WP}   & 50~\invab & 8.4$\times10^{-10}$ \\  
                       & BaBar~\cite{BaBar:2009wtb}   & 451~\invfb & 1.9$\times10^{-7}$ &                            &           &                     \\\hline
$e^-e^+e^-$	           & Belle~\cite{Hayasaka:2010np} & 782~\invfb & 2.7$\times10^{-8}$ & Belle II~\cite{Belle2WP}   & 50~\invab & 4.7$\times10^{-10}$ \\  
                       & BaBar~\cite{BaBar:2010axs}   & 468~\invfb & 2.9$\times10^{-8}$ &                            &           &                     \\\hline
$\mu^-e^+e^-$	       & Belle~\cite{Hayasaka:2010np} & 782~\invfb & 1.8$\times10^{-8}$ & Belle II~\cite{Belle2WP}   & 50~\invab & 2.9$\times10^{-10}$ \\  
                       & BaBar~\cite{BaBar:2010axs}   & 468~\invfb & 2.2$\times10^{-8}$ &                            &           &                     \\\hline
$e^-\mu^+\mu^-$ 	   & Belle~\cite{Hayasaka:2010np} & 782~\invfb & 2.7$\times10^{-8}$ & Belle II~\cite{Belle2WP}   & 50~\invab & 4.5$\times10^{-10}$ \\  
                       & BaBar~\cite{BaBar:2010axs}   & 468~\invfb & 3.2$\times10^{-8}$ &                            &           &                     \\\hline
$\mu^-\mu^+\mu^-$	   & Belle~\cite{Hayasaka:2010np} & 782~\invfb & 2.1$\times10^{-8}$ & Belle II~\cite{Belle2WP}   & 50~\invab & 3.6$\times10^{-10}$ \\  
                       & BaBar~\cite{BaBar:2010axs}   & 468~\invfb & 3.3$\times10^{-8}$ &                            &           &                     \\  
                       & LHCb~\cite{LHCb:2014kws}     &   3~\invfb & 4.6$\times10^{-8}$ & LHCb~\cite{LHCb:2018roe}   &300~\invfb & ${\cal{O}}(10^{-9})$\\  
                       & CMS~\cite{cmst3m}       &  33~\invfb & 8.0$\times10^{-8}$ & CMS~\cite{cmsupt3m}        &  3~\invab & 3.7$\times10^{-9}$  \\  
                       & ATLAS~\cite{ATLAS:2016jts}   &  20~\invfb & 3.8$\times10^{-7}$ & ATLAS~\cite{atlast3m}      &  3~\invab & 1.0$\times10^{-9}$  \\
                       &                              &            &                    & STCF~\cite{STCF:clfv}      &  1~\invab & 1.4$\times10^{-9}$  \\
                       &                              &            &                    & FCC-ee~\cite{FCC:2018evy,Dam:2018rfz}  
                                                                                                                     & 150~\invab&${\cal{O}}(10^{-10})$\\\hline
$e^+\mu^-\mu^-$        & Belle~\cite{Hayasaka:2010np} & 782~\invfb & 1.7$\times10^{-8}$ & Belle II~\cite{Belle2WP}   & 50~\invab & 2.6$\times10^{-10}$ \\  
                       & BaBar~\cite{BaBar:2010axs}   & 468~\invfb & 2.6$\times10^{-8}$ &                            &           &                     \\\hline
$\mu^+e^-e^-$          & Belle~\cite{Hayasaka:2010np} & 782~\invfb & 1.5$\times10^{-8}$ & Belle II~\cite{Belle2WP}   & 50~\invab & 2.3$\times10^{-10}$ \\  
                       & BaBar~\cite{BaBar:2010axs}   & 468~\invfb & 1.8$\times10^{-8}$ &                            &           &                     \\\hline
$e^-\pi^+\pi^-$	       & Belle~\cite{Belle:2012unr}   & 854~\invfb & 2.3$\times10^{-8}$ & Belle II~\cite{Belle2WP}   & 50~\invab & 5.8$\times10^{-10}$ \\ 
                       & BaBar~\cite{BaBar:2005yvr}   & 221~\invfb & 1.2$\times10^{-7}$ &                            &           &                     \\\hline
$\mu^-\pi^+\pi^-$	   & Belle~\cite{Belle:2012unr}   & 854~\invfb & 2.1$\times10^{-8}$ & Belle II~\cite{Belle2WP}   & 50~\invab & 5.6$\times10^{-10}$ \\ 
                       & BaBar~\cite{BaBar:2005yvr}   & 221~\invfb & 2.9$\times10^{-7}$ &                            &           &                     \\
                       &                              &            &                    & STCF~\cite{STCF:clfv}      &  1~\invab &${\cal{O}}(10^{-9})$ \\\hline
$e^-\pi^+K^-$	       & Belle~\cite{Belle:2012unr}   & 854~\invfb & 3.7$\times10^{-8}$ & Belle II~\cite{Belle2WP}   & 50~\invab & 7.1$\times10^{-10}$ \\ 
                       & BaBar~\cite{BaBar:2005yvr}   & 221~\invfb & 3.2$\times10^{-7}$ &                            &           &                     \\\hline
$\mu^-\pi^+K^-$	       & Belle~\cite{Belle:2012unr}   & 854~\invfb & 8.6$\times10^{-8}$ & Belle II~\cite{Belle2WP}   & 50~\invab & 1.2$\times10^{-9}$  \\ 
                       & BaBar~\cite{BaBar:2005yvr}   & 221~\invfb & 2.6$\times10^{-7}$ &                            &           &                     \\
                       &                              &            &                    & STCF~\cite{STCF:clfv}      &  1~\invab &${\cal{O}}(10^{-9})$ \\\hline
$e^-K^+\pi^{-}$	       & Belle~\cite{Belle:2012unr}   & 854~\invfb & 3.1$\times10^{-8}$ & Belle II~\cite{Belle2WP}   & 50~\invab & 7.8$\times10^{-10}$ \\ 
                       & BaBar~\cite{BaBar:2005yvr}   & 221~\invfb & 1.7$\times10^{-7}$ &                            &           &                     \\\hline
$\mu^-K^+\pi^{-}$	   & Belle~\cite{Belle:2012unr}   & 854~\invfb & 4.5$\times10^{-8}$ & Belle II~\cite{Belle2WP}   & 50~\invab & 1.2$\times10^{-9}$  \\ 
                       & BaBar~\cite{BaBar:2005yvr}   & 221~\invfb & 3.2$\times10^{-7}$ &                            &           &                     \\\hline
$e^-K^+K^-$	           & Belle~\cite{Belle:2012unr}   & 854~\invfb & 3.4$\times10^{-8}$ & Belle II~\cite{Belle2WP}   & 50~\invab & 6.5$\times10^{-10}$ \\ 
                       & BaBar~\cite{BaBar:2005yvr}   & 221~\invfb & 1.4$\times10^{-7}$ &                            &           &                     \\\hline
$\mu^-K^+K^-$	       & Belle~\cite{Belle:2012unr}   & 854~\invfb & 4.4$\times10^{-8}$ & Belle II~\cite{Belle2WP}   & 50~\invab & 1.1$\times10^{-9}$  \\ 
                       & BaBar~\cite{BaBar:2005yvr}   & 221~\invfb & 2.5$\times10^{-7}$ &                            &           &                     \\
                       &                              &            &                    & STCF~\cite{STCF:clfv}      &  1~\invab &${\cal{O}}(10^{-9})$ \\\hline
$e^-K_S^0K_S^0$	       & Belle~\cite{Belle:2010rxj}   & 671~\invfb & 7.1$\times10^{-8}$ & Belle II~\cite{Belle2WP}   & 50~\invab & 9.7$\times10^{-10}$ \\\hline
$\mu^-K_S^0K_S^0$	   & Belle~\cite{Belle:2010rxj}   & 671~\invfb & 8.0$\times10^{-8}$ & Belle II~\cite{Belle2WP}   & 50~\invab & 1.1$\times10^{-9}$  \\\hline
$e^+\pi^-\pi^-$        & Belle~\cite{Belle:2012unr}   & 854~\invfb & 2.0$\times10^{-8}$ & Belle II~\cite{Belle2WP}   & 50~\invab & 4.6$\times10^{-10}$ \\ 
                       & BaBar~\cite{BaBar:2005yvr}   & 221~\invfb & 2.7$\times10^{-7}$ &                            &           &                     \\\hline
$\mu^+\pi^-\pi^-$      & Belle~\cite{Belle:2012unr}   & 854~\invfb & 3.9$\times10^{-8}$ & Belle II~\cite{Belle2WP}   & 50~\invab & 4.5$\times10^{-10}$ \\ 
                       & BaBar~\cite{BaBar:2005yvr}   & 221~\invfb & 7.0$\times10^{-8}$ &                            &           &                     \\\hline
$e^+\pi^-K^-$          & Belle~\cite{Belle:2012unr}   & 854~\invfb & 3.2$\times10^{-8}$ & Belle II~\cite{Belle2WP}   & 50~\invab & 7.7$\times10^{-10}$ \\ 
                       & BaBar~\cite{BaBar:2005yvr}   & 221~\invfb & 1.8$\times10^{-7}$ &                            &           &                     \\\hline
$\mu^+\pi^-K^-$        & Belle~\cite{Belle:2012unr}   & 854~\invfb & 4.8$\times10^{-8}$ & Belle II~\cite{Belle2WP}   & 50~\invab & 1.2$\times10^{-9}$  \\ 
                       & BaBar~\cite{BaBar:2005yvr}   & 221~\invfb & 2.2$\times10^{-7}$ &                            &           &                     \\\hline
$e^+K^-K^-$            & Belle~\cite{Belle:2012unr}   & 854~\invfb & 3.3$\times10^{-8}$ & Belle II~\cite{Belle2WP}   & 50~\invab & 5.8$\times10^{-10}$ \\ 
                       & BaBar~\cite{BaBar:2005yvr}   & 221~\invfb & 1.5$\times10^{-7}$ &                            &           &                     \\\hline
$\mu^+K^-K^-$          & Belle~\cite{Belle:2012unr}   & 854~\invfb & 4.7$\times10^{-8}$ & Belle II~\cite{Belle2WP}   & 50~\invab & 9.7$\times10^{-10}$ \\ 
                       & BaBar~\cite{BaBar:2005yvr}   & 221~\invfb & 4.8$\times10^{-7}$ &                            &           &                     \\\hline
$\pi^-\bar{\Lambda}$   & Belle~\cite{Belle:2005exq}   & 154~\invfb & 1.4$\times10^{-7}$ & Belle II~\cite{Belle2WP}   & 50~\invab & 5.5$\times10^{-10}$ \\\hline
$\pi^-\Lambda$	       & Belle~\cite{Belle:2005exq}   & 154~\invfb & 7.2$\times10^{-8}$ & Belle II~\cite{Belle2WP}   & 50~\invab & 5.4$\times10^{-10}$ \\\hline
$\bar{p}^-e^+e^-$	   & Belle~\cite{Belle:2020lfn}   & 921~\invfb & 3.0$\times10^{-8}$ & Belle II~\cite{Belle2WP}   & 50~\invab & 4.0$\times10^{-10}$ \\\hline
$\bar{p}^-e^+\mu^-$	   & Belle~\cite{Belle:2020lfn}   & 921~\invfb & 2.0$\times10^{-8}$ & Belle II~\cite{Belle2WP}   & 50~\invab & 4.4$\times10^{-10}$ \\\hline
$\bar{p}^-\mu^+e^-$	   & Belle~\cite{Belle:2020lfn}   & 921~\invfb & 1.8$\times10^{-8}$ & Belle II~\cite{Belle2WP}   & 50~\invab & 4.4$\times10^{-10}$ \\\hline
$\bar{p}^-\mu^+\mu^-$  & Belle~\cite{Belle:2020lfn}   & 921~\invfb & 1.8$\times10^{-8}$ & Belle II~\cite{Belle2WP}   & 50~\invab & 7.4$\times10^{-10}$ \\
                       & LHCb~\cite{LHCb:2013fsr}     &   1~\invfb & 3.3$\times10^{-7}$ &                            &           &                     \\\hline
$p^+e^-e^-$            & Belle~\cite{Belle:2020lfn}   & 921~\invfb & 3.0$\times10^{-8}$ & Belle II~\cite{Belle2WP}   & 50~\invab & 3.6$\times10^{-10}$ \\\hline
$p^+\mu^-\mu^-$        & Belle~\cite{Belle:2020lfn}   & 921~\invfb & 4.0$\times10^{-8}$ & Belle II~\cite{Belle2WP}   & 50~\invab & 8.3$\times10^{-10}$ \\
                       & LHCb~\cite{LHCb:2013fsr}     &   1~\invfb & 4.4$\times10^{-7}$ &                            &           &                     \\\hline
\end{longtable}
}

\begin{figure}[!h]
    \centering
    \includegraphics[width=\linewidth]{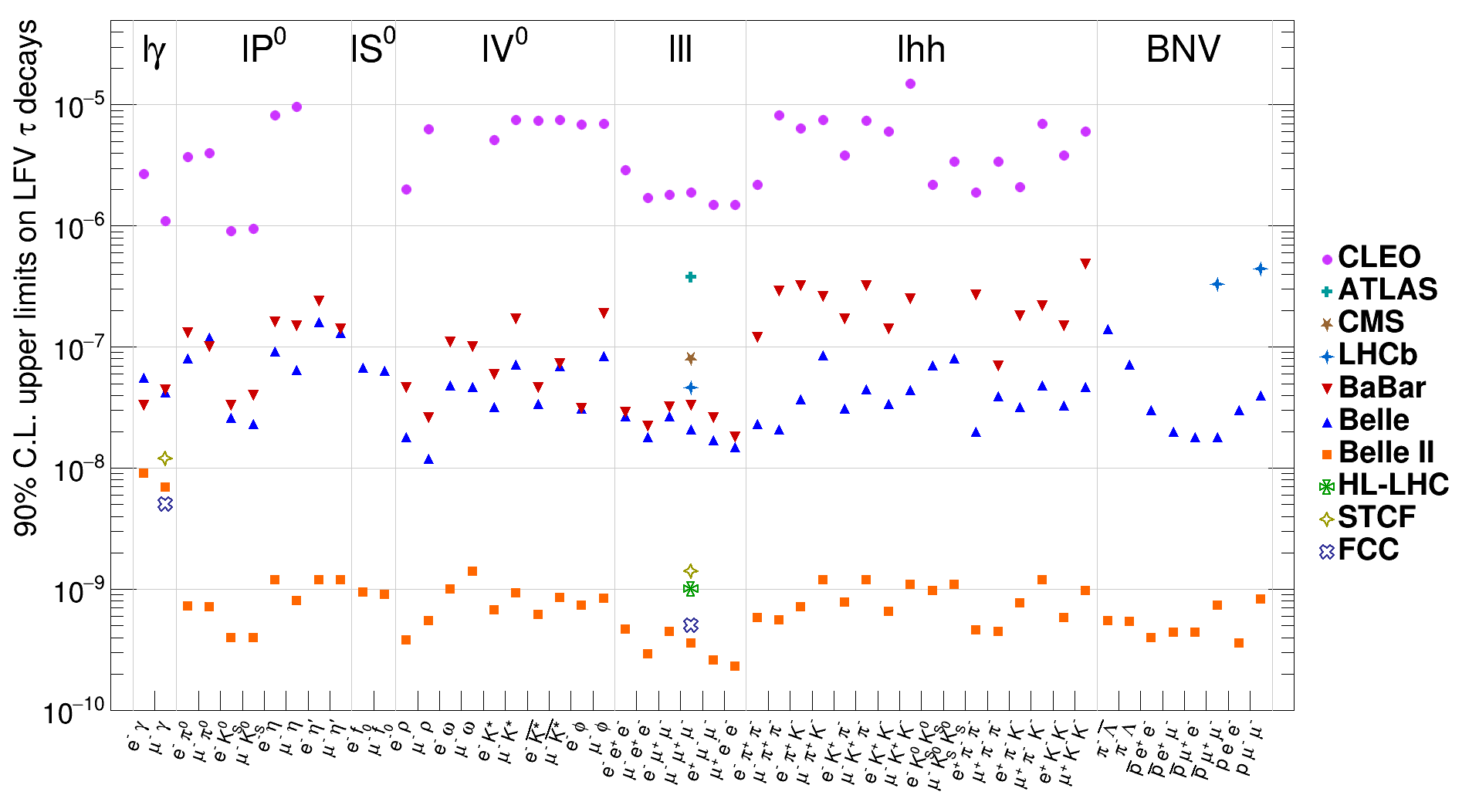} 
    \caption{Summary of upper limits on LFV  processes in $\tau$ decays.}
    \label{fig:TauLFV_all}
\end{figure}

A summary of all observed limits at past and current experiemnts, and projection of expected limits at future experiments on the observed bounds on LFV in the 52 benchmark $\tau$ decay channels are shown in Figure~\ref{fig:TauLFV_all}.

A summary of observed limits on LFV processes in heavy particles decaying into final states containing a $\tau$ lepton from different experimental efforts are presented in Table~\ref{tab:taufinalstate}.
\begin{table}[!h]
\begin{center}
\caption{Bounds on selected LFV decays with $\tau$ in the final state are shown at 90\% CL, except for limits on those decays marked with a $(*)$, which are quoted at 95\% CL.}
\label{tab:taufinalstate}
\begin{tabular}{r|l|r}
\toprule
  Channel            & Upper limit    & Experiment [Ref.] \\
\midrule
$J/\psi \to e^\pm \tau^\mp$      & $7.5 \times 10^{-8}$ & BES III~\cite{BESIII:2021slj}\\
$J/\psi \to \mu^\pm \tau^\mp$    & $2.0 \times 10^{-6}$ & BES~\cite{BES:2004jiw} \\
\hline
$B^0 \to e^\pm \tau^\mp$         & $2.8 \times 10^{-5}$ & BaBar~\cite{Aubert:2008cu}\\
$B^0 \to \mu^\pm \tau^\mp$       & $2.2 \times 10^{-5}$ & BaBar~\cite{Aubert:2008cu}\\
                                 & $1.2 \times 10^{-5}$  & LHCb~\cite{Aaij:2019okb} \\
\hline
$B^+ \to \pi^+ e^\pm \tau^\mp$    & $7.5 \times 10^{-5}$ & BaBar~\cite{BaBar:2012azg} \\
$B^+ \to \pi^+ \mu^\pm \tau^\mp$  & $7.2 \times 10^{-5}$ & BaBar~\cite{BaBar:2012azg} \\
$B^+ \to K^+ e^\pm \tau^\mp$      & $3.0 \times 10^{-5}$ & BaBar~\cite{BaBar:2012azg} \\
$B^+ \to K^+ \mu^\pm \tau^\mp$    & $4.8 \times 10^{-5}$ & BaBar~\cite{BaBar:2012azg} \\
$B^+ \to K^+ \mu^- \tau^+$        & $3.9 \times 10^{-5}$ & LHCb~\cite{LHCb:2020khb} \\
\hline
$B_s^0 \to \mu^\pm \tau^\mp$      & $3.4 \times 10^{-5}$  & LHCb~\cite{Aaij:2019okb} \\
\hline
$\Upsilon(1S) \to e^\pm \tau^\mp$ & $2.7\times 10^{-6}$  & Belle~\cite{Belle:2022cce} \\
$\Upsilon(1S) \to \mu^\pm \tau^\mp$& $2.7\times 10^{-6}$  & Belle~\cite{Belle:2022cce} \\
\hline
$\Upsilon(2S) \to e^\pm \tau^\mp$  & $3.2\times 10^{-6}$ & BaBar~\cite{Lees:2010jk} \\
$\Upsilon(2S) \to \mu^\pm \tau^\mp$& $3.3\times 10^{-6}$ & BaBar~\cite{Lees:2010jk} \\
\hline
$\Upsilon(3S) \to e^\pm \tau^\mp$  & $4.2\times 10^{-6}$ & BaBar~\cite{Lees:2010jk} \\
$\Upsilon(3S) \to \mu^\pm \tau^\mp$& $3.1\times 10^{-6}$ & BaBar~\cite{Lees:2010jk} \\
\hline
$Z \to   e^\pm \tau^\mp$           & $5.0\times 10^{-6}$ (*) & ATLAS~\cite{EXOT-2020-28} \\
$Z \to   \mu^\pm \tau^\mp$         & $6.5\times 10^{-6}$ (*) & ATLAS~\cite{EXOT-2020-28} \\
\hline
$H \to   e^\pm \tau^\mp$          & 0.47\% (*) & ATLAS~\cite{ATLAS:2019pmk} \\
                                  &  0.22\% (*) & CMS~\cite{CMS:2021rsq}\\
$H \to   \mu^\pm \tau^\mp$        & 0.28\% (*) & ATLAS~\cite{ATLAS:2019pmk} \\
                                  &  0.15\% (*) & CMS~\cite{CMS:2021rsq}\\
                                  &  26\% (*) &LHCb~\cite{LHCb:2018ukt}\\
\bottomrule
\end{tabular}
\end{center}
\end{table}

\clearpage

\section{Multi-probe analysis of \texorpdfstring{$\tau$}{tau} CLFV}
\label{sect:global}
\subsection{\texorpdfstring{$\tau \rightarrow e$ transitions}{tau->e transitions}}

\begin{figure}[!htbp]
\centering
\includegraphics[width=\columnwidth]{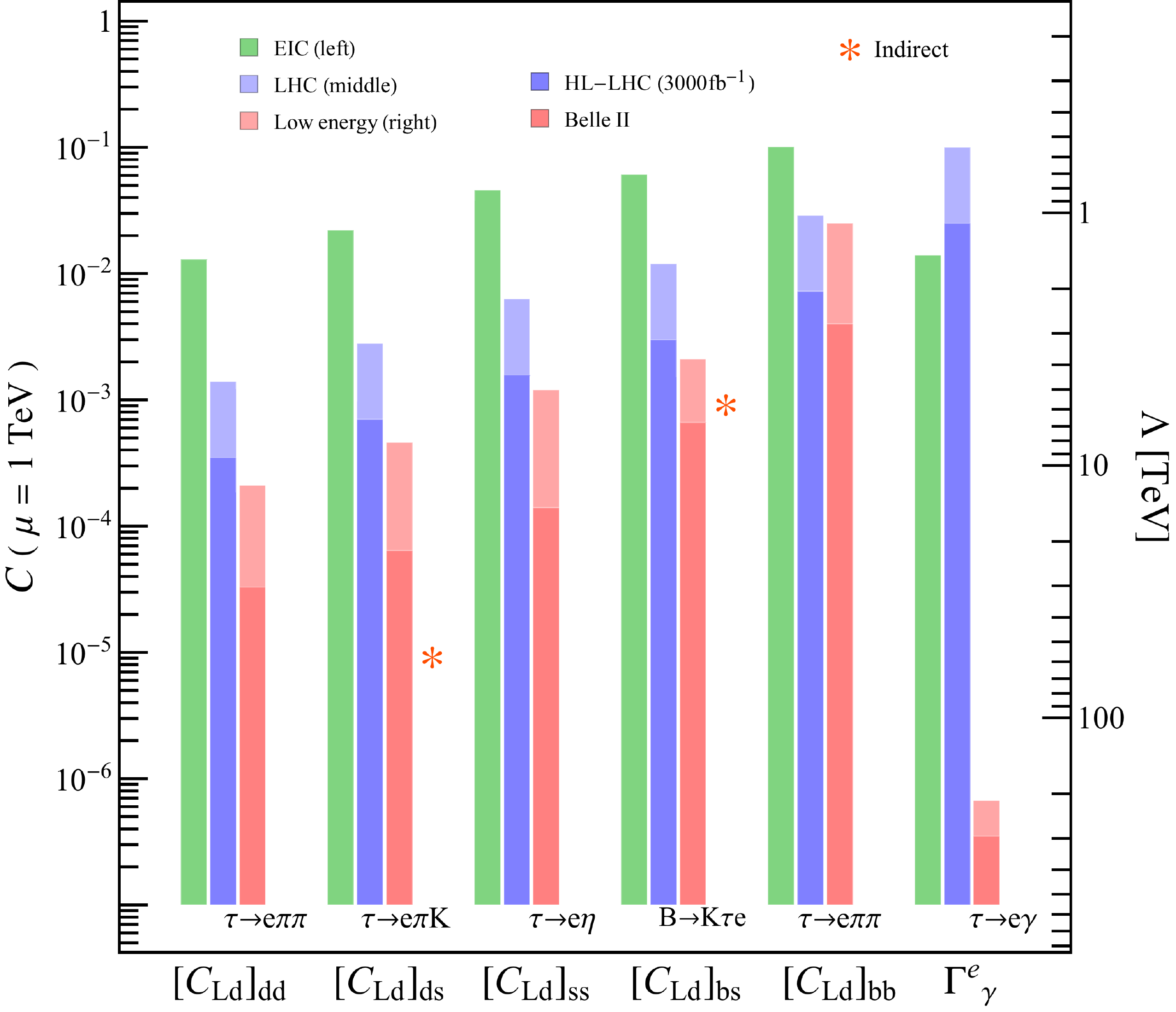}
\caption{Upper limits on $[C_{Ld}]_{\tau e}$ and $\Gamma^e_{\gamma}$ operators from the EIC (green, left), LHC (blue, middle) and low-energy $\tau$ and $B$ meson decays (pink, right). The rightmost vertical axis depicts the lower limit on the scale of new physics $\Lambda$.
The light pink and blue bars denote existing limits from $\tau$ and $B$ decays from the B-factories and other low energy experiments, and from LFV Drell Yan at the LHC, respectively. 
The darker blue and pink bars overlaid on the lighter ones are the
expected sensitivity at the HL-LHC and Belle II. 
Indirect bounds originating from charged-current 
decays and meson decays to neutrinos
are indicated by an asterisk in orange.  }
\label{fig:barchart}
\end{figure}

We present here constraints on CLFV $e$-$\tau$ operators from low- and high-energy experiments based on the SMEFT analysis in \cite{Cirigliano:2021img}, which we updated to include the projected bounds from Belle II in Table \ref{tab:taudecays}, 
the projected luminosity of the HL-LHC, 
and a more realistic estimate of the EIC sensitivity, along the lines discussed in Section \ref{EIC}.

As an example, in Figure \ref{fig:barchart} we show the limits on a down-type four-fermion operator, $C_{Ld}$, which couples left-handed leptons to right-handed quarks,
and on the photon dipole operator $\Gamma^e_{\gamma}$. These operators are defined as
\begin{equation}
\label{eq:OpDefTauE}
{\cal L}_{\rm eff} \supset [C_{Ld}]_{ij}O_{Ld}=[C_{Ld}]_{ij}\frac{4G_\text{F}}{\sqrt{2}}\bar{\ell}_{\tau}\gamma^{\mu}{\ell}_e\bar{d}_i\gamma_{\mu}d_j,
\qquad
{\cal L}_{\rm eff} \supset 
\Gamma^e_{\gamma}O_{\gamma}^e=\Gamma^e_{\gamma}\frac{e}{2v}\bar{\tau}_L\sigma^{\mu\nu}e_RF_{\mu\nu},
\end{equation}
where $[C_{Ld}]_{ij}$ is an arbitrary matrix in quark-flavor space, and the factors of $G_\text{F}$ and $v$ are inserted to make the Wilson coefficients dimensionless.
To obtain the bounds in Fig. \ref{fig:barchart}, we assume that a single operator at a time is turned on at the high scale $\Lambda \sim 1$ TeV, we consider its renormalization group evolution (RGE) to the scales probed at the LHC and EIC and then further evolve it down to a low-energy scale $\mu \sim 2$ GeV. 
In this way, operators with heavy quarks such as $[C_{Ld}]_{bb}$ generate contributions to light-quark operators that can be probed in $\tau$ decays, e.g.\ via $\tau \rightarrow e \pi\pi$. 

The leftmost and rightmost vertical axes in Fig. \ref{fig:barchart} depict the upper bounds on the LFV operator and lower bounds on the scale $\Lambda$ obtained by taking $4G_\text{F}C/\sqrt{2}=1/\Lambda^2$. 
While the green (left) bars correspond to the EIC-expected sensitivity, the blue (middle) and pink (right) bars represent the limits from the LHC and low-energy LFV $\tau$ and $B$ meson decays.  
We next discuss in details how the limits were obtained. 

The {\bf  light pink bars} denote existing low-energy bounds,
and are labeled by the decay mode that gives the strongest limit. The relevant $\tau$ decay channels are listed in Table \ref{tab:taudecays}, and are dominated by Belle and BaBar. 
Operators that are both LFV and quark-flavor-changing, such as $\left[C_{Ld}\right]_{bd}$ and $\left[C_{Ld}\right]_{bs}$,
are constrained by $B \rightarrow \tau e$,
$B \rightarrow \pi \tau e$ and $B \rightarrow K \tau e$ \cite{Zyla:2020zbs}. These channels are currently dominated by BaBar \cite{BaBar:2008pet,BaBar:2012azg}, but will be further studied at Belle II and LHCb. 
Heavy quark operators
( $\left[C_{Ld}\right]_{bb}$)
can also be probed via
$\Upsilon(nS) \rightarrow e \tau$. The limits that can be inferred from Refs.
 \cite{BaBar:2010vxb,Belle:2022cce} are however weaker than $\tau$ decays.
Bounds on selected lepton flavor violating decays of heavy particles decaying into $\tau$ are shown in Table \ref{tab:taufinalstate}.
Finally, the asterisk mark in orange represents bounds from charged current processes (e.g.\ $\pi \rightarrow e \nu_\tau$), or meson decays 
to two neutrinos ($K\to \pi \bar{\nu}_e \nu_\tau$
and $B\to K \bar{\nu}_e \nu_\tau$). For certain SMEFT operators, these processes are correlated to $\tau$-$e$
transitions by gauge invariance. Since  
the flavor of the neutrino is not resolved and these processes have SM background, we dub the resulting bounds as ``indirect". In the case of the $C_{Ld}$ operator, 
$[C_{Ld}]_{ds}$ and 
$[C_{Ld}]_{sd}$ would induce large corrections
to $K\to \pi \bar{\nu}\nu$
and are constrained to be less that $10^{-5}$ by 
the NA62 and KOTO experiments \cite{NA62:2020fhy,KOTO:2018dsc}. This limit is stronger than the direct limit from $\tau \rightarrow e K \pi$.
$[C_{Ld}]_{bd,\, db}$
and $[C_{Ld}]_{bs,\, sb}$ 
are constrained to be $\mathcal O(10^{-3})$ by $B \rightarrow \pi \nu\nu$
and $B \rightarrow K \nu\nu$,
with the strongest limit coming from Belle \cite{Belle:2017oht} and BaBar \cite{BaBar:2013npw}.

The {\bf dark pink bars} are obtained using the projected sensitivity of  Belle II, shown in Table \ref{tab:taudecays}.
With $50$ ab$^{-1}$, Belle II will probe the BRs of $\tau\to e$ decays at the $\mathcal O(10^{-9})$-$\mathcal O(10^{-10})$ level, improving the current limits on SMEFT coefficients by a factor of 5 to 10. 
While we have referred here to
the projected sensitivity of Belle II, STCF and FCC-ee could also give competitive limits as discussed in previous sections.

The {\bf light and dark blue bars} correspond to bounds from the LHC. For the photon dipole operator and for semileptonic four-fermion operators, such as $C_{Ld}$, the strongest LHC constraints arise from searches of LFV in the Drell-Yan process, $p p \rightarrow \tau e$. Four-fermion operators affect in particular the tail of the dilepton invariant mass distribution, and are thus strongly constrained by searches at high invariant mass.  
\footnote{In addition to the $\tau$ decay channels 
discussed in Section \ref{HLHC} and to high-invariant-mass Drell-Yan, ATLAS and CMS have carried out searches of $\tau$ CLFV in $Z$ decays \cite{ATLAS:2021bdj,ATLAS:2020zlz}, Higgs decays \cite{ATLAS:2019pmk,CMS:2021rsq} and 
top decays \cite{ATLAS:2018avw}, which constrain SMEFT operators not shown in Fig. \ref{fig:barchart}.}
To obtain the bounds in Fig. \ref{fig:barchart} we recast the analysis of  Ref. \cite{ATLAS:2018mrn},
which used $36.1~{\rm fb}^{-1}$ of data binned in six invariant mass bin up to $m_{e\tau} = 3$ TeV, in terms of SMEFT operators.
Our results agree well with the similar analysis of Ref. \cite{Angelescu:2020uug}.
The dark blue band show  future HL-LHC limits, assuming a luminosity of $3$ ab$^{-1}$. If the sensitivity at the HL-LHC would just scale with the luminosity, one would expect an improvement by a factor of 10 with respect to the light blue bands. We however expect that there will be quite large background systematic uncertainties. Therefore, for the  HL-LHC sensitivity,  we rescale the current bounds by a factor of 4. 

The {\bf dark green bands} correspond to the EIC sensitivity. 
This is estimated
by assuming a center-of-mass energy $\sqrt{s} = 141$ GeV and an integrated luminosity ${\cal L} =100~{\rm fb}^{-1}$, and by considering two $\tau$ decay modes $\tau^-\to \mu^- \nu_{\tau} \bar{\nu}_{\mu}$ and $\tau^-\to \pi^-\pi^+\pi^- n \pi^0 \nu_\tau$ (n=0,1). At this $\sqrt{s}$, the dipole operator
induces a cross section of $\sigma = 44(5) (v/\Lambda)^4$ pb,
while four-fermion operators give cross sections in the range $\sigma = \{1, 100\} (v/\Lambda)^4 $ pb, depending on the quark-flavor and Lorentz structure of the operator.   
The major backgrounds from SM processes at the EIC include neutral current and charged current DIS. For the muonic channel, we could obtain a background-free signal when considering hard kinematic cuts for the final state particles. The typical cut efficiencies for $C_{Ld}$ operators could range from 1\% to 6\%, depending on the quark flavor. Owing to the fact that kinematic distributions from $\Gamma_\gamma^e$ are similar to the DIS background, the cut efficiency for the photon dipole operator is very small, i.e.\ $\sim 0.1\%$. For the 3-prong decay mode, we only consider some soft kinematic cuts, as a result, the cut efficiencies for the signals are not sensitive to the Lorentz and quark-flavor structure of the SMEFT operators. In this case, the background events cannot be ignored after we include all the kinematic cuts and the background is dominated by the charged-current DIS process.  
The typical cut efficiency of the signal in the 3-prong mode is around 5.2\%. The expected limits from EIC for operators $C_{Ld}$ and $\Gamma_\gamma^e$ can be found in Fig.~\ref{fig:barchart}(green bar).

Putting everything together, we see that,
in a single coupling analysis, for most operators the current strongest limits originate from low-energy LFV $\tau$ and $B$ meson decays. In the case of $[C_{Ld}]_{bb}$, 
since the contribution to $\tau$ decays is obtained 
at one electroweak loop,
the LHC gives a comparable limit to $\tau\to e \pi^+\pi^-$. 
In the future, 
low-energy, the HL-LHC and the EIC will play a complementary role, especially if the EIC efficiency can be improved by combining various $\tau$ decay modes
and by softening the cuts in the muon channel.

\begin{figure}
    \centering
    \includegraphics[width=0.7\textwidth]{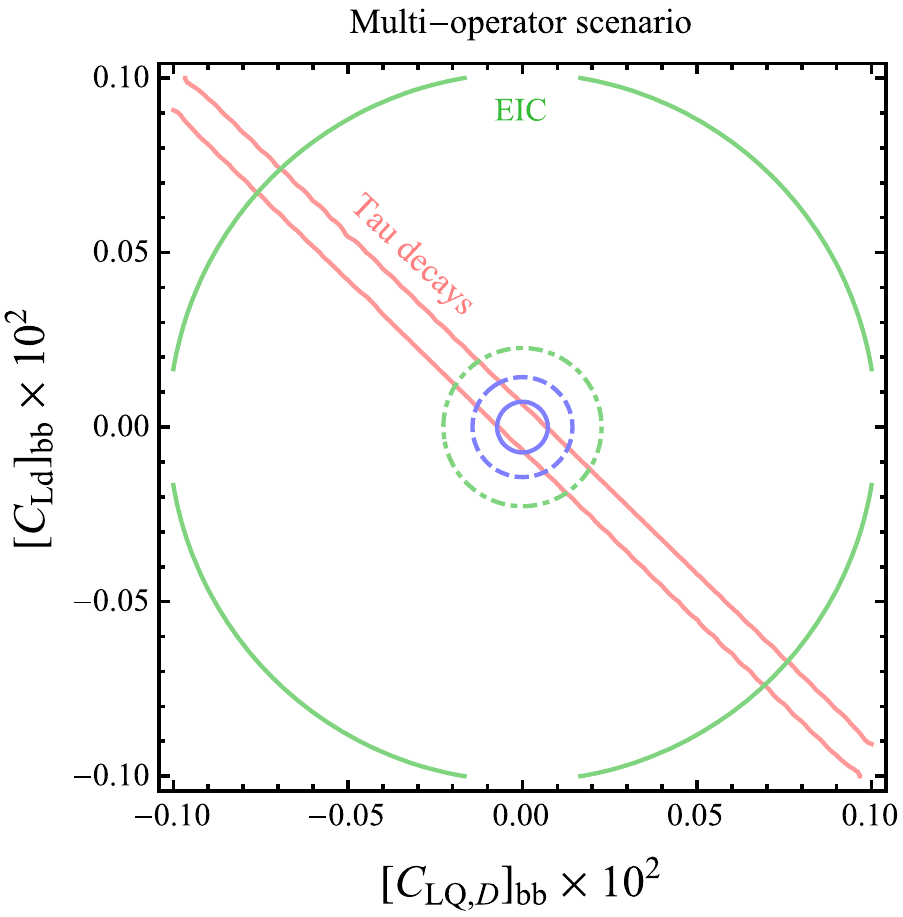}
    \caption{
    90\% C.L. limits in the $[C_{LQ,D} ]_{bb}$- $[C_{Ld}]_{bb}$
plane, after marginalizing over five more SMEFT couplings (see discussion in the text). The pink lines are limits from $\tau$ decays. The blue and green solid lines are bounds
from the LHC and EIC  respectively. 
The EIC limits are obtained using a combination
of the $\tau$ muonic and 3-prong decay modes, as discussed in the text.
The green dashed line denotes an EIC projected limit assuming a 20\% efficiency in the muonic decay channel. 
The blue dashed line assumes the $t$-channel exchange of a particle with $M = 1$ TeV at the LHC. 
}
    \label{fig:2dplot}
\end{figure}

Fig. \ref{fig:barchart} considers the extremely simplified case of the dominance of a single SMEFT operator, which is rarely realized in concrete BSM models. In a bottom-up approach, the existence of free directions, not probed by existing experiments,  needs to be assessed in a global fit to the SMEFT operator basis discussed in Ref. \cite{Cirigliano:2021img,Husek:2020fru}.
While such a fit does not yet exist, Fig. \ref{fig:2dplot} considers a scenario in which 
seven operator coefficients are turned on:
the three flavor-diagonal components of $C_{Ld}$, the three flavor-diagonal components of $C_{LQ, D}$,  a semileptonic operator coupling left-handed down-type quarks and leptons, and one LFV coupling of the $Z$ boson to left-handed leptons. Fig. \ref{fig:2dplot} shows the bounds on the $bb$ components, marginalizing over the other five coefficients. We see that, in this case,  low-energy experiments only probe one combination of heavy flavor couplings. The free direction needs to be closed 
by collider experiments or by $\Upsilon$
decays. With the projections discussed in this white paper, the HL-LHC has an advantage over the EIC and quarkonium decays, with the caveat that, when analyzing the tail of the Drell-Yan distribution in SMEFT, 
one should make sure to be working in the regime of validity of the EFT.  
For example, the HL-LHC limits can be weakened by more than factor of 2 if the SMEFT operators are replaced by the $t$-channel exchange of a BSM particle with mass of about 1 TeV (e.g.\ a leptoquark), which would not appear as a bump in the dilepton invariant mass spectrum. On the other hand, the LHC would be even more sensitive to a $s$-channel exchange.
Fig. \ref{fig:2dplot} shows that the EIC can reach similar sensitivities as the HL-LHC. The solid line denotes the projected limits using the hard kinematic cuts in the muonic decay mode described above, which, for heavy-quark operators, significantly reduce the signal \cite{Cirigliano:2021img}. The green dashed line shows the EIC potential in the assumption that one can reach similar efficiencies as for light quark operators, $\epsilon \sim 20\%$. It is thus important, before the EIC will start taking data, to carry out detailed studies and optimize the $\tau$ tagging efficiency.
The scenario in Fig. \ref{fig:2dplot}
is still very far from a global analysis; the importance of having several experiments in different energy regimes and with comparable sensitivity will be even greater in a full-fledged global fit.

In a top-down approach, the SMEFT formalism developed in Refs. \cite{Cirigliano:2021img,Husek:2020fru}
can be applied to any model with heavy 
degrees of freedom, which will induce a correlated subset of the operators in SMEFT basis.  
A few specific leptoquark models were considered in Refs. \cite{Gonderinger:2010yn,Cirigliano:2021img,Husek:2021isa}. These studies once again illustrate the need for complementary probes of CLFV.

\subsection{\texorpdfstring{$\tau \rightarrow \mu$ transitions}{tau->mu transitions}}

Following a bottom-up approach, a step in the direction of the aforementioned global analysis of the whole SMEFT operator basis addressing CLFV $\tau$-involved processes was taken in Ref.~\cite{Husek:2020fru}, where the authors focused on low-energy $\tau\to\mu(e)$ transitions.
Here, we present the main features of this analysis --- including the employed statistical tool --- as well as the constraints on the SMEFT Wilson coefficients (WCs) stemming from most of the limits shown in Table~\ref{tab:taudecays}.
We refer the reader to Ref.~\cite{Husek:2020fru} for further details on the analysis and the theoretical aspects behind it.

The full set of $D=6$ operators appearing in the Lagrangian in Eq.~\eqref{eq:SMEFT} that contribute to the CLFV $\tau$-involved processes considered here is given in Table~\ref{tab:SMEFTCLFVTauOps}, following the basis given in Ref.~\cite{Grzadkowski:2010es}.
\begin{table}[!t]
\capstart
\begin{center}
\renewcommand{\arraystretch}{1.4}
\begin{tabular}{c|c||c|c}
\toprule
WC & Operator & WC & Operator \\
\midrule
 $C_{LQ}^{(1)}$ & $\left( \bar{L}_p \gamma_{\mu} L_r \right) \left( \bar{Q}_s \gamma^{\mu} Q_t \right)$ & $C_{e \varphi}$  &
 $\left( \varphi^{\dagger} \varphi \right) \left( \bar{L}_p e_r \varphi \right)$   \\
 $C_{LQ}^{(3)}$ & $\left( \bar{L}_p \gamma_{\mu} \sigma^I L_r \right) \left( \bar{Q}_s \gamma^{\mu} \sigma^I Q_t \right)$ &
 $C_{\varphi e}$  &
 $\left( \varphi^{\dagger}  i \overset{\leftrightarrow}{D}_{\mu} \varphi \right) \left( e_p \gamma^{\mu} e_r  \right)$   \\
 $C_{eu}$ & $\left( \bar{e}_p \gamma_{\mu} e_r \right) \left( \bar{u}_s \gamma^{\mu} u_t \right)$ & $C_{\varphi L}^{(1)}$ &
 $\left( \varphi^{\dagger} i \overset{\leftrightarrow}{D}_{\mu} \varphi \right) \left( \bar{L}_p \gamma^{\mu} L_r \right)$ \\
 $C_{ed}$ &  $\left( \bar{e}_p \gamma_{\mu} e_r \right) \left( \bar{d}_s \gamma^{\mu} d_t \right)$ & $C_{\varphi L}^{(3)}$ &
 $\left( \varphi^{\dagger} i \overset{\leftrightarrow}{D}_{I\mu} \varphi \right) \left( \bar{L}_p \sigma_I \gamma^{\mu} L_r \right)$ \\
 $C_{Lu}$ & $\left( \bar{L}_p \gamma_{\mu} L_r \right) \left( \bar{u}_s \gamma^{\mu} u_t \right)$ & $C_{eW}$ & $ \left( \bar{L}_p \sigma^{\mu \nu} e_r \right) \sigma_I \varphi W^I_{\mu \nu}$ \\
 $C_{Ld}$ &  $\left( \bar{L}_p \gamma_{\mu} L_r \right) \left( \bar{d}_s \gamma^{\mu} d_t \right)$ & $C_{eB}$ & $ \left( \bar{L}_p \sigma^{\mu \nu} e_r \right) \varphi B_{\mu \nu}$ \\ 
 $C_{Qe}$ &  $\left( \bar{Q}_p \gamma_{\mu} Q_r \right) \left( \bar{e}_s \gamma^{\mu} e_t \right)$ & & \\
 $C_{LedQ}$ & $\left( \bar{L}^j_p e_r \right) \left( \bar{d}_s Q^j_t \right)$ & & \\ 
 $C_{LeQu}^{(1)}$ & $\left( \bar{L}_p^j e_r \right) \varepsilon_{jk}  \left( \bar{Q}_s^k u_t \right)$ & & \\
 $C_{LeQu}^{(3)}$ & $\left( \bar{L}_p^j \sigma_{\mu \nu} e_r \right)  \varepsilon_{jk}  \left( \bar{Q}_s^k \sigma^{\mu \nu} u_t \right)$ & & \\
\bottomrule
\end{tabular}
\end{center}
\caption{\label{tab:SMEFTCLFVTauOps}
$D=6$ operators appearing in the Lagrangian (\ref{eq:SMEFT}) and contributing to the CLFV processes that we studied in Ref.~\cite{Husek:2020fru}. The four-fermion operators are shown on the left-hand side, while those involving the Higgs doublet $\varphi$ and the gauge bosons are on the right. The notation (up to small apparent changes) is the one from Ref.~\cite{Grzadkowski:2010es}. For the family indices we use $p$, $r$, $s$ and $t$, while $j$ and $k$ are isospin indices. For $I=1,2,3$, $\sigma_I$ are the Pauli matrices, with $\varepsilon = i \sigma_2$, and $\sigma^{\mu\nu}\equiv\frac i2[\gamma^\mu,\gamma^\nu]$. $\Lambda$ is then the scale where the new dynamics arises. The operators share the same notation with the associated couplings, substituting simply $C \rightarrow {\cal O}$, i.e.\ ${\cal O}_{LQ}^{(1)}$ and so on.}
\end{table}
However, it turned out that for the work in question, a slight modification of that basis was more suitable.
We proceed now to comment on these modifications.
First, it was found that the operators ${\cal O}_{\varphi L}^{(1)}$ and ${\cal O}_{\varphi L}^{(3)}$ lead to the same contribution to the studied processes.
Therefore, the analysis was not sensitive to the associated WCs separately but only to their combination, namely
\begin{equation}
C_{\varphi L}^{(1) \, \prime} \equiv C_{\varphi L}^{(1)}+C_{\varphi L}^{(3)} \, .
\end{equation}  
Similarly, the contributions stemming from ${\cal O}_{eB}$ and ${\cal O}_{eW}$ are equal up to factors of $c_\text{W}\equiv\cos\theta_\text{W}$ and $s_\text{W}\equiv\sin\theta_\text{W}$, with $\theta_\text{W}$ being the weak mixing angle, and so only an appropriate combination of WCs is effectively present.
Moreover, both operators contribute through a photon and $Z$ exchange to the studied processes. Hence, to disentangle these contributions, a `rotation' of both WCs was performed and their particular combinations $C_{\gamma}$ and $C_{Z}$ were defined as
\begin{equation}
\label{eq:RotationCgamma}
\begin{pmatrix}
C_{\gamma} \\
C_{Z}
\end{pmatrix}=
\begin{pmatrix}
c_\text{W} & -s_\text{W}\\
s_\text{W} & c_\text{W}
\end{pmatrix}
\begin{pmatrix}
C_{eB} \\
C_{eW}
\end{pmatrix}.
\end{equation}
Accordingly, the parameters $C_{\gamma}$ and $C_{Z}$ are then constrained instead of $C_{e B}$ and $C_{eW}$.
In order to take into account the dominant (QCD) running affecting these processes, the Wilson coefficients accompanying the scalar quark densities are redefined as
\begin{equation}
\label{eq:scalar_redefinition}
C_{LedQ}=\frac{m_{i}}{m_{\tau}}\,C_{LedQ }^{\,\prime}\,,\qquad\qquad
C_{LeQu}^{(1)}=\frac{m_{i}}{m_{\tau}}\,C_{LeQu}^{(1) \prime}\,,
\end{equation}
so that one arrives at scale-independent $C_{LedQ}^{\,\prime}$ and $C_{LeQu}^{(1)\prime}$.
Above, $m_{i}$ stands for a quark mass stemming from the associated quark current.
Finally, the set of 15 independent WCs considered in the general analysis thus reads
\begin{equation}
\Big\{C_{LQ}^{(1)},\,C_{LQ}^{(3)},\,C_{eu},\,C_{ed},\,C_{Lu},\,C_{Ld},\,C_{Qe},\,C_{LedQ}^{\, \prime},\,C_{LeQu}^{(1)\,\prime},\,C_{LeQu}^{(3)},\,C_{\varphi L}^{(1)\,\prime},\,C_{\varphi e},\,C_{\gamma},\,C_{Z},\,C_{e \varphi}\Big\}\,.
\end{equation}

\begin{figure}
    \centering
    \includegraphics[width=\columnwidth]{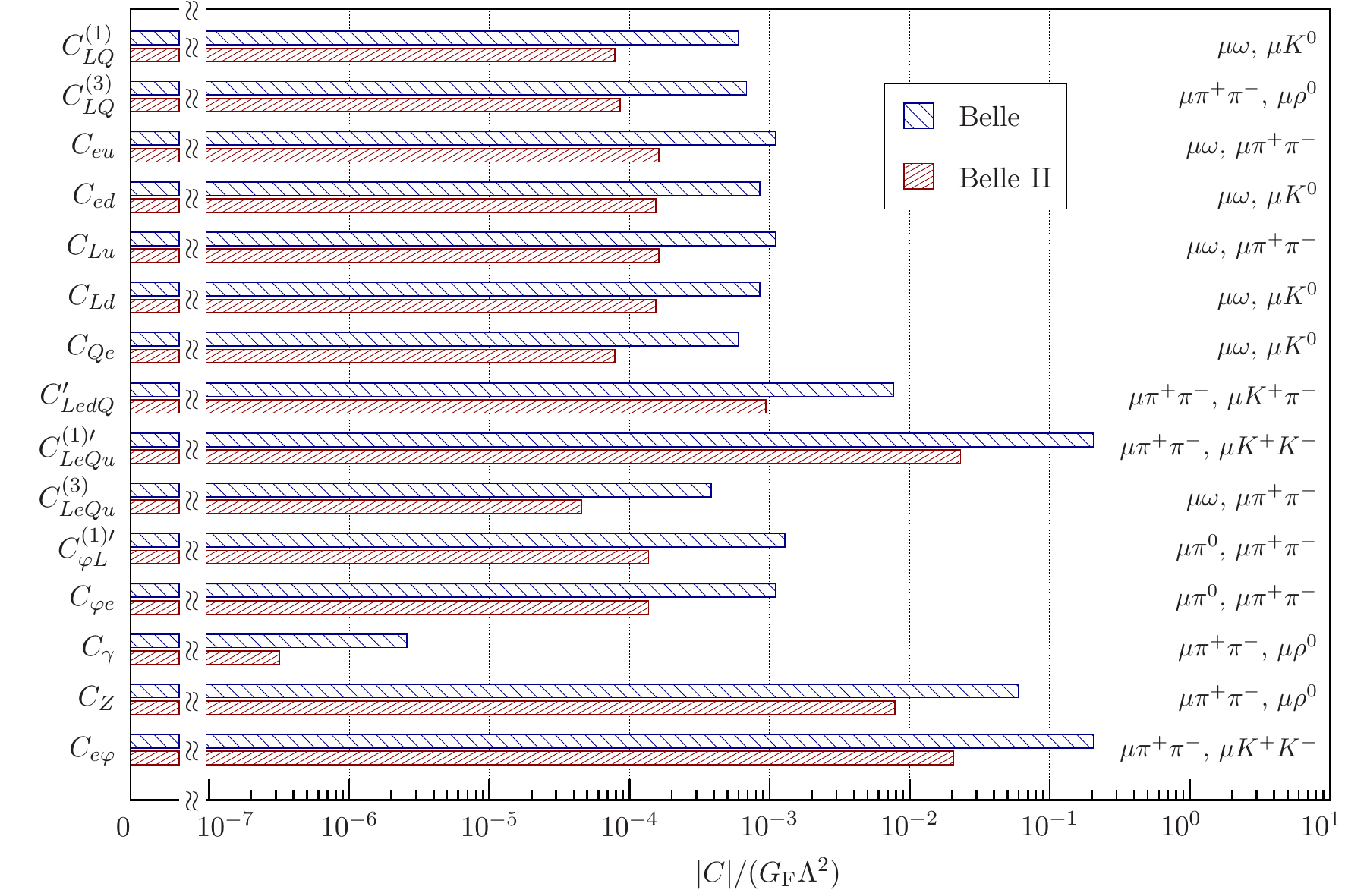}
    \caption{Allowed values for $C/(G_\text{F}\Lambda^2)$ based on the current Belle and expected Belle II limits, stemming from the {\em individual} analysis for hadronic tau decays, given at the 99\% confidence level.
    The two most sensitive channels for the given WC are shown.}
    \label{fig:6_ind}
\end{figure}

\begin{figure}
    \centering
    \includegraphics[width=\columnwidth]{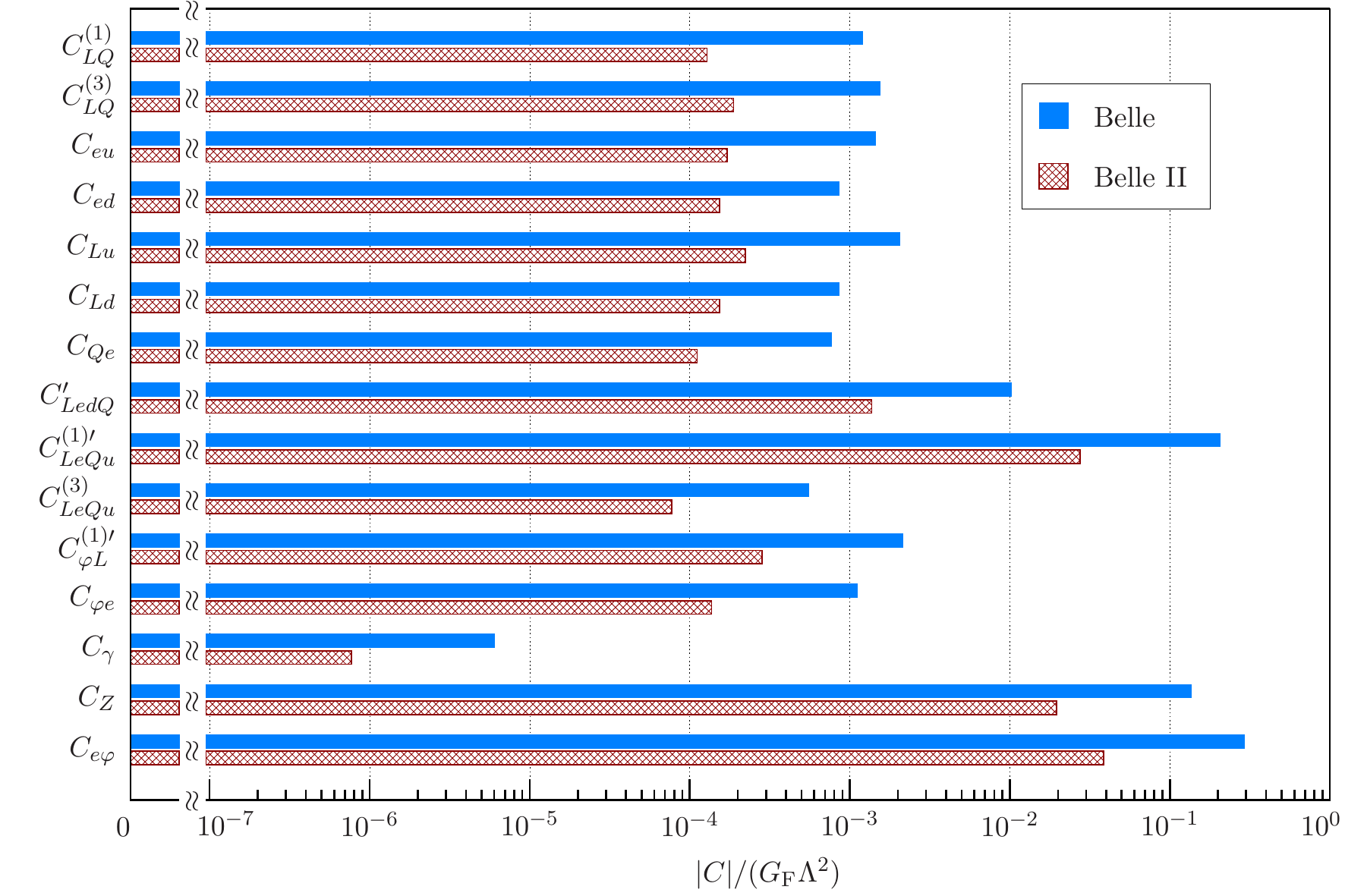}
    \caption{Allowed values for $C/(G_\text{F}\Lambda^2)$ based on the current Belle and expected Belle II limits, stemming from the {\em marginalized} analysis for hadronic tau decays, given at the 99\% confidence level.}
    \label{fig:6_marg}
\end{figure}

\begin{figure}
    \centering
    \includegraphics[width=\columnwidth]{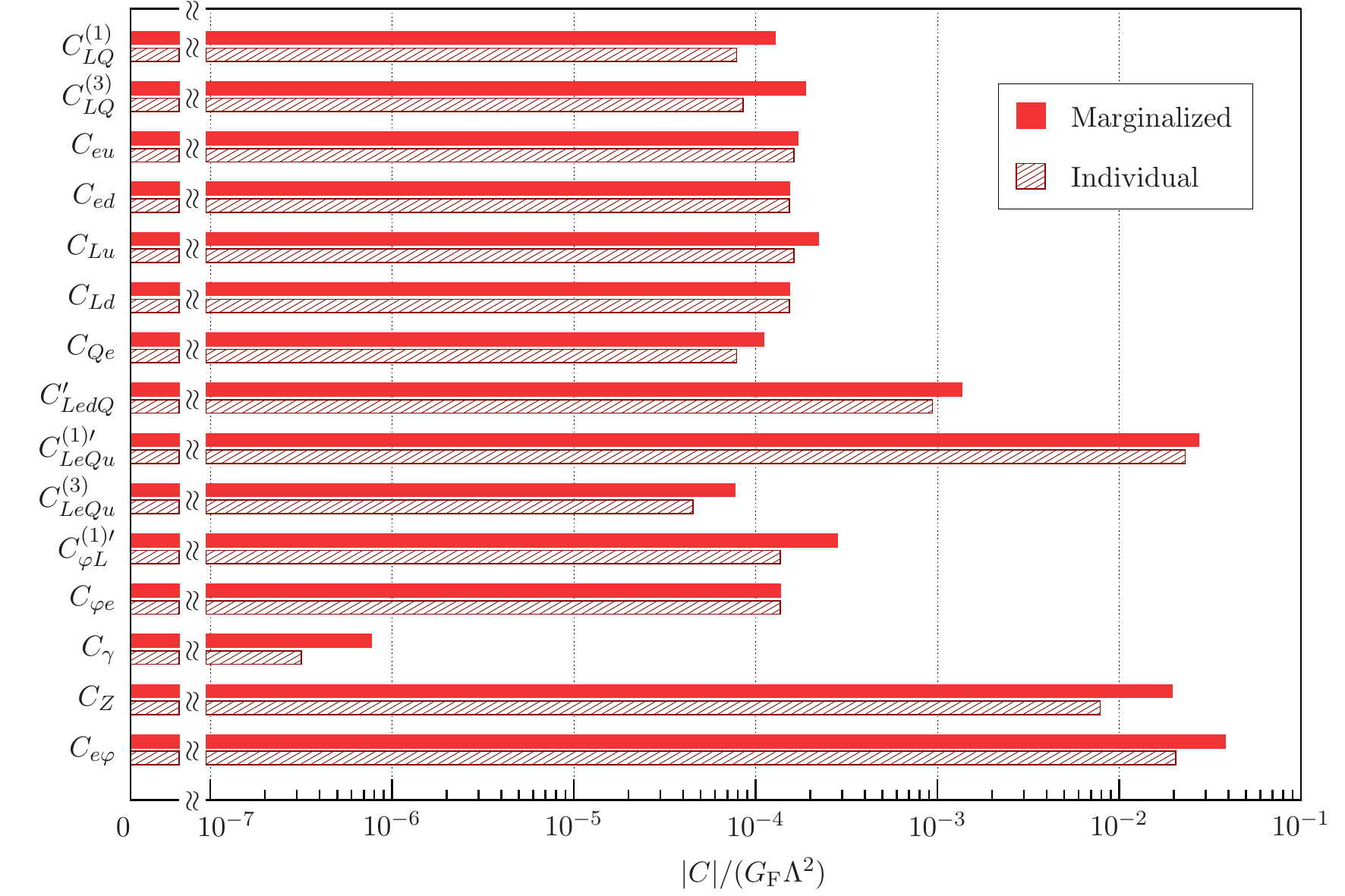}
    \caption{Allowed values for $C/(G_\text{F}\Lambda^2)$ based on the expected Belle II limits, comparing the individual and marginalized analyses for hadronic tau decays, given at the 99\% confidence level.}
    \label{fig:6_Belle_2}
\end{figure}

The goal is then to translate the available information on a set of CLFV observables into relations and constraints on the Wilson coefficients and the characteristic energy scale for the new degrees of freedom represented as $\Lambda$ in Eq.~\eqref{eq:SMEFT}.
During the analysis, the dimensionful ratios $C/\Lambda^2$ were fitted.
However, we present here  results for the dimensionless ratios $C/(G_\text{F} \Lambda^2)$ in the spirit of Eq.~\eqref{eq:OpDefTauE}.
The statistical analysis is then performed with the help of HEPfit~\cite{deBlas:2019okz}, an open-source tool embedded with a Bayesian statistical framework that uses a Markov chain Monte Carlo routine.
In this way the complete WC parameter space can be sampled.
One then obtains allowed values for the WCs at different confidence levels, as well as the correlations among all of them.
While working within a Bayesian framework, the priors chosen for the WCs (i.e.\ their initial probability distributions) are of special importance.
Here the flat distributions for the WCs were used since there is no apparent reason to favour some values over others.

The set of observables used in the analysis was
\begin{align*}
&\tau\to\ell P: & P&=\pi^{0},K^{0}, \eta,\eta^{\prime}\,,\\
&\tau\to\ell P_1P_2: & P_1P_2&=\pi^{+}\pi^{-},K^{0}\bar{K}^{0},K^{+}K^{-},\pi^{+}K^{-},K^{+}\pi^{-}\,,\\
&\tau\to\ell V: & V&=\rho^{0}(770),\omega(782),\phi(1020),K^{*0}(892),\bar{K}^{*0}(892)\,.
\end{align*}
The corresponding results are presented in Figs.~\ref{fig:6_ind}, \ref{fig:6_marg}, and \ref{fig:6_Belle_2}.
In Fig.~\ref{fig:6_ind} we compare the bounds on the Wilson coefficients obtained from an individual analysis, i.e.\ when only one of the operators and accompanying Wilson coefficients is considered to contribute to the observables at a time.
The two $\tau$-decay channels which restrict the given Wilson coefficient the most are shown.
This scenario gives the most stringent bounds since correlations among the parameters are omitted.
A more realistic scenario is given in Fig.~\ref{fig:6_marg}, where we present the bounds set on 
$C/(G_\text{F} \Lambda^2)$ 
from a marginalized analysis, i.e.\ when all Wilson coefficients are varied simultaneously.
This kind of analysis gives a more accurate picture since it also takes into account the possible correlations among the parameters, which in turn tend to relax the bounds set on them.
The improvement entailed by the expected Belle II limits over current Belle data can be readily seen from this figure.
It represents a more realistic description of nature since most of BSM theories provide us with several extra degrees of freedom, which in turn contribute to different Wilson coefficients. Finally, to directly compare how the correlations among the Wilson coefficients affect the imposed bounds, in Fig.~\ref{fig:6_Belle_2} we show the results from both the individual and marginalized analyses based on Belle II expected data. 

In the work~\cite{Husek:2020fru}, the main focus was on the hadronic $\tau$ decays and the corresponding Wilson coefficients involved.
Accordingly, the $\tau\to\mu\gamma$ as well as the $\tau \to 3 \mu$, two golden channels to study CLFV $\tau$-involved processes, were not considered.
However, these two modes are usually enhanced close to observable rates in several BSM extensions.
Hence, the results found in Ref.~\cite{Husek:2020fru} have been recently applied in Ref.~\cite{Husek:2021isa} to constrain the most general leptoquark framework (considering all possible leptoquarks at the same time), where the computation of the $\tau \to \mu \gamma$ process has been performed and a subsequent bound on the Yukawa couplings of leptoquarks to matter has been set.
It was shown --- even though leptoquarks contribute at the loop level to this observable --- that current and expected experimental sensitivities for this process may contribute to constrain combinations of Yukawa couplings that otherwise would remain much less restricted.

To demonstrate the sensitivity of the golden channel $\tau\to\mu\gamma$ to new physics, let us compute the (only) contribution from the SMEFT operators in Table~\ref{tab:SMEFTCLFVTauOps} to this process, i.e.\ the contribution from $C_\gamma{\cal O}_\gamma$ (see Eq.~\eqref{eq:RotationCgamma}).
The decay width is given by
\begin{equation}
\Gamma (\tau \to \mu \gamma)
=\frac{v^2}{4\pi}\,m_\tau^3\bigg(1-\frac{m_\ell^2}{m_\tau^2}\bigg)^3 \frac{C^2}{\Lambda^4}\,.
\end{equation}
The bounds from Table~\ref{tab:taudecays} (with 90\% confidence level) then translate into (applying the individual-analysis approach)
\begin{equation}
\frac{|C_\gamma|}{G_\text{F}\Lambda^{2}}\lesssim\bigg\{\;
\begin{matrix*}[l]
1.7 \times 10^{-7}\;[\text{Belle}]\,,\\
6.6 \times 10^{-8}\;[\text{Belle II}]\,,
\end{matrix*}
\end{equation}
which entail the following bounds for the probed $\Lambda$ once a natural value of order 1 is set for the Wilson coefficients:
\begin{equation}
\Lambda_{C_\gamma\approx1}\gtrsim\bigg\{\hspace{-1ex}
\begin{matrix*}[l]
\hphantom{11}720\,\text{TeV}\;[\text{Belle}]\,,\\
\hphantom{7}1100\,\text{TeV}\;[\text{Belle II}]\,.
\end{matrix*}
\end{equation}

\section{Conclusion}

Observation of lepton flavor violation in the charged lepton sector would completely change our understanding of Nature and herald a new era of discovery in elementary particle physics. 
We are entering a very interesting era in the searches for lepton flavor violation in the tau lepton sector, as the current limits will improve by few orders of magnitude in the next decades.
The current and future experiments will explore uncharted territory and therefore have significant discovery potential. 
They will probe new physics in the multi-TeV scale or conversely strongly constrain the flavor structure of 
TeV-scale extensions of the Standard Model. 
Either way these searches will shed new light on the nature of fundamental interactions.  

\section{Acknowledgments}

This project was supported from funds from various funding agencies.
Sw.~B. is supported by the U.S. Department of Energy under research Grant No. DE-SC0022350.
K.~F., E.~M. and B.~Y. are supported  by the US Department of Energy through
the Office of Nuclear Physics  and  the  
LDRD program at Los Alamos National Laboratory. Los Alamos National Laboratory is operated by Triad National Security, LLC, for the National Nuclear Security Administration of U.S.\ Department of Energy (Contract No. 89233218CNA000001).
V.C. is supported by the U.S. DOE under Grant No. DE-FG02-00ER41132.
J.P. and K.M.-P. are supported by Grant No.\ MCIN/AEI/FPA2017-84445-P and by MCIN/AEI/10.13039/501100011033 Grant No.\ PID2020-114473GB-I00, by PROMETEO/2017/053 and PROMETEO/2021/071 (GV).
T.H. is supported by the Swedish Research Council grants contract numbers 2016-05996 and 2019-03779.
L.~F. is supported by Grant MCIN/AEI/RTI2018-094270-B-I00, Spanish Ministry of Innovation and Research and ERDF.
M.H.V. and A.R. are supported by BMBF and HGF, Germany.
H.P.P. and X.R.Z.  are supported by National Natural Science Foundation of China under project No. 11625523 and 12122509, and by international partnership program of the Chinese Academy of Sciences Grant No. 211134KYSB20200057.

\setboolean{inbibliography}{true}
\addcontentsline{toc}{section}{References}
\bibliographystyle{bibtex_style}
\bibliography{references,bibliography}

\ifx\mcitethebibliography\mciteundefinedmacro
\PackageError{LHCb.bst}{mciteplus.sty has not been loaded}
{This bibstyle requires the use of the mciteplus package.}\fi
\providecommand{\href}[2]{#2}
\begin{mcitethebibliography}{100}
\mciteSetBstSublistMode{n}
\mciteSetBstMaxWidthForm{subitem}{\alph{mcitesubitemcount})}
\mciteSetBstSublistLabelBeginEnd{\mcitemaxwidthsubitemform\space}
{\relax}{\relax}

\bibitem{Petcov:1976ff}
S.~T. Petcov, \ifthenelse{\boolean{articletitles}}{\emph{{The Processes $\mu
  \to e + \gamma$, $\mu \to e + e + \bar e$, $\nu^\prime \to \nu+ \gamma$ in
  the Weinberg-Salam Model with Neutrino Mixing}}, }{}Sov.\ J.\ Nucl.\ Phys.\
  \textbf{25} (1977) 340, [Erratum: Sov. J. Nucl. Phys. 25, 698 (1977),
  Erratum: Yad. Fiz. 25, 1336 (1977)]\relax
\mciteBstWouldAddEndPuncttrue
\mciteSetBstMidEndSepPunct{\mcitedefaultmidpunct}
{\mcitedefaultendpunct}{\mcitedefaultseppunct}\relax
\EndOfBibitem
\bibitem{Marciano:1977wx}
W.~J. Marciano and A.~I. Sanda,
  \ifthenelse{\boolean{articletitles}}{\emph{{Exotic Decays of the Muon and
  Heavy Leptons in Gauge Theories}},
  }{}\href{https://doi.org/10.1016/0370-2693(77)90377-X}{Phys.\ Lett.\ B
  \textbf{67} (1977) 303}\relax
\mciteBstWouldAddEndPuncttrue
\mciteSetBstMidEndSepPunct{\mcitedefaultmidpunct}
{\mcitedefaultendpunct}{\mcitedefaultseppunct}\relax
\EndOfBibitem
\bibitem{Lee:1977qz}
B.~W. Lee, S.~Pakvasa, R.~E. Shrock, and H.~Sugawara,
  \ifthenelse{\boolean{articletitles}}{\emph{{Muon and Electron Number
  Nonconservation in a $V-A$ Gauge Model}},
  }{}\href{https://doi.org/10.1103/PhysRevLett.38.937}{Phys.\ Rev.\ Lett.\
  \textbf{38} (1977) 937}, [Erratum: Phys. Rev. Lett. 38, 1230 (1977)]\relax
\mciteBstWouldAddEndPuncttrue
\mciteSetBstMidEndSepPunct{\mcitedefaultmidpunct}
{\mcitedefaultendpunct}{\mcitedefaultseppunct}\relax
\EndOfBibitem
\bibitem{Lee:1977tib}
B.~W. Lee and R.~E. Shrock, \ifthenelse{\boolean{articletitles}}{\emph{{Natural
  Suppression of Symmetry Violation in Gauge Theories: Muon - Lepton and
  Electron Lepton Number Nonconservation}},
  }{}\href{https://doi.org/10.1103/PhysRevD.16.1444}{Phys.\ Rev.\ D \textbf{16}
  (1977) 1444}\relax
\mciteBstWouldAddEndPuncttrue
\mciteSetBstMidEndSepPunct{\mcitedefaultmidpunct}
{\mcitedefaultendpunct}{\mcitedefaultseppunct}\relax
\EndOfBibitem
\bibitem{Lee:1984kr}
I.-H. Lee, \ifthenelse{\boolean{articletitles}}{\emph{{Lepton Number Violation
  in Softly Broken Supersymmetry}},
  }{}\href{https://doi.org/10.1016/0370-2693(84)91885-9}{Phys.\ Lett.\ B
  \textbf{138} (1984) 121}\relax
\mciteBstWouldAddEndPuncttrue
\mciteSetBstMidEndSepPunct{\mcitedefaultmidpunct}
{\mcitedefaultendpunct}{\mcitedefaultseppunct}\relax
\EndOfBibitem
\bibitem{Lee:1984tn}
I.-H. Lee, \ifthenelse{\boolean{articletitles}}{\emph{{Lepton Number Violation
  in Softly Broken Supersymmetry. 2.}},
  }{}\href{https://doi.org/10.1016/0550-3213(84)90117-2}{Nucl.\ Phys.\ B
  \textbf{246} (1984) 120}\relax
\mciteBstWouldAddEndPuncttrue
\mciteSetBstMidEndSepPunct{\mcitedefaultmidpunct}
{\mcitedefaultendpunct}{\mcitedefaultseppunct}\relax
\EndOfBibitem
\bibitem{Borzumati:1986qx}
F.~Borzumati and A.~Masiero, \ifthenelse{\boolean{articletitles}}{\emph{{Large
  Muon and electron Number Violations in Supergravity Theories}},
  }{}\href{https://doi.org/10.1103/PhysRevLett.57.961}{Phys.\ Rev.\ Lett.\
  \textbf{57} (1986) 961}\relax
\mciteBstWouldAddEndPuncttrue
\mciteSetBstMidEndSepPunct{\mcitedefaultmidpunct}
{\mcitedefaultendpunct}{\mcitedefaultseppunct}\relax
\EndOfBibitem
\bibitem{Barbieri:1995tw}
R.~Barbieri, L.~J. Hall, and A.~Strumia,
  \ifthenelse{\boolean{articletitles}}{\emph{{Violations of lepton flavor and
  CP in supersymmetric unified theories}},
  }{}\href{https://doi.org/10.1016/0550-3213(95)00208-A}{Nucl.\ Phys.\ B
  \textbf{445} (1995) 219},
  \href{http://arxiv.org/abs/hep-ph/9501334}{{\normalfont\ttfamily
  arXiv:hep-ph/9501334}}\relax
\mciteBstWouldAddEndPuncttrue
\mciteSetBstMidEndSepPunct{\mcitedefaultmidpunct}
{\mcitedefaultendpunct}{\mcitedefaultseppunct}\relax
\EndOfBibitem
\bibitem{Altmannshofer:2013lfa}
W.~Altmannshofer, R.~Harnik, and J.~Zupan,
  \ifthenelse{\boolean{articletitles}}{\emph{{Low Energy Probes of PeV Scale
  Sfermions}}, }{}\href{https://doi.org/10.1007/JHEP11(2013)202}{JHEP
  \textbf{11} (2013) 202},
  \href{http://arxiv.org/abs/1308.3653}{{\normalfont\ttfamily
  arXiv:1308.3653}}\relax
\mciteBstWouldAddEndPuncttrue
\mciteSetBstMidEndSepPunct{\mcitedefaultmidpunct}
{\mcitedefaultendpunct}{\mcitedefaultseppunct}\relax
\EndOfBibitem
\bibitem{Abada:2008ea}
A.~Abada {\em et~al.}, \ifthenelse{\boolean{articletitles}}{\emph{{$\mu \to e
  \gamma$ and $\tau \to l \gamma$ decays in the fermion triplet seesaw model}},
  }{}\href{https://doi.org/10.1103/PhysRevD.78.033007}{Phys.\ Rev.\ D
  \textbf{78} (2008) 033007},
  \href{http://arxiv.org/abs/0803.0481}{{\normalfont\ttfamily
  arXiv:0803.0481}}\relax
\mciteBstWouldAddEndPuncttrue
\mciteSetBstMidEndSepPunct{\mcitedefaultmidpunct}
{\mcitedefaultendpunct}{\mcitedefaultseppunct}\relax
\EndOfBibitem
\bibitem{Abada:2007ux}
A.~Abada {\em et~al.}, \ifthenelse{\boolean{articletitles}}{\emph{{Low energy
  effects of neutrino masses}},
  }{}\href{https://doi.org/10.1088/1126-6708/2007/12/061}{JHEP \textbf{12}
  (2007) 061}, \href{http://arxiv.org/abs/0707.4058}{{\normalfont\ttfamily
  arXiv:0707.4058}}\relax
\mciteBstWouldAddEndPuncttrue
\mciteSetBstMidEndSepPunct{\mcitedefaultmidpunct}
{\mcitedefaultendpunct}{\mcitedefaultseppunct}\relax
\EndOfBibitem
\bibitem{Alonso:2012ji}
R.~Alonso, M.~Dhen, M.~B. Gavela, and T.~Hambye,
  \ifthenelse{\boolean{articletitles}}{\emph{{Muon conversion to electron in
  nuclei in type-I seesaw models}},
  }{}\href{https://doi.org/10.1007/JHEP01(2013)118}{JHEP \textbf{01} (2013)
  118}, \href{http://arxiv.org/abs/1209.2679}{{\normalfont\ttfamily
  arXiv:1209.2679}}\relax
\mciteBstWouldAddEndPuncttrue
\mciteSetBstMidEndSepPunct{\mcitedefaultmidpunct}
{\mcitedefaultendpunct}{\mcitedefaultseppunct}\relax
\EndOfBibitem
\bibitem{Cirigliano:2005ck}
V.~Cirigliano, B.~Grinstein, G.~Isidori, and M.~B. Wise,
  \ifthenelse{\boolean{articletitles}}{\emph{{Minimal flavor violation in the
  lepton sector}},
  }{}\href{https://doi.org/10.1016/j.nuclphysb.2005.08.037}{Nucl.\ Phys.\ B
  \textbf{728} (2005) 121},
  \href{http://arxiv.org/abs/hep-ph/0507001}{{\normalfont\ttfamily
  arXiv:hep-ph/0507001}}\relax
\mciteBstWouldAddEndPuncttrue
\mciteSetBstMidEndSepPunct{\mcitedefaultmidpunct}
{\mcitedefaultendpunct}{\mcitedefaultseppunct}\relax
\EndOfBibitem
\bibitem{Raidal:2008jk}
M.~Raidal {\em et~al.}, \ifthenelse{\boolean{articletitles}}{\emph{{Flavour
  physics of leptons and dipole moments}},
  }{}\href{https://doi.org/10.1140/epjc/s10052-008-0715-2}{Eur.\ Phys.\ J.\ C
  \textbf{57} (2008) 13},
  \href{http://arxiv.org/abs/0801.1826}{{\normalfont\ttfamily
  arXiv:0801.1826}}\relax
\mciteBstWouldAddEndPuncttrue
\mciteSetBstMidEndSepPunct{\mcitedefaultmidpunct}
{\mcitedefaultendpunct}{\mcitedefaultseppunct}\relax
\EndOfBibitem
\bibitem{deGouvea:2013zba}
A.~de~Gouvea and P.~Vogel, \ifthenelse{\boolean{articletitles}}{\emph{{Lepton
  Flavor and Number Conservation, and Physics Beyond the Standard Model}},
  }{}\href{https://doi.org/10.1016/j.ppnp.2013.03.006}{Prog.\ Part.\ Nucl.\
  Phys.\  \textbf{71} (2013) 75},
  \href{http://arxiv.org/abs/1303.4097}{{\normalfont\ttfamily
  arXiv:1303.4097}}\relax
\mciteBstWouldAddEndPuncttrue
\mciteSetBstMidEndSepPunct{\mcitedefaultmidpunct}
{\mcitedefaultendpunct}{\mcitedefaultseppunct}\relax
\EndOfBibitem
\bibitem{Bernstein:2013hba}
R.~H. Bernstein and P.~S. Cooper,
  \ifthenelse{\boolean{articletitles}}{\emph{{Charged Lepton Flavor Violation:
  An Experimenter's Guide}},
  }{}\href{https://doi.org/10.1016/j.physrep.2013.07.002}{Phys.\ Rept.\
  \textbf{532} (2013) 27},
  \href{http://arxiv.org/abs/1307.5787}{{\normalfont\ttfamily
  arXiv:1307.5787}}\relax
\mciteBstWouldAddEndPuncttrue
\mciteSetBstMidEndSepPunct{\mcitedefaultmidpunct}
{\mcitedefaultendpunct}{\mcitedefaultseppunct}\relax
\EndOfBibitem
\bibitem{Calibbi:2017uvl}
L.~Calibbi and G.~Signorelli,
  \ifthenelse{\boolean{articletitles}}{\emph{{Charged Lepton Flavour Violation:
  An Experimental and Theoretical Introduction}},
  }{}\href{https://doi.org/10.1393/ncr/i2018-10144-0}{Riv.\ Nuovo Cim.\
  \textbf{41} (2018) 71},
  \href{http://arxiv.org/abs/1709.00294}{{\normalfont\ttfamily
  arXiv:1709.00294}}\relax
\mciteBstWouldAddEndPuncttrue
\mciteSetBstMidEndSepPunct{\mcitedefaultmidpunct}
{\mcitedefaultendpunct}{\mcitedefaultseppunct}\relax
\EndOfBibitem
\bibitem{TheMEG:2016wtm}
MEG, A.~M. Baldini {\em et~al.},
  \ifthenelse{\boolean{articletitles}}{\emph{{Search for the lepton flavour
  violating decay $\mu ^+ \rightarrow \mathrm {e}^+ \gamma $ with the full
  dataset of the MEG experiment}},
  }{}\href{https://doi.org/10.1140/epjc/s10052-016-4271-x}{Eur.\ Phys.\ J.\
  \textbf{C76} (2016) 434},
  \href{http://arxiv.org/abs/1605.05081}{{\normalfont\ttfamily
  arXiv:1605.05081}}\relax
\mciteBstWouldAddEndPuncttrue
\mciteSetBstMidEndSepPunct{\mcitedefaultmidpunct}
{\mcitedefaultendpunct}{\mcitedefaultseppunct}\relax
\EndOfBibitem
\bibitem{Tanabashi:2018oca}
Particle Data Group, M.~Tanabashi {\em et~al.},
  \ifthenelse{\boolean{articletitles}}{\emph{{Review of Particle Physics}},
  }{}\href{https://doi.org/10.1103/PhysRevD.98.030001}{Phys.\ Rev.\
  \textbf{D98} (2018) 030001}\relax
\mciteBstWouldAddEndPuncttrue
\mciteSetBstMidEndSepPunct{\mcitedefaultmidpunct}
{\mcitedefaultendpunct}{\mcitedefaultseppunct}\relax
\EndOfBibitem
\bibitem{Weinberg:1979sa}
S.~Weinberg, \ifthenelse{\boolean{articletitles}}{\emph{{Baryon and Lepton
  Nonconserving Processes}},
  }{}\href{https://doi.org/10.1103/PhysRevLett.43.1566}{Phys.\ Rev.\ Lett.\
  \textbf{43} (1979) 1566}\relax
\mciteBstWouldAddEndPuncttrue
\mciteSetBstMidEndSepPunct{\mcitedefaultmidpunct}
{\mcitedefaultendpunct}{\mcitedefaultseppunct}\relax
\EndOfBibitem
\bibitem{Wilczek:1979hc}
F.~Wilczek and A.~Zee, \ifthenelse{\boolean{articletitles}}{\emph{{Operator
  Analysis of Nucleon Decay}},
  }{}\href{https://doi.org/10.1103/PhysRevLett.43.1571}{Phys.\ Rev.\ Lett.\
  \textbf{43} (1979) 1571}\relax
\mciteBstWouldAddEndPuncttrue
\mciteSetBstMidEndSepPunct{\mcitedefaultmidpunct}
{\mcitedefaultendpunct}{\mcitedefaultseppunct}\relax
\EndOfBibitem
\bibitem{Buchmuller:1985jz}
W.~Buchm{\"u}ller and D.~Wyler,
  \ifthenelse{\boolean{articletitles}}{\emph{{Effective Lagrangian Analysis of
  New Interactions and Flavor Conservation}},
  }{}\href{https://doi.org/10.1016/0550-3213(86)90262-2}{Nucl.\ Phys.\ B
  \textbf{268} (1986) 621}\relax
\mciteBstWouldAddEndPuncttrue
\mciteSetBstMidEndSepPunct{\mcitedefaultmidpunct}
{\mcitedefaultendpunct}{\mcitedefaultseppunct}\relax
\EndOfBibitem
\bibitem{Grzadkowski:2010es}
B.~Grzadkowski, M.~Iskrzynski, M.~Misiak, and J.~Rosiek,
  \ifthenelse{\boolean{articletitles}}{\emph{{Dimension-Six Terms in the
  Standard Model Lagrangian}},
  }{}\href{https://doi.org/10.1007/JHEP10(2010)085}{JHEP \textbf{1010} (2010)
  085}, \href{http://arxiv.org/abs/1008.4884}{{\normalfont\ttfamily
  arXiv:1008.4884}}\relax
\mciteBstWouldAddEndPuncttrue
\mciteSetBstMidEndSepPunct{\mcitedefaultmidpunct}
{\mcitedefaultendpunct}{\mcitedefaultseppunct}\relax
\EndOfBibitem
\bibitem{Jenkins:2013zja}
E.~E. Jenkins, A.~V. Manohar, and M.~Trott,
  \ifthenelse{\boolean{articletitles}}{\emph{{Renormalization Group Evolution
  of the Standard Model Dimension Six Operators I: Formalism and lambda
  Dependence}}, }{}\href{https://doi.org/10.1007/JHEP10(2013)087}{JHEP
  \textbf{10} (2013) 087},
  \href{http://arxiv.org/abs/1308.2627}{{\normalfont\ttfamily
  arXiv:1308.2627}}\relax
\mciteBstWouldAddEndPuncttrue
\mciteSetBstMidEndSepPunct{\mcitedefaultmidpunct}
{\mcitedefaultendpunct}{\mcitedefaultseppunct}\relax
\EndOfBibitem
\bibitem{Jenkins:2013wua}
E.~E. Jenkins, A.~V. Manohar, and M.~Trott,
  \ifthenelse{\boolean{articletitles}}{\emph{{Renormalization Group Evolution
  of the Standard Model Dimension Six Operators II: Yukawa Dependence}},
  }{}\href{https://doi.org/10.1007/JHEP01(2014)035}{JHEP \textbf{01} (2014)
  035}, \href{http://arxiv.org/abs/1310.4838}{{\normalfont\ttfamily
  arXiv:1310.4838}}\relax
\mciteBstWouldAddEndPuncttrue
\mciteSetBstMidEndSepPunct{\mcitedefaultmidpunct}
{\mcitedefaultendpunct}{\mcitedefaultseppunct}\relax
\EndOfBibitem
\bibitem{Alonso:2013hga}
R.~Alonso, E.~E. Jenkins, A.~V. Manohar, and M.~Trott,
  \ifthenelse{\boolean{articletitles}}{\emph{{Renormalization Group Evolution
  of the Standard Model Dimension Six Operators III: Gauge Coupling Dependence
  and Phenomenology}}, }{}\href{https://doi.org/10.1007/JHEP04(2014)159}{JHEP
  \textbf{04} (2014) 159},
  \href{http://arxiv.org/abs/1312.2014}{{\normalfont\ttfamily
  arXiv:1312.2014}}\relax
\mciteBstWouldAddEndPuncttrue
\mciteSetBstMidEndSepPunct{\mcitedefaultmidpunct}
{\mcitedefaultendpunct}{\mcitedefaultseppunct}\relax
\EndOfBibitem
\bibitem{Crivellin:2013hpa}
A.~Crivellin, S.~Najjari, and J.~Rosiek,
  \ifthenelse{\boolean{articletitles}}{\emph{{Lepton Flavor Violation in the
  Standard Model with general Dimension-Six Operators}},
  }{}\href{https://doi.org/10.1007/JHEP04(2014)167}{JHEP \textbf{04} (2014)
  167}, \href{http://arxiv.org/abs/1312.0634}{{\normalfont\ttfamily
  arXiv:1312.0634}}\relax
\mciteBstWouldAddEndPuncttrue
\mciteSetBstMidEndSepPunct{\mcitedefaultmidpunct}
{\mcitedefaultendpunct}{\mcitedefaultseppunct}\relax
\EndOfBibitem
\bibitem{Husek:2021isa}
T.~Husek, K.~Mons\'alvez-Pozo, and J.~Portol\'es,
  \ifthenelse{\boolean{articletitles}}{\emph{{Constraints on leptoquarks from
  lepton-flavour-violating tau-lepton processes}},
  }{}\href{http://arxiv.org/abs/2111.06872}{{\normalfont\ttfamily
  arXiv:2111.06872}}\relax
\mciteBstWouldAddEndPuncttrue
\mciteSetBstMidEndSepPunct{\mcitedefaultmidpunct}
{\mcitedefaultendpunct}{\mcitedefaultseppunct}\relax
\EndOfBibitem
\bibitem{Cirigliano:2021img}
V.~Cirigliano {\em et~al.}, \ifthenelse{\boolean{articletitles}}{\emph{{Charged
  Lepton Flavor Violation at the EIC}},
  }{}\href{https://doi.org/10.1007/JHEP03(2021)256}{JHEP \textbf{03} (2021)
  256}, \href{http://arxiv.org/abs/2102.06176}{{\normalfont\ttfamily
  arXiv:2102.06176}}\relax
\mciteBstWouldAddEndPuncttrue
\mciteSetBstMidEndSepPunct{\mcitedefaultmidpunct}
{\mcitedefaultendpunct}{\mcitedefaultseppunct}\relax
\EndOfBibitem
\bibitem{Antusch:2020vul}
S.~Antusch, A.~Hammad, and A.~Rashed,
  \ifthenelse{\boolean{articletitles}}{\emph{{Searching for charged lepton
  flavor violation at $ep$ colliders}},
  }{}\href{http://arxiv.org/abs/2010.08907}{{\normalfont\ttfamily
  arXiv:2010.08907}}\relax
\mciteBstWouldAddEndPuncttrue
\mciteSetBstMidEndSepPunct{\mcitedefaultmidpunct}
{\mcitedefaultendpunct}{\mcitedefaultseppunct}\relax
\EndOfBibitem
\bibitem{Husek:2020fru}
T.~Husek, K.~Mons\'alvez-Pozo, and J.~Portol\'es,
  \ifthenelse{\boolean{articletitles}}{\emph{{Lepton-flavour violation in
  hadronic tau decays and $\mu$--$\tau$ conversion in nuclei}},
  }{}\href{https://doi.org/10.1007/JHEP01(2021)059}{JHEP \textbf{01} (2021)
  059}, \href{http://arxiv.org/abs/2009.10428}{{\normalfont\ttfamily
  arXiv:2009.10428}}\relax
\mciteBstWouldAddEndPuncttrue
\mciteSetBstMidEndSepPunct{\mcitedefaultmidpunct}
{\mcitedefaultendpunct}{\mcitedefaultseppunct}\relax
\EndOfBibitem
\bibitem{Gninenko:2018num}
S.~Gninenko {\em et~al.}, \ifthenelse{\boolean{articletitles}}{\emph{{Deep
  inelastic $e-\tau$ and $\mu-\tau$ conversion in the NA64 experiment at the
  CERN SPS}}, }{}\href{https://doi.org/10.1103/PhysRevD.98.015007}{Phys.\ Rev.\
  D \textbf{98} (2018) 015007},
  \href{http://arxiv.org/abs/1804.05550}{{\normalfont\ttfamily
  arXiv:1804.05550}}\relax
\mciteBstWouldAddEndPuncttrue
\mciteSetBstMidEndSepPunct{\mcitedefaultmidpunct}
{\mcitedefaultendpunct}{\mcitedefaultseppunct}\relax
\EndOfBibitem
\bibitem{Aaboud:2016hmk}
ATLAS, M.~Aaboud {\em et~al.},
  \ifthenelse{\boolean{articletitles}}{\emph{{Search for new phenomena in
  different-flavour high-mass dilepton final states in pp collisions at
  $\sqrt{s}=13$ TeV with the ATLAS detector}},
  }{}\href{https://doi.org/10.1140/epjc/s10052-016-4385-1}{Eur.\ Phys.\ J.\ C
  \textbf{76} (2016) 541},
  \href{http://arxiv.org/abs/1607.08079}{{\normalfont\ttfamily
  arXiv:1607.08079}}\relax
\mciteBstWouldAddEndPuncttrue
\mciteSetBstMidEndSepPunct{\mcitedefaultmidpunct}
{\mcitedefaultendpunct}{\mcitedefaultseppunct}\relax
\EndOfBibitem
\bibitem{Abada:2016vzu}
A.~Abada, V.~De~Romeri, J.~Orloff, and A.~M. Teixeira,
  \ifthenelse{\boolean{articletitles}}{\emph{{In-flight cLFV conversion:
  ${e-\mu }$ , ${e-\tau }$ and ${\mu -\tau }$ in minimal extensions of the
  standard model with sterile fermions}},
  }{}\href{https://doi.org/10.1140/epjc/s10052-017-4864-z}{Eur.\ Phys.\ J.\ C
  \textbf{77} (2017) 304},
  \href{http://arxiv.org/abs/1612.05548}{{\normalfont\ttfamily
  arXiv:1612.05548}}\relax
\mciteBstWouldAddEndPuncttrue
\mciteSetBstMidEndSepPunct{\mcitedefaultmidpunct}
{\mcitedefaultendpunct}{\mcitedefaultseppunct}\relax
\EndOfBibitem
\bibitem{Takeuchi:2017btl}
M.~Takeuchi, Y.~Uesaka, and M.~Yamanaka,
  \ifthenelse{\boolean{articletitles}}{\emph{{Higgs mediated CLFV processes
  $\mu (e) N \to \tau N X$ via gluon operators}},
  }{}\href{https://doi.org/10.1016/j.physletb.2017.06.054}{Phys.\ Lett.\ B
  \textbf{772} (2017) 279},
  \href{http://arxiv.org/abs/1705.01059}{{\normalfont\ttfamily
  arXiv:1705.01059}}\relax
\mciteBstWouldAddEndPuncttrue
\mciteSetBstMidEndSepPunct{\mcitedefaultmidpunct}
{\mcitedefaultendpunct}{\mcitedefaultseppunct}\relax
\EndOfBibitem
\bibitem{Hazard:2016fnc}
D.~E. Hazard and A.~A. Petrov,
  \ifthenelse{\boolean{articletitles}}{\emph{{Lepton flavor violating
  quarkonium decays}},
  }{}\href{https://doi.org/10.1103/PhysRevD.94.074023}{Phys.\ Rev.\ D
  \textbf{94} (2016) 074023},
  \href{http://arxiv.org/abs/1607.00815}{{\normalfont\ttfamily
  arXiv:1607.00815}}\relax
\mciteBstWouldAddEndPuncttrue
\mciteSetBstMidEndSepPunct{\mcitedefaultmidpunct}
{\mcitedefaultendpunct}{\mcitedefaultseppunct}\relax
\EndOfBibitem
\bibitem{Celis:2014asa}
A.~Celis, V.~Cirigliano, and E.~Passemar,
  \ifthenelse{\boolean{articletitles}}{\emph{{Model-discriminating power of
  lepton flavor violating $\tau$ decays}},
  }{}\href{https://doi.org/10.1103/PhysRevD.89.095014}{Phys.\ Rev.\ D
  \textbf{89} (2014) 095014},
  \href{http://arxiv.org/abs/1403.5781}{{\normalfont\ttfamily
  arXiv:1403.5781}}\relax
\mciteBstWouldAddEndPuncttrue
\mciteSetBstMidEndSepPunct{\mcitedefaultmidpunct}
{\mcitedefaultendpunct}{\mcitedefaultseppunct}\relax
\EndOfBibitem
\bibitem{Celis:2013xja}
A.~Celis, V.~Cirigliano, and E.~Passemar,
  \ifthenelse{\boolean{articletitles}}{\emph{{Lepton flavor violation in the
  Higgs sector and the role of hadronic $\tau$-lepton decays}},
  }{}\href{https://doi.org/10.1103/PhysRevD.89.013008}{Phys.\ Rev.\ D
  \textbf{89} (2014) 013008},
  \href{http://arxiv.org/abs/1309.3564}{{\normalfont\ttfamily
  arXiv:1309.3564}}\relax
\mciteBstWouldAddEndPuncttrue
\mciteSetBstMidEndSepPunct{\mcitedefaultmidpunct}
{\mcitedefaultendpunct}{\mcitedefaultseppunct}\relax
\EndOfBibitem
\bibitem{Petrov:2013vka}
A.~A. Petrov and D.~V. Zhuridov,
  \ifthenelse{\boolean{articletitles}}{\emph{{Lepton flavor-violating
  transitions in effective field theory and gluonic operators}},
  }{}\href{https://doi.org/10.1103/PhysRevD.89.033005}{Phys.\ Rev.\ D
  \textbf{89} (2014) 033005},
  \href{http://arxiv.org/abs/1308.6561}{{\normalfont\ttfamily
  arXiv:1308.6561}}\relax
\mciteBstWouldAddEndPuncttrue
\mciteSetBstMidEndSepPunct{\mcitedefaultmidpunct}
{\mcitedefaultendpunct}{\mcitedefaultseppunct}\relax
\EndOfBibitem
\bibitem{Daub:2012mu}
J.~T. Daub {\em et~al.}, \ifthenelse{\boolean{articletitles}}{\emph{{Improving
  the Hadron Physics of Non-Standard-Model Decays: Example Bounds on R-parity
  Violation}}, }{}\href{https://doi.org/10.1007/JHEP01(2013)179}{JHEP
  \textbf{01} (2013) 179},
  \href{http://arxiv.org/abs/1212.4408}{{\normalfont\ttfamily
  arXiv:1212.4408}}\relax
\mciteBstWouldAddEndPuncttrue
\mciteSetBstMidEndSepPunct{\mcitedefaultmidpunct}
{\mcitedefaultendpunct}{\mcitedefaultseppunct}\relax
\EndOfBibitem
\bibitem{Han:2010sa}
T.~Han, I.~Lewis, and M.~Sher,
  \ifthenelse{\boolean{articletitles}}{\emph{{$\mu\tau$ Production at Hadron
  Colliders}}, }{}\href{https://doi.org/10.1007/JHEP03(2010)090}{JHEP
  \textbf{03} (2010) 090},
  \href{http://arxiv.org/abs/1001.0022}{{\normalfont\ttfamily
  arXiv:1001.0022}}\relax
\mciteBstWouldAddEndPuncttrue
\mciteSetBstMidEndSepPunct{\mcitedefaultmidpunct}
{\mcitedefaultendpunct}{\mcitedefaultseppunct}\relax
\EndOfBibitem
\bibitem{Gonderinger:2010yn}
M.~Gonderinger and M.~J. Ramsey-Musolf,
  \ifthenelse{\boolean{articletitles}}{\emph{{Electron-to-Tau Lepton Flavor
  Violation at the Electron-Ion Collider}},
  }{}\href{https://doi.org/10.1007/JHEP11(2010)045}{JHEP \textbf{11} (2010)
  045}, \href{http://arxiv.org/abs/1006.5063}{{\normalfont\ttfamily
  arXiv:1006.5063}}, [Erratum: JHEP 05, 047 (2012)]\relax
\mciteBstWouldAddEndPuncttrue
\mciteSetBstMidEndSepPunct{\mcitedefaultmidpunct}
{\mcitedefaultendpunct}{\mcitedefaultseppunct}\relax
\EndOfBibitem
\bibitem{Dassinger:2007ru}
B.~M. Dassinger, T.~Feldmann, T.~Mannel, and S.~Turczyk,
  \ifthenelse{\boolean{articletitles}}{\emph{{Model-independent analysis of
  lepton flavour violating tau decays}},
  }{}\href{https://doi.org/10.1088/1126-6708/2007/10/039}{JHEP \textbf{10}
  (2007) 039}, \href{http://arxiv.org/abs/0707.0988}{{\normalfont\ttfamily
  arXiv:0707.0988}}\relax
\mciteBstWouldAddEndPuncttrue
\mciteSetBstMidEndSepPunct{\mcitedefaultmidpunct}
{\mcitedefaultendpunct}{\mcitedefaultseppunct}\relax
\EndOfBibitem
\bibitem{Matsuzaki:2007hh}
A.~Matsuzaki and A.~I. Sanda,
  \ifthenelse{\boolean{articletitles}}{\emph{{Analysis of lepton flavor
  violating $\tau^\pm\to \mu^\pm \mu^\pm \mu^\mp$ decays}},
  }{}\href{https://doi.org/10.1103/PhysRevD.77.073003}{Phys.\ Rev.\ D
  \textbf{77} (2008) 073003},
  \href{http://arxiv.org/abs/0711.0792}{{\normalfont\ttfamily
  arXiv:0711.0792}}\relax
\mciteBstWouldAddEndPuncttrue
\mciteSetBstMidEndSepPunct{\mcitedefaultmidpunct}
{\mcitedefaultendpunct}{\mcitedefaultseppunct}\relax
\EndOfBibitem
\bibitem{Black:2002wh}
D.~Black, T.~Han, H.-J. He, and M.~Sher,
  \ifthenelse{\boolean{articletitles}}{\emph{{$\tau$--$\mu$ flavor violation as
  a probe of the scale of new physics}},
  }{}\href{https://doi.org/10.1103/PhysRevD.66.053002}{Phys.\ Rev.\ D
  \textbf{66} (2002) 053002},
  \href{http://arxiv.org/abs/hep-ph/0206056}{{\normalfont\ttfamily
  arXiv:hep-ph/0206056}}\relax
\mciteBstWouldAddEndPuncttrue
\mciteSetBstMidEndSepPunct{\mcitedefaultmidpunct}
{\mcitedefaultendpunct}{\mcitedefaultseppunct}\relax
\EndOfBibitem
\bibitem{Banerjee:2007is}
S.~Banerjee, B.~Pietrzyk, J.~M. Roney, and Z.~Was,
  \ifthenelse{\boolean{articletitles}}{\emph{{Tau and muon pair production
  cross-sections in electron-positron annihilations at $\sqrt{s} = 10.58$
  GeV}}, }{}\href{https://doi.org/10.1103/PhysRevD.77.054012}{Phys.\ Rev.\ D
  \textbf{77} (2008) 054012},
  \href{http://arxiv.org/abs/0706.3235}{{\normalfont\ttfamily
  arXiv:0706.3235}}\relax
\mciteBstWouldAddEndPuncttrue
\mciteSetBstMidEndSepPunct{\mcitedefaultmidpunct}
{\mcitedefaultendpunct}{\mcitedefaultseppunct}\relax
\EndOfBibitem
\bibitem{LopezCastro:2012udb}
G.~Lopez~Castro and N.~Quintero,
  \ifthenelse{\boolean{articletitles}}{\emph{{Lepton number violating four-body
  tau lepton decays}},
  }{}\href{https://doi.org/10.1103/PhysRevD.85.076006}{Phys.\ Rev.\ D
  \textbf{85} (2012) 076006},
  \href{http://arxiv.org/abs/1203.0537}{{\normalfont\ttfamily
  arXiv:1203.0537}}, [Erratum: Phys. Rev. D~86, 079904 (2012)]\relax
\mciteBstWouldAddEndPuncttrue
\mciteSetBstMidEndSepPunct{\mcitedefaultmidpunct}
{\mcitedefaultendpunct}{\mcitedefaultseppunct}\relax
\EndOfBibitem
\bibitem{Hou:2005iu}
W.-S. Hou, M.~Nagashima, and A.~Soddu,
  \ifthenelse{\boolean{articletitles}}{\emph{{Baryon number violation involving
  higher generations}},
  }{}\href{https://doi.org/10.1103/PhysRevD.72.095001}{Phys.\ Rev.\ D
  \textbf{72} (2005) 095001},
  \href{http://arxiv.org/abs/hep-ph/0509006}{{\normalfont\ttfamily
  arXiv:hep-ph/0509006}}\relax
\mciteBstWouldAddEndPuncttrue
\mciteSetBstMidEndSepPunct{\mcitedefaultmidpunct}
{\mcitedefaultendpunct}{\mcitedefaultseppunct}\relax
\EndOfBibitem
\bibitem{Pacheco:2021djh}
I.~Pacheco and P.~Roig, \ifthenelse{\boolean{articletitles}}{\emph{{Lepton
  flavor violation in the Littlest Higgs Model with T parity realizing an
  inverse seesaw}}, }{}\href{https://doi.org/10.1007/JHEP02(2022)054}{JHEP
  \textbf{02} (2022) 054},
  \href{http://arxiv.org/abs/2110.03711}{{\normalfont\ttfamily
  arXiv:2110.03711}}\relax
\mciteBstWouldAddEndPuncttrue
\mciteSetBstMidEndSepPunct{\mcitedefaultmidpunct}
{\mcitedefaultendpunct}{\mcitedefaultseppunct}\relax
\EndOfBibitem
\bibitem{Babu:2002et}
K.~S. Babu and C.~Kolda, \ifthenelse{\boolean{articletitles}}{\emph{{Higgs
  mediated $\tau \to 3 \mu$ in the supersymmetric seesaw model}},
  }{}\href{https://doi.org/10.1103/PhysRevLett.89.241802}{Phys.\ Rev.\ Lett.\
  \textbf{89} (2002) 241802},
  \href{http://arxiv.org/abs/hep-ph/0206310}{{\normalfont\ttfamily
  arXiv:hep-ph/0206310}}\relax
\mciteBstWouldAddEndPuncttrue
\mciteSetBstMidEndSepPunct{\mcitedefaultmidpunct}
{\mcitedefaultendpunct}{\mcitedefaultseppunct}\relax
\EndOfBibitem
\bibitem{Sher:2002ew}
M.~Sher, \ifthenelse{\boolean{articletitles}}{\emph{{$\tau \to \mu \eta$ in
  supersymmetric models}},
  }{}\href{https://doi.org/10.1103/PhysRevD.66.057301}{Phys.\ Rev.\ D
  \textbf{66} (2002) 057301},
  \href{http://arxiv.org/abs/hep-ph/0207136}{{\normalfont\ttfamily
  arXiv:hep-ph/0207136}}\relax
\mciteBstWouldAddEndPuncttrue
\mciteSetBstMidEndSepPunct{\mcitedefaultmidpunct}
{\mcitedefaultendpunct}{\mcitedefaultseppunct}\relax
\EndOfBibitem
\bibitem{Brignole:2004ah}
A.~Brignole and A.~Rossi, \ifthenelse{\boolean{articletitles}}{\emph{{Anatomy
  and phenomenology of mu-tau lepton flavor violation in the MSSM}},
  }{}\href{https://doi.org/10.1016/j.nuclphysb.2004.08.037}{Nucl.\ Phys.\ B
  \textbf{701} (2004) 3},
  \href{http://arxiv.org/abs/hep-ph/0404211}{{\normalfont\ttfamily
  arXiv:hep-ph/0404211}}\relax
\mciteBstWouldAddEndPuncttrue
\mciteSetBstMidEndSepPunct{\mcitedefaultmidpunct}
{\mcitedefaultendpunct}{\mcitedefaultseppunct}\relax
\EndOfBibitem
\bibitem{Goto:2007ee}
T.~Goto, Y.~Okada, T.~Shindou, and M.~Tanaka,
  \ifthenelse{\boolean{articletitles}}{\emph{{Patterns of flavor signals in
  supersymmetric models}},
  }{}\href{https://doi.org/10.1103/PhysRevD.77.095010}{Phys.\ Rev.\ D
  \textbf{77} (2008) 095010},
  \href{http://arxiv.org/abs/0711.2935}{{\normalfont\ttfamily
  arXiv:0711.2935}}\relax
\mciteBstWouldAddEndPuncttrue
\mciteSetBstMidEndSepPunct{\mcitedefaultmidpunct}
{\mcitedefaultendpunct}{\mcitedefaultseppunct}\relax
\EndOfBibitem
\bibitem{Belle2WP}
Belle II Collaboration, \ifthenelse{\boolean{articletitles}}{\emph{{Snowmass
  White Paper - Belle II physics reach and plans for the next decade and
  beyond}}, }{}2022\relax
\mciteBstWouldAddEndPuncttrue
\mciteSetBstMidEndSepPunct{\mcitedefaultmidpunct}
{\mcitedefaultendpunct}{\mcitedefaultseppunct}\relax
\EndOfBibitem
\bibitem{Amhis:2019ckw}
Heavy Flavor Averaging Group, Y.~S. Amhis {\em et~al.},
  \ifthenelse{\boolean{articletitles}}{\emph{{Averages of $b$-hadron,
  $c$-hadron, and $\tau$-lepton properties as of 2018}},
  }{}\href{http://arxiv.org/abs/1909.12524}{{\normalfont\ttfamily
  arXiv:1909.12524}}\relax
\mciteBstWouldAddEndPuncttrue
\mciteSetBstMidEndSepPunct{\mcitedefaultmidpunct}
{\mcitedefaultendpunct}{\mcitedefaultseppunct}\relax
\EndOfBibitem
\bibitem{Belle-II:2018jsg}
Belle-II, W.~Altmannshofer {\em et~al.},
  \ifthenelse{\boolean{articletitles}}{\emph{{The Belle II Physics Book}},
  }{}\href{https://doi.org/10.1093/ptep/ptz106}{PTEP \textbf{2019} (2019)
  123C01}, \href{http://arxiv.org/abs/1808.10567}{{\normalfont\ttfamily
  arXiv:1808.10567}}, [Erratum: PTEP 2020, 029201 (2020)]\relax
\mciteBstWouldAddEndPuncttrue
\mciteSetBstMidEndSepPunct{\mcitedefaultmidpunct}
{\mcitedefaultendpunct}{\mcitedefaultseppunct}\relax
\EndOfBibitem
\bibitem{LHCb:2011zfl}
LHCb, R.~Aaij {\em et~al.},
  \ifthenelse{\boolean{articletitles}}{\emph{{Measurement of $J/\psi$
  production in $pp$ collisions at $\sqrt{s}=7$~TeV}},
  }{}\href{https://doi.org/10.1140/epjc/s10052-011-1645-y}{Eur.\ Phys.\ J.\ C
  \textbf{71} (2011) 1645},
  \href{http://arxiv.org/abs/1103.0423}{{\normalfont\ttfamily
  arXiv:1103.0423}}\relax
\mciteBstWouldAddEndPuncttrue
\mciteSetBstMidEndSepPunct{\mcitedefaultmidpunct}
{\mcitedefaultendpunct}{\mcitedefaultseppunct}\relax
\EndOfBibitem
\bibitem{LHCb:2013xam}
LHCb, R.~Aaij {\em et~al.}, \ifthenelse{\boolean{articletitles}}{\emph{{Prompt
  charm production in pp collisions at $\sqrt{s}=7$~TeV}},
  }{}\href{https://doi.org/10.1016/j.nuclphysb.2013.02.010}{Nucl.\ Phys.\ B
  \textbf{871} (2013) 1},
  \href{http://arxiv.org/abs/1302.2864}{{\normalfont\ttfamily
  arXiv:1302.2864}}\relax
\mciteBstWouldAddEndPuncttrue
\mciteSetBstMidEndSepPunct{\mcitedefaultmidpunct}
{\mcitedefaultendpunct}{\mcitedefaultseppunct}\relax
\EndOfBibitem
\bibitem{ParticleDataGroup:2014cgo}
Particle Data Group, K.~A. Olive {\em et~al.},
  \ifthenelse{\boolean{articletitles}}{\emph{{Review of Particle Physics}},
  }{}\href{https://doi.org/10.1088/1674-1137/38/9/090001}{Chin.\ Phys.\ C
  \textbf{38} (2014) 090001}\relax
\mciteBstWouldAddEndPuncttrue
\mciteSetBstMidEndSepPunct{\mcitedefaultmidpunct}
{\mcitedefaultendpunct}{\mcitedefaultseppunct}\relax
\EndOfBibitem
\bibitem{LHCb:2013fsr}
LHCb, R.~Aaij {\em et~al.},
  \ifthenelse{\boolean{articletitles}}{\emph{{Searches for violation of lepton
  flavour and baryon number in tau lepton decays at LHCb}},
  }{}\href{https://doi.org/10.1016/j.physletb.2013.05.063}{Phys.\ Lett.\ B
  \textbf{724} (2013) 36},
  \href{http://arxiv.org/abs/1304.4518}{{\normalfont\ttfamily
  arXiv:1304.4518}}\relax
\mciteBstWouldAddEndPuncttrue
\mciteSetBstMidEndSepPunct{\mcitedefaultmidpunct}
{\mcitedefaultendpunct}{\mcitedefaultseppunct}\relax
\EndOfBibitem
\bibitem{LHCb:2014kws}
LHCb, R.~Aaij {\em et~al.}, \ifthenelse{\boolean{articletitles}}{\emph{{Search
  for the lepton flavour violating decay \ensuremath{\tau}$^{-}$
  \textrightarrow{} \ensuremath{\mu}$^{-}$ \ensuremath{\mu}$^{+}$
  \ensuremath{\mu}$^{-}$}},
  }{}\href{https://doi.org/10.1007/JHEP02(2015)121}{JHEP \textbf{02} (2015)
  121}, \href{http://arxiv.org/abs/1409.8548}{{\normalfont\ttfamily
  arXiv:1409.8548}}\relax
\mciteBstWouldAddEndPuncttrue
\mciteSetBstMidEndSepPunct{\mcitedefaultmidpunct}
{\mcitedefaultendpunct}{\mcitedefaultseppunct}\relax
\EndOfBibitem
\bibitem{Aaij:2019okb}
LHCb, R.~Aaij {\em et~al.}, \ifthenelse{\boolean{articletitles}}{\emph{{Search
  for the lepton-flavour-violating decays $B^{0}_{s}\to\tau^{\pm}\mu^{\mp}$ and
  $B^{0}\to\tau^{\pm}\mu^{\mp}$}},
  }{}\href{https://doi.org/10.1103/PhysRevLett.123.211801}{Phys.\ Rev.\ Lett.\
  \textbf{123} (2019) 211801},
  \href{http://arxiv.org/abs/1905.06614}{{\normalfont\ttfamily
  arXiv:1905.06614}}\relax
\mciteBstWouldAddEndPuncttrue
\mciteSetBstMidEndSepPunct{\mcitedefaultmidpunct}
{\mcitedefaultendpunct}{\mcitedefaultseppunct}\relax
\EndOfBibitem
\bibitem{LHCb:2020khb}
LHCb, R.~Aaij {\em et~al.}, \ifthenelse{\boolean{articletitles}}{\emph{{Search
  for the lepton flavour violating decay $B^+ \rightarrow K^+ \mu^- \tau^+$
  using $B_{s2}^{*0}$ decays}},
  }{}\href{https://doi.org/10.1007/JHEP06(2020)129}{JHEP \textbf{06} (2020)
  129}, \href{http://arxiv.org/abs/2003.04352}{{\normalfont\ttfamily
  arXiv:2003.04352}}\relax
\mciteBstWouldAddEndPuncttrue
\mciteSetBstMidEndSepPunct{\mcitedefaultmidpunct}
{\mcitedefaultendpunct}{\mcitedefaultseppunct}\relax
\EndOfBibitem
\bibitem{LHCb:2018ukt}
LHCb, R.~Aaij {\em et~al.}, \ifthenelse{\boolean{articletitles}}{\emph{{Search
  for lepton-flavour-violating decays of Higgs-like bosons}},
  }{}\href{https://doi.org/10.1140/epjc/s10052-018-6386-8}{Eur.\ Phys.\ J.\ C
  \textbf{78} (2018) 1008},
  \href{http://arxiv.org/abs/1808.07135}{{\normalfont\ttfamily
  arXiv:1808.07135}}\relax
\mciteBstWouldAddEndPuncttrue
\mciteSetBstMidEndSepPunct{\mcitedefaultmidpunct}
{\mcitedefaultendpunct}{\mcitedefaultseppunct}\relax
\EndOfBibitem
\bibitem{ATLAS:2019pmk}
{ATLAS Collaboration}, \ifthenelse{\boolean{articletitles}}{\emph{{Searches for
  lepton-flavour-violating decays of the Higgs boson in $\sqrt{s}=13$ TeV pp
  collisions with the ATLAS detector}},
  }{}\href{https://doi.org/10.1016/j.physletb.2019.135069}{Phys.\ Lett.\ B
  \textbf{800} (2020) 135069},
  \href{http://arxiv.org/abs/1907.06131}{{\normalfont\ttfamily
  arXiv:1907.06131}}\relax
\mciteBstWouldAddEndPuncttrue
\mciteSetBstMidEndSepPunct{\mcitedefaultmidpunct}
{\mcitedefaultendpunct}{\mcitedefaultseppunct}\relax
\EndOfBibitem
\bibitem{CMS:2021rsq}
{CMS Collaboration}, \ifthenelse{\boolean{articletitles}}{\emph{{Search for
  lepton-flavor violating decays of the Higgs boson in the $\mu\tau$ and
  e$\tau$ final states in proton-proton collisions at $\sqrt{s}$ = 13 TeV}},
  }{}\href{https://doi.org/10.1103/PhysRevD.104.032013}{Phys.\ Rev.\ D
  \textbf{104} (2021) 032013},
  \href{http://arxiv.org/abs/2105.03007}{{\normalfont\ttfamily
  arXiv:2105.03007}}\relax
\mciteBstWouldAddEndPuncttrue
\mciteSetBstMidEndSepPunct{\mcitedefaultmidpunct}
{\mcitedefaultendpunct}{\mcitedefaultseppunct}\relax
\EndOfBibitem
\bibitem{cmst3m}
{CMS Collaboration}, \ifthenelse{\boolean{articletitles}}{\emph{{Search for the
  lepton flavor violating decay $\tau$ $\to$ 3$\mu$ in proton-proton collisions
  at $\sqrt{s} =$ 13 TeV}},
  }{}\href{https://doi.org/10.1007/JHEP01(2021)163}{JHEP \textbf{01} (2021)
  163}, \href{http://arxiv.org/abs/2007.05658}{{\normalfont\ttfamily
  arXiv:2007.05658}}\relax
\mciteBstWouldAddEndPuncttrue
\mciteSetBstMidEndSepPunct{\mcitedefaultmidpunct}
{\mcitedefaultendpunct}{\mcitedefaultseppunct}\relax
\EndOfBibitem
\bibitem{ATLAS:2016jts}
{ATLAS Collaboration}, \ifthenelse{\boolean{articletitles}}{\emph{{Probing
  lepton flavour violation via neutrinoless $\tau\longrightarrow 3\mu$ decays
  with the ATLAS detector}},
  }{}\href{https://doi.org/10.1140/epjc/s10052-016-4041-9}{Eur.\ Phys.\ J.\ C
  \textbf{76} (2016) 232},
  \href{http://arxiv.org/abs/1601.03567}{{\normalfont\ttfamily
  arXiv:1601.03567}}\relax
\mciteBstWouldAddEndPuncttrue
\mciteSetBstMidEndSepPunct{\mcitedefaultmidpunct}
{\mcitedefaultendpunct}{\mcitedefaultseppunct}\relax
\EndOfBibitem
\bibitem{EXOT-2020-28}
{ATLAS Collaboration}, \ifthenelse{\boolean{articletitles}}{\emph{{Search for
  lepton-flavor-violation in \(Z\)-boson decays with \(\tau\)-leptons with the
  ATLAS detector}},
  }{}\href{https://doi.org/10.1103/PhysRevLett.127.271801}{Phys.\ Rev.\ Lett
  \textbf{127} (2021) 271801},
  \href{http://arxiv.org/abs/2105.12491}{{\normalfont\ttfamily
  arXiv:2105.12491}}\relax
\mciteBstWouldAddEndPuncttrue
\mciteSetBstMidEndSepPunct{\mcitedefaultmidpunct}
{\mcitedefaultendpunct}{\mcitedefaultseppunct}\relax
\EndOfBibitem
\bibitem{EXOT-2018-36}
{ATLAS Collaboration},
  \ifthenelse{\boolean{articletitles}}{\emph{{Charged-lepton-flavour violation
  at the LHC: a search for \(Z\to e\tau/\mu\tau\) decays with the ATLAS
  detector}}, }{}\href{https://doi.org/10.1038/s41567-021-01225-z}{Nature
  Phys.\  \textbf{17} (2021) 819},
  \href{http://arxiv.org/abs/2010.02566}{{\normalfont\ttfamily
  arXiv:2010.02566}}\relax
\mciteBstWouldAddEndPuncttrue
\mciteSetBstMidEndSepPunct{\mcitedefaultmidpunct}
{\mcitedefaultendpunct}{\mcitedefaultseppunct}\relax
\EndOfBibitem
\bibitem{HIGG-2015-09}
{ATLAS Collaboration}, \ifthenelse{\boolean{articletitles}}{\emph{{Search for
  lepton-flavour-violating decays of the Higgs and \(Z\) bosons with the ATLAS
  detector}}, }{}\href{https://doi.org/10.1140/epjc/s10052-017-4624-0}{Eur.\
  Phys.\ J.\ C \textbf{77} (2017) 70},
  \href{http://arxiv.org/abs/1604.07730}{{\normalfont\ttfamily
  arXiv:1604.07730}}\relax
\mciteBstWouldAddEndPuncttrue
\mciteSetBstMidEndSepPunct{\mcitedefaultmidpunct}
{\mcitedefaultendpunct}{\mcitedefaultseppunct}\relax
\EndOfBibitem
\bibitem{STCF}
H.~p.\ Peng, \ifthenelse{\boolean{articletitles}}{\emph{{Experimental Program
  for Super Tau-Charm Facility}},
  }{}\href{https://doi.org/https://indico.ihep.ac.cn/event/12805/overview}{talk
  at FPCP 2021, Shanghai, China, 7-11 June, 2021 }\relax
\mciteBstWouldAddEndPuncttrue
\mciteSetBstMidEndSepPunct{\mcitedefaultmidpunct}
{\mcitedefaultendpunct}{\mcitedefaultseppunct}\relax
\EndOfBibitem
\bibitem{Luo:2019gri}
Q.~Luo, \ifthenelse{\boolean{articletitles}}{\emph{{Progress of Preliminary
  Work for the Accelerators of a 2--7~GeV Super Tau Charm Facility at China}},
  }{}\relax
\mciteBstWouldAddEndPuncttrue
\mciteSetBstMidEndSepPunct{\mcitedefaultmidpunct}
{\mcitedefaultendpunct}{\mcitedefaultseppunct}\relax
\EndOfBibitem
\bibitem{STCF:clfv}
X.~R. Zhou, \ifthenelse{\boolean{articletitles}}{\emph{{Tau LFV decays: Super
  Tau Charm Factory, Snowmass-2021 CLFV workshop}}, }{}Snowmass-2021 CLFV
  workshop, USA, Jul 23, 2020, "\url{https://indico.\ fnal.\
  gov/event/44457/}"\relax
\mciteBstWouldAddEndPuncttrue
\mciteSetBstMidEndSepPunct{\mcitedefaultmidpunct}
{\mcitedefaultendpunct}{\mcitedefaultseppunct}\relax
\EndOfBibitem
\bibitem{Shi:2020nrf}
X.-D. Shi, X.-R. Zhou, X.-S. Qin, and H.-P. Peng,
  \ifthenelse{\boolean{articletitles}}{\emph{{A fast simulation package for
  STCF detector}},
  }{}\href{https://doi.org/10.1088/1748-0221/16/03/P03029}{JINST \textbf{16}
  (2021) P03029}, \href{http://arxiv.org/abs/2011.01654}{{\normalfont\ttfamily
  arXiv:2011.01654}}\relax
\mciteBstWouldAddEndPuncttrue
\mciteSetBstMidEndSepPunct{\mcitedefaultmidpunct}
{\mcitedefaultendpunct}{\mcitedefaultseppunct}\relax
\EndOfBibitem
\bibitem{LHCb:2018roe}
LHCb, R.~Aaij {\em et~al.}, \ifthenelse{\boolean{articletitles}}{\emph{{Physics
  case for an LHCb Upgrade II - Opportunities in flavour physics, and beyond,
  in the HL-LHC era}},
  }{}\href{http://arxiv.org/abs/1808.08865}{{\normalfont\ttfamily
  arXiv:1808.08865}}\relax
\mciteBstWouldAddEndPuncttrue
\mciteSetBstMidEndSepPunct{\mcitedefaultmidpunct}
{\mcitedefaultendpunct}{\mcitedefaultseppunct}\relax
\EndOfBibitem
\bibitem{cmsupt3m}
{CMS Collaboration}, \ifthenelse{\boolean{articletitles}}{\emph{{ The Phase-2
  Upgrade of the CMS Muon Detectors}}, }{}{CERN-LHCC-2017-012} (2017),
  \url{https://cds.cern.ch/record/2283189}\relax
\mciteBstWouldAddEndPuncttrue
\mciteSetBstMidEndSepPunct{\mcitedefaultmidpunct}
{\mcitedefaultendpunct}{\mcitedefaultseppunct}\relax
\EndOfBibitem
\bibitem{atlast3m}
{ATLAS Collaboration}, \ifthenelse{\boolean{articletitles}}{\emph{{Prospects
  for lepton flavour violation measurements in $\tau\rightarrow 3\mu$ decays
  with the ATLAS detector at the HL-LHC}}, }{}{ATL-PHYS-PUB-2018-032} (2018),
  \url{https://cds.cern.ch/record/2647956}\relax
\mciteBstWouldAddEndPuncttrue
\mciteSetBstMidEndSepPunct{\mcitedefaultmidpunct}
{\mcitedefaultendpunct}{\mcitedefaultseppunct}\relax
\EndOfBibitem
\bibitem{ZEUS:2005nsy}
ZEUS, S.~Chekanov {\em et~al.},
  \ifthenelse{\boolean{articletitles}}{\emph{{Search for lepton-flavor
  violation at HERA}},
  }{}\href{https://doi.org/10.1140/epjc/s2005-02399-1}{Eur.\ Phys.\ J.\ C
  \textbf{44} (2005) 463},
  \href{http://arxiv.org/abs/hep-ex/0501070}{{\normalfont\ttfamily
  arXiv:hep-ex/0501070}}\relax
\mciteBstWouldAddEndPuncttrue
\mciteSetBstMidEndSepPunct{\mcitedefaultmidpunct}
{\mcitedefaultendpunct}{\mcitedefaultseppunct}\relax
\EndOfBibitem
\bibitem{H1:2007dum}
H1, A.~Aktas {\em et~al.}, \ifthenelse{\boolean{articletitles}}{\emph{{Search
  for lepton flavour violation in ep collisions at HERA}},
  }{}\href{https://doi.org/10.1140/epjc/s10052-007-0440-2}{Eur.\ Phys.\ J.\ C
  \textbf{52} (2007) 833},
  \href{http://arxiv.org/abs/hep-ex/0703004}{{\normalfont\ttfamily
  arXiv:hep-ex/0703004}}\relax
\mciteBstWouldAddEndPuncttrue
\mciteSetBstMidEndSepPunct{\mcitedefaultmidpunct}
{\mcitedefaultendpunct}{\mcitedefaultseppunct}\relax
\EndOfBibitem
\bibitem{ZEUS2012}
ZEUS Collaboration, H.~Abramowicz and et.\ al.\,
  \ifthenelse{\boolean{articletitles}}{\emph{Search for first-generation
  leptoquarks at hera},
  }{}\href{https://doi.org/10.1103/PhysRevD.86.012005}{Phys.\ Rev.\ D
  \textbf{86} (2012) 012005}\relax
\mciteBstWouldAddEndPuncttrue
\mciteSetBstMidEndSepPunct{\mcitedefaultmidpunct}
{\mcitedefaultendpunct}{\mcitedefaultseppunct}\relax
\EndOfBibitem
\bibitem{ZEUS2019}
ZEUS Collaboration, H.~Abramowicz and et.\ al.\,
  \ifthenelse{\boolean{articletitles}}{\emph{Limits on contact interactions and
  leptoquarks at hera},
  }{}\href{https://doi.org/10.1103/PhysRevD.99.092006}{Phys.\ Rev.\ D
  \textbf{99} (2019) 092006}\relax
\mciteBstWouldAddEndPuncttrue
\mciteSetBstMidEndSepPunct{\mcitedefaultmidpunct}
{\mcitedefaultendpunct}{\mcitedefaultseppunct}\relax
\EndOfBibitem
\bibitem{BaBar:2009hkt}
BaBar, B.~Aubert {\em et~al.},
  \ifthenelse{\boolean{articletitles}}{\emph{{Searches for Lepton Flavor
  Violation in the Decays $\tau^+\to e^+\gamma$ and $\tau^+\to\mu^+\gamma$}},
  }{}\href{https://doi.org/10.1103/PhysRevLett.104.021802}{Phys.\ Rev.\ Lett.\
  \textbf{104} (2010) 021802},
  \href{http://arxiv.org/abs/0908.2381}{{\normalfont\ttfamily
  arXiv:0908.2381}}\relax
\mciteBstWouldAddEndPuncttrue
\mciteSetBstMidEndSepPunct{\mcitedefaultmidpunct}
{\mcitedefaultendpunct}{\mcitedefaultseppunct}\relax
\EndOfBibitem
\bibitem{ATLAS:2018mrn}
ATLAS, M.~Aaboud {\em et~al.},
  \ifthenelse{\boolean{articletitles}}{\emph{{Search for lepton-flavor
  violation in different-flavor, high-mass final states in $pp$ collisions at
  $\sqrt s=13 $ TeV with the ATLAS detector}},
  }{}\href{https://doi.org/10.1103/PhysRevD.98.092008}{Phys.\ Rev.\ D
  \textbf{98} (2018) 092008},
  \href{http://arxiv.org/abs/1807.06573}{{\normalfont\ttfamily
  arXiv:1807.06573}}\relax
\mciteBstWouldAddEndPuncttrue
\mciteSetBstMidEndSepPunct{\mcitedefaultmidpunct}
{\mcitedefaultendpunct}{\mcitedefaultseppunct}\relax
\EndOfBibitem
\bibitem{Bellagamba:2001fk}
L.~Bellagamba, \ifthenelse{\boolean{articletitles}}{\emph{{LQGENEP: A
  leptoquark generator for e p scattering}},
  }{}\href{https://doi.org/10.1016/S0010-4655(01)00295-8}{Comput.\ Phys.\
  Commun.\  \textbf{141} (2001) 83}\relax
\mciteBstWouldAddEndPuncttrue
\mciteSetBstMidEndSepPunct{\mcitedefaultmidpunct}
{\mcitedefaultendpunct}{\mcitedefaultseppunct}\relax
\EndOfBibitem
\bibitem{Abada:2019lih}
Future Circular Collider Study, A.~Abada {\em et~al.},
  \ifthenelse{\boolean{articletitles}}{\emph{{FCC Physics Opportunities}:
  {Future Circular Collider Conceptual Design Report Volume 1}},
  }{}\href{https://doi.org/10.1140/epjc/s10052-019-6904-3}{Eur.\ Phys.\ J.\ C
  \textbf{79} (2019) 474}\relax
\mciteBstWouldAddEndPuncttrue
\mciteSetBstMidEndSepPunct{\mcitedefaultmidpunct}
{\mcitedefaultendpunct}{\mcitedefaultseppunct}\relax
\EndOfBibitem
\bibitem{FCC:2018evy}
Future Circular Collider Study, A.~Abada {\em et~al.},
  \ifthenelse{\boolean{articletitles}}{\emph{{FCC-ee: The Lepton Collider}:
  {Future Circular Collider Conceptual Design Report Volume 2}},
  }{}\href{https://doi.org/10.1140/epjst/e2019-900045-4}{Eur.\ Phys.\ J.\ ST
  \textbf{228} (2019) 261}\relax
\mciteBstWouldAddEndPuncttrue
\mciteSetBstMidEndSepPunct{\mcitedefaultmidpunct}
{\mcitedefaultendpunct}{\mcitedefaultseppunct}\relax
\EndOfBibitem
\bibitem{Blondel:2021ema}
A.~Blondel and P.~Janot, \ifthenelse{\boolean{articletitles}}{\emph{{FCC-ee
  overview: new opportunities create new challenges, {\rm in A future Higgs and
  Electroweak factory (FCC): Challenges towards discovery, EPJ+ special issue,
  Focus on FCC-ee}}},
  }{}\href{http://arxiv.org/abs/2106.13885}{{\normalfont\ttfamily
  arXiv:2106.13885}}\relax
\mciteBstWouldAddEndPuncttrue
\mciteSetBstMidEndSepPunct{\mcitedefaultmidpunct}
{\mcitedefaultendpunct}{\mcitedefaultseppunct}\relax
\EndOfBibitem
\bibitem{ALEPH:2005ab}
ALEPH, DELPHI, L3, OPAL, SLD, LEP Electroweak Working Group, SLD Electroweak
  Group, SLD Heavy Flavour Group, S.~Schael {\em et~al.},
  \ifthenelse{\boolean{articletitles}}{\emph{{Precision electroweak
  measurements on the $Z$ resonance}},
  }{}\href{https://doi.org/10.1016/j.physrep.2005.12.006}{Phys.\ Rept.\
  \textbf{427} (2006) 257},
  \href{http://arxiv.org/abs/hep-ex/0509008}{{\normalfont\ttfamily
  arXiv:hep-ex/0509008}}\relax
\mciteBstWouldAddEndPuncttrue
\mciteSetBstMidEndSepPunct{\mcitedefaultmidpunct}
{\mcitedefaultendpunct}{\mcitedefaultseppunct}\relax
\EndOfBibitem
\bibitem{Pich:2020qna}
A.~Pich, \ifthenelse{\boolean{articletitles}}{\emph{{Challenges for tau physics
  at the TeraZ, {\rm in A future Higgs and Electroweak factory (FCC):
  Challenges towards discovery, EPJ+ special issue, Focus on FCC-ee}}},
  }{}\href{http://arxiv.org/abs/2012.07099}{{\normalfont\ttfamily
  arXiv:2012.07099}}\relax
\mciteBstWouldAddEndPuncttrue
\mciteSetBstMidEndSepPunct{\mcitedefaultmidpunct}
{\mcitedefaultendpunct}{\mcitedefaultseppunct}\relax
\EndOfBibitem
\bibitem{Dam:2018rfz}
M.~Dam, \ifthenelse{\boolean{articletitles}}{\emph{{Tau-lepton Physics at the
  FCC-ee circular e$^+$e$^-$ Collider}},
  }{}\href{https://doi.org/10.21468/SciPostPhysProc.1.041}{SciPost Phys.\
  Proc.\  \textbf{1} (2019) 041},
  \href{http://arxiv.org/abs/1811.09408}{{\normalfont\ttfamily
  arXiv:1811.09408}}\relax
\mciteBstWouldAddEndPuncttrue
\mciteSetBstMidEndSepPunct{\mcitedefaultmidpunct}
{\mcitedefaultendpunct}{\mcitedefaultseppunct}\relax
\EndOfBibitem
\bibitem{Aleksa:2021ztd}
M.~Aleksa {\em et~al.}, \ifthenelse{\boolean{articletitles}}{\emph{{Calorimetry
  at FCC-ee}}, }{}\href{https://doi.org/10.1140/epjp/s13360-021-02034-2}{Eur.\
  Phys.\ J.\ Plus \textbf{136} (2021) 1066},
  \href{http://arxiv.org/abs/2109.00391}{{\normalfont\ttfamily
  arXiv:2109.00391}}\relax
\mciteBstWouldAddEndPuncttrue
\mciteSetBstMidEndSepPunct{\mcitedefaultmidpunct}
{\mcitedefaultendpunct}{\mcitedefaultseppunct}\relax
\EndOfBibitem
\bibitem{Belle:2021ysv}
Belle, A.~Abdesselam {\em et~al.},
  \ifthenelse{\boolean{articletitles}}{\emph{{Search for
  lepton-flavor-violating tau-lepton decays to $\ell\gamma$ at Belle}},
  }{}\href{https://doi.org/10.1007/JHEP10(2021)019}{JHEP \textbf{10} (2021)
  19}, \href{http://arxiv.org/abs/2103.12994}{{\normalfont\ttfamily
  arXiv:2103.12994}}\relax
\mciteBstWouldAddEndPuncttrue
\mciteSetBstMidEndSepPunct{\mcitedefaultmidpunct}
{\mcitedefaultendpunct}{\mcitedefaultseppunct}\relax
\EndOfBibitem
\bibitem{Belle:2007cio}
Belle, Y.~Miyazaki {\em et~al.},
  \ifthenelse{\boolean{articletitles}}{\emph{{Search for lepton flavor
  violating tau- decays into $\ell^- \eta$, $\ell^- \eta^\prime$ and $\ell^-
  \pi^0$}}, }{}\href{https://doi.org/10.1016/j.physletb.2007.03.027}{Phys.\
  Lett.\ B \textbf{648} (2007) 341},
  \href{http://arxiv.org/abs/hep-ex/0703009}{{\normalfont\ttfamily
  arXiv:hep-ex/0703009}}\relax
\mciteBstWouldAddEndPuncttrue
\mciteSetBstMidEndSepPunct{\mcitedefaultmidpunct}
{\mcitedefaultendpunct}{\mcitedefaultseppunct}\relax
\EndOfBibitem
\bibitem{BaBar:2006jhm}
BaBar, B.~Aubert {\em et~al.},
  \ifthenelse{\boolean{articletitles}}{\emph{{Search for Lepton Flavor
  Violating Decays $\tau^\pm \to \ell^\pm \pi^0$, $\ell^\pm \eta$, $\ell^\pm
  \eta^\prime$}},
  }{}\href{https://doi.org/10.1103/PhysRevLett.98.061803}{Phys.\ Rev.\ Lett.\
  \textbf{98} (2007) 061803},
  \href{http://arxiv.org/abs/hep-ex/0610067}{{\normalfont\ttfamily
  arXiv:hep-ex/0610067}}\relax
\mciteBstWouldAddEndPuncttrue
\mciteSetBstMidEndSepPunct{\mcitedefaultmidpunct}
{\mcitedefaultendpunct}{\mcitedefaultseppunct}\relax
\EndOfBibitem
\bibitem{Belle:2010rxj}
Belle, Y.~Miyazaki {\em et~al.},
  \ifthenelse{\boolean{articletitles}}{\emph{{Search for Lepton Flavor
  Violating tau- Decays into ${\ell^-K^0_S}$ and ${\ell^-K^0_SK^0_S}$}},
  }{}\href{https://doi.org/10.1016/j.physletb.2010.07.012}{Phys.\ Lett.\ B
  \textbf{692} (2010) 4},
  \href{http://arxiv.org/abs/1003.1183}{{\normalfont\ttfamily
  arXiv:1003.1183}}\relax
\mciteBstWouldAddEndPuncttrue
\mciteSetBstMidEndSepPunct{\mcitedefaultmidpunct}
{\mcitedefaultendpunct}{\mcitedefaultseppunct}\relax
\EndOfBibitem
\bibitem{BaBar:2009qra}
BaBar, B.~Aubert {\em et~al.},
  \ifthenelse{\boolean{articletitles}}{\emph{{Search for Lepton Flavor
  Violating Decays $\tau^- \to \ell^- K^0_S$ with the BABAR Experiment}},
  }{}\href{https://doi.org/10.1103/PhysRevD.79.012004}{Phys.\ Rev.\ D
  \textbf{79} (2009) 012004},
  \href{http://arxiv.org/abs/0812.3804}{{\normalfont\ttfamily
  arXiv:0812.3804}}\relax
\mciteBstWouldAddEndPuncttrue
\mciteSetBstMidEndSepPunct{\mcitedefaultmidpunct}
{\mcitedefaultendpunct}{\mcitedefaultseppunct}\relax
\EndOfBibitem
\bibitem{Belle:2008pdf}
Belle, Y.~Miyazaki {\em et~al.},
  \ifthenelse{\boolean{articletitles}}{\emph{{Search for
  Lepton-Flavor-Violating tau Decays into Lepton and f0(980) Meson}},
  }{}\href{https://doi.org/10.1016/j.physletb.2009.01.058}{Phys.\ Lett.\ B
  \textbf{672} (2009) 317},
  \href{http://arxiv.org/abs/0810.3519}{{\normalfont\ttfamily
  arXiv:0810.3519}}\relax
\mciteBstWouldAddEndPuncttrue
\mciteSetBstMidEndSepPunct{\mcitedefaultmidpunct}
{\mcitedefaultendpunct}{\mcitedefaultseppunct}\relax
\EndOfBibitem
\bibitem{Belle:2011ogy}
Belle, Y.~Miyazaki {\em et~al.},
  \ifthenelse{\boolean{articletitles}}{\emph{{Search for
  Lepton-Flavor-Violating tau Decays into a Lepton and a Vector Meson}},
  }{}\href{https://doi.org/10.1016/j.physletb.2011.04.011}{Phys.\ Lett.\ B
  \textbf{699} (2011) 251},
  \href{http://arxiv.org/abs/1101.0755}{{\normalfont\ttfamily
  arXiv:1101.0755}}\relax
\mciteBstWouldAddEndPuncttrue
\mciteSetBstMidEndSepPunct{\mcitedefaultmidpunct}
{\mcitedefaultendpunct}{\mcitedefaultseppunct}\relax
\EndOfBibitem
\bibitem{BaBar:2009wtb}
BaBar, B.~Aubert {\em et~al.},
  \ifthenelse{\boolean{articletitles}}{\emph{{Improved limits on lepton flavor
  violating tau decays to $\ell^-$ $\phi$, $\ell^-$ $\rho$, $\ell^-$ $K^{\ast
  0}$ and $\ell^-$ $\bar{K}^{\ast0}$}},
  }{}\href{https://doi.org/10.1103/PhysRevLett.103.021801}{Phys.\ Rev.\ Lett.\
  \textbf{103} (2009) 021801},
  \href{http://arxiv.org/abs/0904.0339}{{\normalfont\ttfamily
  arXiv:0904.0339}}\relax
\mciteBstWouldAddEndPuncttrue
\mciteSetBstMidEndSepPunct{\mcitedefaultmidpunct}
{\mcitedefaultendpunct}{\mcitedefaultseppunct}\relax
\EndOfBibitem
\bibitem{BaBar:2007amy}
BaBar, B.~Aubert {\em et~al.},
  \ifthenelse{\boolean{articletitles}}{\emph{{Search for lepton flavor
  violating decays $\tau^\pm \to \ell^\pm \omega$ $(\ell = e, \mu)$}},
  }{}\href{https://doi.org/10.1103/PhysRevLett.100.071802}{Phys.\ Rev.\ Lett.\
  \textbf{100} (2008) 071802},
  \href{http://arxiv.org/abs/0711.0980}{{\normalfont\ttfamily
  arXiv:0711.0980}}\relax
\mciteBstWouldAddEndPuncttrue
\mciteSetBstMidEndSepPunct{\mcitedefaultmidpunct}
{\mcitedefaultendpunct}{\mcitedefaultseppunct}\relax
\EndOfBibitem
\bibitem{Hayasaka:2010np}
K.~Hayasaka {\em et~al.}, \ifthenelse{\boolean{articletitles}}{\emph{{Search
  for Lepton Flavor Violating Tau Decays into Three Leptons with 719 Million
  Produced $\tau^+\tau^-$ Pairs}},
  }{}\href{https://doi.org/10.1016/j.physletb.2010.03.037}{Phys.\ Lett.\ B
  \textbf{687} (2010) 139},
  \href{http://arxiv.org/abs/1001.3221}{{\normalfont\ttfamily
  arXiv:1001.3221}}\relax
\mciteBstWouldAddEndPuncttrue
\mciteSetBstMidEndSepPunct{\mcitedefaultmidpunct}
{\mcitedefaultendpunct}{\mcitedefaultseppunct}\relax
\EndOfBibitem
\bibitem{BaBar:2010axs}
BaBar, J.~P. Lees {\em et~al.},
  \ifthenelse{\boolean{articletitles}}{\emph{{Limits on tau Lepton-Flavor
  Violating Decays in three charged leptons}},
  }{}\href{https://doi.org/10.1103/PhysRevD.81.111101}{Phys.\ Rev.\ D
  \textbf{81} (2010) 111101},
  \href{http://arxiv.org/abs/1002.4550}{{\normalfont\ttfamily
  arXiv:1002.4550}}\relax
\mciteBstWouldAddEndPuncttrue
\mciteSetBstMidEndSepPunct{\mcitedefaultmidpunct}
{\mcitedefaultendpunct}{\mcitedefaultseppunct}\relax
\EndOfBibitem
\bibitem{Belle:2012unr}
Belle, Y.~Miyazaki {\em et~al.},
  \ifthenelse{\boolean{articletitles}}{\emph{{Search for
  Lepton-Flavor-Violating and Lepton-Number-Violating $\tau \to \ell h
  h^\prime$ Decay Modes}},
  }{}\href{https://doi.org/10.1016/j.physletb.2013.01.032}{Phys.\ Lett.\ B
  \textbf{719} (2013) 346},
  \href{http://arxiv.org/abs/1206.5595}{{\normalfont\ttfamily
  arXiv:1206.5595}}\relax
\mciteBstWouldAddEndPuncttrue
\mciteSetBstMidEndSepPunct{\mcitedefaultmidpunct}
{\mcitedefaultendpunct}{\mcitedefaultseppunct}\relax
\EndOfBibitem
\bibitem{BaBar:2005yvr}
BaBar, B.~Aubert {\em et~al.},
  \ifthenelse{\boolean{articletitles}}{\emph{{Search for lepton-flavor and
  lepton-number violation in the decay $\tau^- \to \ell^\mp h^\pm h^{\prime
  -}$}}, }{}\href{https://doi.org/10.1103/PhysRevLett.95.191801}{Phys.\ Rev.\
  Lett.\  \textbf{95} (2005) 191801},
  \href{http://arxiv.org/abs/hep-ex/0506066}{{\normalfont\ttfamily
  arXiv:hep-ex/0506066}}\relax
\mciteBstWouldAddEndPuncttrue
\mciteSetBstMidEndSepPunct{\mcitedefaultmidpunct}
{\mcitedefaultendpunct}{\mcitedefaultseppunct}\relax
\EndOfBibitem
\bibitem{Belle:2005exq}
Belle, Y.~Miyazaki {\em et~al.},
  \ifthenelse{\boolean{articletitles}}{\emph{{Search for lepton and baryon
  number violating $\tau^-$ decays into $\bar{\Lambda} \pi^-$ and $\Lambda
  \pi^-$}}, }{}\href{https://doi.org/10.1016/j.physletb.2005.10.024}{Phys.\
  Lett.\ B \textbf{632} (2006) 51},
  \href{http://arxiv.org/abs/hep-ex/0508044}{{\normalfont\ttfamily
  arXiv:hep-ex/0508044}}\relax
\mciteBstWouldAddEndPuncttrue
\mciteSetBstMidEndSepPunct{\mcitedefaultmidpunct}
{\mcitedefaultendpunct}{\mcitedefaultseppunct}\relax
\EndOfBibitem
\bibitem{Belle:2020lfn}
Belle, D.~Sahoo {\em et~al.},
  \ifthenelse{\boolean{articletitles}}{\emph{{Search for lepton-number- and
  baryon-number-violating tau decays at Belle}},
  }{}\href{https://doi.org/10.1103/PhysRevD.102.111101}{Phys.\ Rev.\ D
  \textbf{102} (2020) 111101},
  \href{http://arxiv.org/abs/2010.15361}{{\normalfont\ttfamily
  arXiv:2010.15361}}\relax
\mciteBstWouldAddEndPuncttrue
\mciteSetBstMidEndSepPunct{\mcitedefaultmidpunct}
{\mcitedefaultendpunct}{\mcitedefaultseppunct}\relax
\EndOfBibitem
\bibitem{BESIII:2021slj}
BESIII, M.~Ablikim {\em et~al.},
  \ifthenelse{\boolean{articletitles}}{\emph{{Search for the charged lepton
  flavor violating decay $J/\psi\to e\tau$}},
  }{}\href{https://doi.org/10.1103/PhysRevD.103.112007}{Phys.\ Rev.\ D
  \textbf{103} (2021) 112007},
  \href{http://arxiv.org/abs/2103.11540}{{\normalfont\ttfamily
  arXiv:2103.11540}}\relax
\mciteBstWouldAddEndPuncttrue
\mciteSetBstMidEndSepPunct{\mcitedefaultmidpunct}
{\mcitedefaultendpunct}{\mcitedefaultseppunct}\relax
\EndOfBibitem
\bibitem{BES:2004jiw}
BES, M.~Ablikim {\em et~al.},
  \ifthenelse{\boolean{articletitles}}{\emph{{Search for the lepton flavor
  violation processes $J / \psi \to \mu \tau$ and $e \tau$}},
  }{}\href{https://doi.org/10.1016/j.physletb.2004.08.005}{Phys.\ Lett.\ B
  \textbf{598} (2004) 172},
  \href{http://arxiv.org/abs/hep-ex/0406018}{{\normalfont\ttfamily
  arXiv:hep-ex/0406018}}\relax
\mciteBstWouldAddEndPuncttrue
\mciteSetBstMidEndSepPunct{\mcitedefaultmidpunct}
{\mcitedefaultendpunct}{\mcitedefaultseppunct}\relax
\EndOfBibitem
\bibitem{Aubert:2008cu}
BaBar, B.~Aubert {\em et~al.},
  \ifthenelse{\boolean{articletitles}}{\emph{{Searches for the decays $B^0 \to
  \ell^\pm \tau^\mp$ and $B^{+} \to \ell^{+} \nu$ ($\ell=e, \mu$) using
  hadronic tag reconstruction}},
  }{}\href{https://doi.org/10.1103/PhysRevD.77.091104}{Phys.\ Rev.\ D
  \textbf{77} (2008) 091104},
  \href{http://arxiv.org/abs/0801.0697}{{\normalfont\ttfamily
  arXiv:0801.0697}}\relax
\mciteBstWouldAddEndPuncttrue
\mciteSetBstMidEndSepPunct{\mcitedefaultmidpunct}
{\mcitedefaultendpunct}{\mcitedefaultseppunct}\relax
\EndOfBibitem
\bibitem{BaBar:2012azg}
BaBar, J.~P. Lees {\em et~al.}, \ifthenelse{\boolean{articletitles}}{\emph{{A
  search for the decay modes $B^{+-} \to h^{+-} \tau^{+-}l$}},
  }{}\href{https://doi.org/10.1103/PhysRevD.86.012004}{Phys.\ Rev.\ D
  \textbf{86} (2012) 012004},
  \href{http://arxiv.org/abs/1204.2852}{{\normalfont\ttfamily
  arXiv:1204.2852}}\relax
\mciteBstWouldAddEndPuncttrue
\mciteSetBstMidEndSepPunct{\mcitedefaultmidpunct}
{\mcitedefaultendpunct}{\mcitedefaultseppunct}\relax
\EndOfBibitem
\bibitem{Belle:2022cce}
Belle, S.~Patra {\em et~al.},
  \ifthenelse{\boolean{articletitles}}{\emph{{Search for charged lepton flavor
  violating decays of $\Upsilon(1S)$}},
  }{}\href{http://arxiv.org/abs/2201.09620}{{\normalfont\ttfamily
  arXiv:2201.09620}}\relax
\mciteBstWouldAddEndPuncttrue
\mciteSetBstMidEndSepPunct{\mcitedefaultmidpunct}
{\mcitedefaultendpunct}{\mcitedefaultseppunct}\relax
\EndOfBibitem
\bibitem{Lees:2010jk}
BaBar, J.~P. Lees {\em et~al.},
  \ifthenelse{\boolean{articletitles}}{\emph{{Search for Charged Lepton Flavor
  Violation in Narrow Upsilon Decays}},
  }{}\href{https://doi.org/10.1103/PhysRevLett.104.151802}{Phys.\ Rev.\ Lett.\
  \textbf{104} (2010) 151802},
  \href{http://arxiv.org/abs/1001.1883}{{\normalfont\ttfamily
  arXiv:1001.1883}}\relax
\mciteBstWouldAddEndPuncttrue
\mciteSetBstMidEndSepPunct{\mcitedefaultmidpunct}
{\mcitedefaultendpunct}{\mcitedefaultseppunct}\relax
\EndOfBibitem
\bibitem{Zyla:2020zbs}
Particle Data Group, P.~A. Zyla {\em et~al.},
  \ifthenelse{\boolean{articletitles}}{\emph{{Review of Particle Physics}},
  }{}\href{https://doi.org/10.1093/ptep/ptaa104}{PTEP \textbf{2020} (2020)
  083C01}\relax
\mciteBstWouldAddEndPuncttrue
\mciteSetBstMidEndSepPunct{\mcitedefaultmidpunct}
{\mcitedefaultendpunct}{\mcitedefaultseppunct}\relax
\EndOfBibitem
\bibitem{BaBar:2008pet}
BaBar, B.~Aubert {\em et~al.},
  \ifthenelse{\boolean{articletitles}}{\emph{{Searches for the decays $B^0 \to
  \ell^\pm \tau^\mp$ and $B^{+} \to \ell^{+} \nu$ (l=e, $\mu$) using hadronic
  tag reconstruction}},
  }{}\href{https://doi.org/10.1103/PhysRevD.77.091104}{Phys.\ Rev.\ D
  \textbf{77} (2008) 091104},
  \href{http://arxiv.org/abs/0801.0697}{{\normalfont\ttfamily
  arXiv:0801.0697}}\relax
\mciteBstWouldAddEndPuncttrue
\mciteSetBstMidEndSepPunct{\mcitedefaultmidpunct}
{\mcitedefaultendpunct}{\mcitedefaultseppunct}\relax
\EndOfBibitem
\bibitem{BaBar:2010vxb}
BaBar, J.~P. Lees {\em et~al.},
  \ifthenelse{\boolean{articletitles}}{\emph{{Search for Charged Lepton Flavor
  Violation in Narrow Upsilon Decays}},
  }{}\href{https://doi.org/10.1103/PhysRevLett.104.151802}{Phys.\ Rev.\ Lett.\
  \textbf{104} (2010) 151802},
  \href{http://arxiv.org/abs/1001.1883}{{\normalfont\ttfamily
  arXiv:1001.1883}}\relax
\mciteBstWouldAddEndPuncttrue
\mciteSetBstMidEndSepPunct{\mcitedefaultmidpunct}
{\mcitedefaultendpunct}{\mcitedefaultseppunct}\relax
\EndOfBibitem
\bibitem{NA62:2020fhy}
NA62, E.~Cortina~Gil {\em et~al.},
  \ifthenelse{\boolean{articletitles}}{\emph{{An investigation of the very rare
  $ {K}^{+}\to {\pi}^{+}\nu \overline{\nu} $ decay}},
  }{}\href{https://doi.org/10.1007/JHEP11(2020)042}{JHEP \textbf{11} (2020)
  042}, \href{http://arxiv.org/abs/2007.08218}{{\normalfont\ttfamily
  arXiv:2007.08218}}\relax
\mciteBstWouldAddEndPuncttrue
\mciteSetBstMidEndSepPunct{\mcitedefaultmidpunct}
{\mcitedefaultendpunct}{\mcitedefaultseppunct}\relax
\EndOfBibitem
\bibitem{KOTO:2018dsc}
KOTO, J.~K. Ahn {\em et~al.},
  \ifthenelse{\boolean{articletitles}}{\emph{{Search for the $K_L \!\to\! \pi^0
  \nu \overline{\nu}$ and $K_L \!\to\! \pi^0 X^0$ decays at the J-PARC KOTO
  experiment}}, }{}\href{https://doi.org/10.1103/PhysRevLett.122.021802}{Phys.\
  Rev.\ Lett.\  \textbf{122} (2019) 021802},
  \href{http://arxiv.org/abs/1810.09655}{{\normalfont\ttfamily
  arXiv:1810.09655}}\relax
\mciteBstWouldAddEndPuncttrue
\mciteSetBstMidEndSepPunct{\mcitedefaultmidpunct}
{\mcitedefaultendpunct}{\mcitedefaultseppunct}\relax
\EndOfBibitem
\bibitem{Belle:2017oht}
Belle, J.~Grygier {\em et~al.},
  \ifthenelse{\boolean{articletitles}}{\emph{{Search for ${\mathbf{B\to
  h\nu\bar{\nu}}}$ decays with semileptonic tagging at Belle}},
  }{}\href{https://doi.org/10.1103/PhysRevD.96.091101}{Phys.\ Rev.\ D
  \textbf{96} (2017) 091101},
  \href{http://arxiv.org/abs/1702.03224}{{\normalfont\ttfamily
  arXiv:1702.03224}}, [Addendum: Phys.Rev.D 97, 099902 (2018)]\relax
\mciteBstWouldAddEndPuncttrue
\mciteSetBstMidEndSepPunct{\mcitedefaultmidpunct}
{\mcitedefaultendpunct}{\mcitedefaultseppunct}\relax
\EndOfBibitem
\bibitem{BaBar:2013npw}
BaBar, J.~P. Lees {\em et~al.},
  \ifthenelse{\boolean{articletitles}}{\emph{{Search for $B \to K^{(*)} \nu
  \overline \nu$ and invisible quarkonium decays}},
  }{}\href{https://doi.org/10.1103/PhysRevD.87.112005}{Phys.\ Rev.\ D
  \textbf{87} (2013) 112005},
  \href{http://arxiv.org/abs/1303.7465}{{\normalfont\ttfamily
  arXiv:1303.7465}}\relax
\mciteBstWouldAddEndPuncttrue
\mciteSetBstMidEndSepPunct{\mcitedefaultmidpunct}
{\mcitedefaultendpunct}{\mcitedefaultseppunct}\relax
\EndOfBibitem
\bibitem{ATLAS:2021bdj}
ATLAS, G.~Aad {\em et~al.}, \ifthenelse{\boolean{articletitles}}{\emph{{Search
  for lepton-flavor-violation in $Z$-boson decays with $\tau$-leptons with the
  ATLAS detector}},
  }{}\href{https://doi.org/10.1103/PhysRevLett.127.271801}{Phys.\ Rev.\ Lett.\
  \textbf{127} (2022) 271801},
  \href{http://arxiv.org/abs/2105.12491}{{\normalfont\ttfamily
  arXiv:2105.12491}}\relax
\mciteBstWouldAddEndPuncttrue
\mciteSetBstMidEndSepPunct{\mcitedefaultmidpunct}
{\mcitedefaultendpunct}{\mcitedefaultseppunct}\relax
\EndOfBibitem
\bibitem{ATLAS:2020zlz}
ATLAS, G.~Aad {\em et~al.}, \ifthenelse{\boolean{articletitles}}{\emph{{Search
  for charged-lepton-flavour violation in $Z$-boson decays with the ATLAS
  detector}}, }{}\href{https://doi.org/10.1038/s41567-021-01225-z}{Nature
  Phys.\  \textbf{17} (2021) 819},
  \href{http://arxiv.org/abs/2010.02566}{{\normalfont\ttfamily
  arXiv:2010.02566}}\relax
\mciteBstWouldAddEndPuncttrue
\mciteSetBstMidEndSepPunct{\mcitedefaultmidpunct}
{\mcitedefaultendpunct}{\mcitedefaultseppunct}\relax
\EndOfBibitem
\bibitem{ATLAS:2018avw}
ATLAS, \ifthenelse{\boolean{articletitles}}{\emph{{Search for charged
  lepton-flavour violation in top-quark decays at the LHC with the ATLAS
  detector}}, }{}\relax
\mciteBstWouldAddEndPuncttrue
\mciteSetBstMidEndSepPunct{\mcitedefaultmidpunct}
{\mcitedefaultendpunct}{\mcitedefaultseppunct}\relax
\EndOfBibitem
\bibitem{Angelescu:2020uug}
A.~Angelescu, D.~A. Faroughy, and O.~Sumensari,
  \ifthenelse{\boolean{articletitles}}{\emph{{Lepton Flavor Violation and
  Dilepton Tails at the LHC}},
  }{}\href{https://doi.org/10.1140/epjc/s10052-020-8210-5}{Eur.\ Phys.\ J.\ C
  \textbf{80} (2020) 641},
  \href{http://arxiv.org/abs/2002.05684}{{\normalfont\ttfamily
  arXiv:2002.05684}}\relax
\mciteBstWouldAddEndPuncttrue
\mciteSetBstMidEndSepPunct{\mcitedefaultmidpunct}
{\mcitedefaultendpunct}{\mcitedefaultseppunct}\relax
\EndOfBibitem
\bibitem{deBlas:2019okz}
J.~de~Blas {\em et~al.},
  \ifthenelse{\boolean{articletitles}}{\emph{{$\texttt{HEPfit}$: a Code for the
  Combination of Indirect and Direct Constraints on High Energy Physics
  Models}}, }{}\href{http://arxiv.org/abs/1910.14012}{{\normalfont\ttfamily
  arXiv:1910.14012}}\relax
\mciteBstWouldAddEndPuncttrue
\mciteSetBstMidEndSepPunct{\mcitedefaultmidpunct}
{\mcitedefaultendpunct}{\mcitedefaultseppunct}\relax
\EndOfBibitem
\end{mcitethebibliography}
\setboolean{inbibliography}{false}

\end{document}